\documentclass[11pt,a4paper]{article}
\usepackage{jheppub_kim}
\usepackage{longtable}

\usepackage{epsfig}
\usepackage{amsmath}
\usepackage{graphicx}
 \usepackage{graphics}
 \usepackage[scriptsize,nooneline,hang]{caption}
\usepackage[hang,nooneline,scriptsize]{subfigure}

\usepackage{array}
\usepackage{makecell}
\begin{document}

\title{Analytical solutions of the geodesic equation in the space-time of a black hole surrounded by perfect fluid in Rastall theory}

\author[a]{Saheb Soroushfar, }
\affiliation[a]{Faculty of Technology and Mining, Yasouj University, Choram 75761-59836, Iran}
%\affiliation[b]{Research Institute for Astronomy and Astrophysics of Maragha (RIAAM), P.O. Box 55134-441, Maragha, Iran}
\author[a]{Maryam Afrooz}
%\affiliation[a]{Faculty of Technology and Mining, Yasouj University, Choram 75761-59836, Iran}
\emailAdd{soroush@yu.ac.ir} 
\emailAdd{afrouzmaryam@gmail.com}

\abstract{In this paper, we investigate 
	the geodesic motion of massive and massless test particles in the vicinity of a black hole space-time surrounded by perfect fluid (quintessence, dust, radiation, cosmological constant and phantom) in Rastall theory. 
	We obtain the full set of analytical solutions of the geodesic equation of motion 
	in the space-time of this black hole.
	For all cases of perfect fluid, we consider some different values of 
	Rastall coupling constant $k\lambda$ so that the equations of motion 
	have integer powers of $\tilde{r}$ and also can be solved analytically. 
	These analytical solutions are presented in the form of 
	elliptic and also hyperelliptic functions. 
	%These analytical solutions are presented in the form of Weierstrass
	%elliptic and Kleinian hyperelliptical sigma functions. 
	In addition, using obtained analytical solution and also figures of effective potential 
	and $L-E^2$ diagrams, we plot some examples of possibles orbits. 
	moreover we use of the angular momentum, conserved energy, electrical charge and also Rastall parameter, 
	to classify the different types of the possible gained orbits. 
	Moreover, we show that when Rastall field structure constant becomes zero ($N=0$) 
	our results are consistent with the analysis of a Reissner-Nordstr\"om black hole,
	however when both Rastall geometric parameter and electric charge vanish $(N=Q=0)$, 
	the metric and results are same as analysis of a Schwarzschild black hole.
	% but the result of vanishing of electrical charge and Rastall geometric parameter 
	%of a black hole ($(N=Q=0)$), are consistent with a Schwarzschild black hole.
	\\
	
	
}

\maketitle

%%%%%%%%%%%%%%%%%%%%%%%%%%%%%%%%%%%%%%%%%%%%%%%%%%%%%%%%%%%%%%%%%
\section{INTRODUCTION}
%implies the existence of black holes
Einstein general relativity (GR) is a geometric gravitational theory which define %describes-
all solar system observations, the dynamic cosmos and gravitation 
as a geometrical curvature, which affects the motion of light ray and test particles in 
space-time \cite{Hackmann:2008zz}.  %Provides a monolithic description of gravity as a geometric feature of space and time or space-time.site
The existence of black hole is one of the important issues in physics which 
predicted by the equations of GR. %\cite{lumerzhal,bthesis}.
Black holes and the metrics that explain the space-time around them are 
very important fields of study and research, because of having a gravitational influence %(effecting) in cosmology
on their surrounding, and also on motion of light ray and test particles \cite{Soroushfar:2015dfz}, 
information about the last step of the star life, and discussion of the dark matter. % \citep{bthesis}. 
% It is important because the motion of matter and light can be used to classify
%an arbitrary space-time, in order to discover its structure. For this purpose, we need to solve
%geodesic equations that describe the motion of particles and light \cite{Soroushfar:2015dfz}.\\
Researchers have posited that the Universe contains dark matter \cite{dm1,dm2} 
and dark energy \cite{de1,de2} which are two important problems of the standard 
present cosmological model which can explain the accelerating expansion of the cosmos \cite{de1}. 
Dark matter is a scalar field (25 percent of energy content in the Universe) 
%Structure formation %lummerzhal testing basic laws...
composed of weakly interacting massive particles that interact through weak force 
whereas dark energy (70 percent of energy content in the Universe) \cite{deau} 
known as an exotic fluid and a type of dynamical quantum vacuum energy or a 
kind of self-repulsive mysterious force with negative pressure \cite{Fernando:2012ue,deau,Radinschi:2012vw}. 
Observational evidence such as Cosmic Microwave Background radiation \cite{CMB}, 
the large-scale structure of the Universe \cite{LSS} and luminosity distance of Supernova Type Ia \cite{de1,SNIa} 
will be known as accelerating expansion phase reasons. 
%Studing the black hole surrounded by dark energy
Dark energy was proposed to interpret %(describe) 
the accelerating rate \cite{Hackmann:2008zz} by a very small positive cosmological 
constant with a state parameter $\omega=-1$ \cite{FERNANDO:2013uxa, 12, 13, Zhaoyi, Radinschi:2012vw, Fernando:2012ue}. 
% \cite{FERNANDO:2013uxa, 12, 13- Radinschi:2012vw, Fernando:2012ue}. %FERNANDO:2013uxa.  
%The simplest explanation for dark energy is a vacuum energy or a very small positive cosmological constant.FERNANDO:2013uxa
%One of the simplest candidates for the dark energy density is the traditional static vacuum energy or cosmological constant with a state parameter $\omega=-1$. However, the observed value is too small in comparison with the theoretical prediction within the framework of general relativity. 
% that acts in the accelerated expansion of the universe [14]. Different theories of dark energy suggest different values of the state parameter $\omega$. %with $\omega<\frac{-1}{3}$ for cosmic accel..
%Existence of cosmological model including dark energy with an 
%equation of state $\omega< -1$ is allowed by recent observations.
Recent observations enable the existence of cosmological model including dark energy 
with an equation of state $\omega< -1$. %%%%%%
Quintessence as a candidate for dark energy with state parameters in the range of 
$-1<\omega_q<-\frac{1}{3}$ \cite{Zhaoyi, Kieslev2002}, 
and phantom field with $\omega=-\frac{4}{3}$, %with a non-constant state equation,% are among the many theories 
%try to explain the nature of DE 
are two exotic matters which try to explain the nature of dark energy \cite{Fernando:2012ue, Kieslev2002}. 
A black hole might be surrounded by regular matter like radiation with $\omega=\frac{1}{3}$ 
and dust with state parameter $\omega=0$ or exotic matter like cosmological constant, 
quintessence and phantom fields or combination of them \cite{Heydarzade:2017wxu}.\\
%Different universe models and also different metrics of space-time are 
%the results of different solutions of the Einstein equations \cite{13}. 
%An effective perfect fluid behaviour of the black hole surrounding field
%has been found by comparing the solutions of Kieslev-like black hole in GR with Rastall theory \cite{Heydarzade:2017wxu}.\\
One of the important results obtained from Einstein fields equations 
is a null divergence of the energy-momentum tensor in the form $T^{\mu\nu}_{\;\;\;\;;\nu}=0$ \cite{Heydarzade:2017wxu}. 
Due to the violation of the usual classical conservation laws and 
verification of the condition $T^{\mu\nu}_{\;\;\;\;;\nu}\neq 0$ 
by particle creation in cosmology \cite{9-16darabi}, 
%When the quantum effects are taking into account, 
a new formulation of %for gravity the classical form of 
the energy-momentum tensor has suggested %by suggesting %introducing 
by quantities related to the curvature of the space-time \cite{Birrell}. 
In 1972 P. Rastall \cite{Rastall1972} proposed a modified theory 
of general relativity with the new a formulation %of gravity, 
of connection between energy-momentum tensor, $T^{\mu\nu}_{\;\;\;\;;\nu}$ 
to the derivative of Ricci scalar, i.e. $T^{\mu\nu}_{\;\;\;\;;\nu}\propto R^{,\nu}$ \cite{Heydarzade:2017wxu}, 
which get back to the Einstien's basic assumptions in the empty Universe \cite{Moradpour} %(flat background)
and represents to the Mach principle \cite{Majernik:2006jg}. 
Rastall assumed in curved space-time the usual conservation laws used in GR are collapsed %break down
\cite{Carames:2015tpa}. %1503-04882
In other words, for a non-minimal way that the matter and geometry fields are joined together, 
%In fact matter fields and the geometry, in a non-minimal way, are paired \cite{Moradpour2016} %1610-03881
$T^{\mu\nu}_{\;\;\;\;;\nu}=\lambda R^{,\nu}$ where 
$\lambda$ called the Rastall free parameter which % quantify-determine-indicate
describes the deviation from the Einstein theory of GR and defined %should be specified
from observations \cite{Rastall1972,Moradpour2016}.\\ %or other parts of physics \cite{Rastall1972}. \\
%%%%%%%%%%%%%%%%%%%%%%%%%%%%%%%
Since only light and particles are detectable, %\cite{Hackmannbook}, % observable %apparent, %A-D18Marcheva
%we need to derive expressions for their paths. So 
studying their orbits %of massive particles and light rays 
in space-time near a black hole, is an important tool 
for %exploring   %   features of gravitational fields 
investigating physical properties and the features of solutions of Einstein field equations
and also for tests of GR. %\cite{Hackmann:2015dwa}. %1506-00804
The coupled geodesic equations will describe the motion of %  coupled 
system by differential equations %dependent
based on the metric of the considered field.  %\cite{Hackmann:2015dwa} %1506-00804
%which evident symmetries enable decoupling of the geodesic equations 
The equations will be decoupled by evident symmetries \cite{Hackmann:2015dwa}. %1506-00804
%A wide range of exactly known solutions of Einstein's field equation possesses certain symmetries, which allow decoupling the geodesic equations. 1506-00804
Analysis of geodesic equation of motion is especially useful for analysing the properties 
of space-time and predict some observational events such as 
perihelion shift, light deflection and gravitational time-delay \cite{Soroushfar:2016yea}. 
In 1916, Schwarzschild discovered the first exact solution to Einstein's equations 
in the case of aspherically symmetric black hole in four-dimensional space-time \cite{Schwarzschild}. %dr Soroshfar (the motion of particles and light described by the geodesic equation)
All analytical solutions of the geodesic equation in a Schwarzschild (AdS) 
space-time and gravitational field have been presented by Hagihara %in form of elliptic functions 
in $1931$ \cite{Hagihara}.\\ % 2008hackmanlummerzhalPRD08
%%%%%%%%%%%%%%%%%%%%%%%%%%%%
Many different space-times in theory of general relativity and also in modified theory, such as 
four-dimensional Schwarzschild-de Sitter \cite{Hackmann:2008tu,Hackmann:2008zz}, 
higher dimensional Schwarzschild, Schwarzschild-(anti-)de Sitter, Reissner-Nordstr\"om, Reissner-Nordstr\"om-(anti-)de Sitter \cite{Chanderaskhar,Hackmann:2008tu,Hackmann:2008zza}, Kerr \cite{Kerr}, 
Kerr-de Sitter \cite{Hackmann:2010zz} and black holes in $f(R)$ gravity \cite{Soroushfar:2015wqa}, 
static and rotating dilaton black hole \cite{Soroushfar:2016yea}, 
$(2+1)$–dimensional charged BTZ \cite{Soroushfar:2015dfz}, 
static cylindrically symmetric conformal gravity \cite{Hoseini:2016nzw}, 
the higher-dimensional Myers-Perry space-time \cite{Enolski:2010if,Kagramanova:2012hw} 
and geodesics in the spactime of a rotating charged 
black hole \cite{Soroushfar:2016esy}, 
have been studied and their geodesic equation solved analytically.\\ %Dr Soroshfar-Jafari
%%%%%%%%%%%%%%%%%The motion of particles is described by the geodesic equation %1506
%The mathematical tools of Weierstrassian elliptic functions demonstrated by Jacobi \cite{Jacobi}, Abel \cite{Abel}, Riemann \cite{Riemann,Riemann2}, Weierstrass \cite{Weierstrass} and Baker \cite{Baker}.\\%soroshfar-sahami
In this paper, we study geodesic equation of motion 
for test particles in the space-time of a black hole surrounded by five perfect fluids 
such as quintessence, dust (energy matter), radiation, cosmological constant and 
phantom fields in Rastall theory. We show our analytical solution here in form of 
elliptic and also hyperelliptic functions. In Section (\ref{sec2}) 
we give a brief review of a black hole surrounded by perfect fluids in Rastall field equations.
In section (\ref{sec3}) we investigate the analytical solution of the equations of motion 
for timelike and null geodesic equations with some possible values of Rastall coupling constant 
for five surrounding fields in five subsections. 
In section (\ref{sec4}), we use the analysis provided in the previous sections for 
geodesic equations and their analytical solutions,
to plot $\tilde{L}-E^2$ diagrams, effective potential and also 
to analyse the possible orbit types, their classification and plot some examples of possible orbits. 
we represent conclude in Section (\ref{conclusions}).
%describe the scenario in which we will solve these equations as well as the methods used.
%%%%%%%%%%%%%%%%%
%A separation of variables in (R(r)) yields hackman thesis
%%%%%%%%%%%%%%%%%%%%%%%%
\section{FIELD EQUATIONS IN RASTALL THEORY OF GRAVITY}\label{sec2}
%Rastall fields equations has been written as $G_{\mu\nu}+k\lambda R=kT_{\mu\nu}$ where k is the Rastall gravitational coupling constant. In the limit of $\lambda\rightarrow 0$ and $k=8\pi G_N$ reobtained to GR field equations.\\
The field equations for a space-time with Ricci scalar $R$, an energy momentum source 
of $T_{\mu \nu}$ and for a space-time metric $g_{\mu \nu}$ in the 
context of gravitational Rastall theory can be written as 
%\begin{equation} T^{\mu \nu};\mu =\lambda R^{,\nu} \end{equation}
\begin{equation}
G_{\mu \nu}+k \lambda g_{\mu \nu}=k T_{\mu \nu} .
\end{equation}
The general spherical symmetric space-time metric with a generic metric function $f_s(r)$ is 
\begin{equation}\label{ds2}
ds^2=-f_s(r)dt^2+\frac{dr^2}{f_s(r)}+r^2(d\theta^2+\sin^2\theta d\phi^2) ,
\end{equation}
with the general metric function in the framework of Rastall theory \cite{Heydarzade:2017wxu}
\begin{equation}\label{eqfr}
f_s(r)=1-\frac{2M}{r}+\frac{Q^2}{r^2}-\frac{N_s}{r^{\frac{1+3\omega_s-6k\lambda(1+\omega_s)}{1-3k\lambda(1+\omega_s)}}} ,
\end{equation}
which it is depended on the Rastall parameters $k$ and $\lambda$, radial coordinate $r$, 
mass $M$, the electric charge of the black hole $Q$, equation of state parameter 
$\omega_s$ and surrounding field structure parameter $N_s$ \cite{Heydarzade:2017wxu}. 
Eqs. (\ref{ds2}) and (\ref{eqfr}) for $ k=8\pi G_{N} $ and $\lambda=0$ convert to 
\begin{equation}\label{metrick}
ds^2=-(1-\frac{2M}{r}+\frac{Q^2}{r^2}-\frac{N_s}{r^{3w_s+1}})dt^2+\frac{dr^2}{1-\frac{2M}{r}+\frac{Q^2}{r^2}-\frac{N_s}{r^{3w_s+1}}}+r^2d \Omega ^2  ,
\end{equation} 
which represent the Reissner-Nordstr\"om black hole surrounded by a surrounding field in GR \cite{kieslev}. %will be obtained which was found by Kieslev as
By comparing the metrics (\ref{ds2})-(\ref{eqfr}) with (\ref{metrick}), some interesting features 
with introducing an ''effective equation of state'' has been studied in detail in \cite{Heydarzade:2017wxu}. 
The subscript "s" indicates the general surrounding field.
%%%%%%%%%%%%%%%%%%%%%%%%%%%%
\section{Geodesics}\label{sec3}
The geodesic differential equation is in general of the form %differential equations %%%%%%%hackmann-Geodesic equations in black hole.. The mathematical tools
\begin{equation}\label{eqgeo}
\frac{d^2x^a}{d\lambda^2}+\Gamma^c_{ba}\frac{dx^a}{d\lambda}\frac{dx^b}{d\lambda}=0 ,
\end{equation}
where $\Gamma^c_{ba}$ are the Christoffel symbols. % and $ds^2=g_{\mu\nu}dx^{\mu}dx^{\nu}$ 
%is the invariant space-time interval.  % along the geodesics. %masters geodesics..
%%%soroshfar-sahami
By using the normalization condition $g_{\mu\nu}\frac{dx^{\mu}}{ds} \frac{dx^{\nu}}{ds}=\epsilon$, 
(where for massive particles $\epsilon=1$ and for light $\epsilon=0$), 
and two constant of motion energy $E$ and the angular momentum $L$ as
\begin{eqnarray}
E=g_{tt}\frac{dt}{ds}=f_s(r)\frac{dt}{ds}\;,\;\;\;\;L=g_{\phi\phi}\frac{d\phi}{ds}=r^2\frac{d\phi}{ds},
\end{eqnarray}
and considering the motion is took place in a equatorial plane, $ \theta=\frac{\pi}{2} $ 
as an initial condition, the equations of the complete explanation of particle's motion %/dynamics
are 
\begin{eqnarray}
&&(\frac{dr}{ds})^2=E^2-f_s(r)(\epsilon+\frac{L^2}{r^2}),\label{a}\\
&&(\frac{dr}{d\phi})^2=\frac{r^4}{L^2}(E^2-f_s(r)(\epsilon+\frac{L^2}{r^2}))=:R(r),\label{b}\\
&&(\frac{dr}{dt})^2=\frac{f^2(r)}{E^2}(E^2-f_s(r)(\epsilon+\frac{L^2}{r^2})).\label{c}
\end{eqnarray}
The effective potential $V_{eff}$ can be get from Eq. (\ref{a}) as 
\begin{eqnarray}
V_{eff}=(1-\frac{2M}{r}+\frac{Q^2}{r^2}-\frac{N_s}{r^{\frac{1+3w_s-6k\lambda(1+w_s)}{1-3k\lambda(1+w_s)}}}
)(\epsilon+\frac{L^2}{r^2}).
\end{eqnarray}
%%%%%%%%%%%%%%%%%%%%
%%%%%%%%%%%%%%%%%%%%%%
We rewrite the equations with new dimensionless %coordinate $\tilde{r}$ and dimensionless 
parameters
\begin{eqnarray}
\tilde{r}=\frac{r}{M},\;\;\;\tilde{L}=\frac{M^2}{L^2},\;\;\;\tilde{Q}=\frac{Q}{M},\;\;\;
\end{eqnarray}
so we have
\begin{eqnarray}
&ds^2=-f_s(\tilde{r})dt^2+\frac{d\tilde{r}^2}{f_s(\tilde{r})}+\tilde{r}^2d \Omega ^2,\label{metrics}\\
&f_s(\tilde{r})=1-\frac{2}{\tilde{r}}+\frac{\tilde{Q}^2}{\tilde{r}^2}-\frac{\tilde{N_s}}{\tilde{r}^{\frac{1+3\omega_s-6k\lambda(1+\omega_s)}{1-3k\lambda(1+w_s)}}},\label{fsr}
\end{eqnarray}
and the Eq. (\ref{b}) with the generic metric of Rastall theory (Eq. (\ref{eqfr})) takes the following form
\begin{eqnarray}\label{r5}
(\frac{d\tilde{r}}{d\phi})^2=\tilde{r}^4\tilde{L}\left(E^2-\left(1-\frac{2}
{\tilde{r}}+\frac{\tilde{Q}^2}{\tilde{r}^2}-\frac{\tilde{N}_s}
{\tilde{r}^{\frac{1+3\omega_s-6k\lambda(1+\omega_s)}
		{1-3k\lambda(1+\omega_s)}}}\right)(\epsilon+\frac{1}{\tilde{L}\tilde{r}^2})\right)=R(\tilde{r}).
\end{eqnarray}
To solve this equation and investigate its results, we study analytical solutions 
of geodesic equations of a black hole surrounded by quintessence, dust, 
cosmological constant, radiation and phantom field. 
%%%%%%%%%%%%%%%%%%%%%%%%%%%%%%%%%%%%% quintessence
\subsection{The black hole surrounded by the quintessence field}
In this section, we obtain the equations of motion for two possible values of $k\lambda$ 
for the quintessence surrounding field. By putting $\omega_s=\omega_q=-\frac{2}{3}$ 
\cite{kieslev}, the Eqs. (\ref{metrics}) and (\ref{fsr}) convert to the following equations
\begin{eqnarray}\label{quin}
&ds^2=-f_q(\tilde{r})dt^2+\frac{d\tilde{r}^2}{f_q(\tilde{r})}+\tilde{r}^2d\Omega ^2,\\
&f_q(\tilde{r})=1-\frac{2}{\tilde{r}}+\frac{\tilde{Q}^2}{r^2}-
\frac{\tilde{N}_q}{\tilde{r}^{\frac{-1-2k\lambda}{1-k\lambda}}}.\label{frq}
\end{eqnarray}
The equation of effective state parameter $\omega_{eff}$ can be obtained by 
comparing the Eqs. (\ref{quin}) and (\ref{frq}) with the original Kieslev metric (Eq. (\ref{metrick})) \cite{Heydarzade:2017wxu}
\begin{equation}\label{veffquint}
\omega_{eff}=\frac{1}{3}(-1-\frac{1+2k\lambda}{1-k\lambda}).
\end{equation}
By considering two values of $ \omega_{eff}\leq -\frac{1}{3}$ and $ w_{eff}\geq -\frac{1}{3}$ 
in Eq. (\ref{metrick}) \cite{Heydarzade:2017wxu},
the range values of $k\lambda$ in Eq. (\ref{veffquint}) are discernible as 
$-\frac{1}{2} \leq k\lambda < 1$ and $k\lambda \leq -\frac{1}{2} \cup k\lambda >1$ respectively. 
Of course, in this paper for all surrounding fields cases, we consider the 
possible values of $k\lambda$, so that $f_s(r)$ in Eq. (\ref{fsr}) have included 
integer powers of $r$ and also Eq. (\ref{r5}) can be solved analytically.
For other values of $k\lambda $, Eq. (\ref{r5}) %will be related to other cases of $\tilde{r}$ and 
have some terms with fractional powers of $\tilde{r}$, which in our ability 
can not be solved analytically but may be solved numerically 
same as applied technique in Ref. \cite{Hartmann:2010rr}.  
\begin{itemize}
	\item For $\omega_{eff}\leq -\frac{1}{3}$ and $k\lambda=\frac{1}{4}$,
\end{itemize}
the metric (\ref{quin}) can be written as
\begin{eqnarray}\label{frqc1}
&ds^2=-f_q(\tilde{r})dt^2+\frac{d\tilde{r}^2}{f_q(\tilde{r})}+\tilde{r}^2d \Omega ^2,\\
&f_q(\tilde{r})=1-\frac{2}{\tilde{r}}+\frac{\tilde{Q}^2}{\tilde{r}^2}-\tilde{N}_q\tilde{r}^2
\end{eqnarray}
and 
\begin{eqnarray}\label{veffqc1}
V_{eff}=(1-\frac{2}{\tilde{r}}+\frac{\tilde{Q}^2}{\tilde{r}^2}-\tilde{N}_q
\tilde{r}^2)(\epsilon+\frac{1}{\tilde{L}\tilde{r}^2}).
\end{eqnarray}
so the Eq. (\ref{r5}) for the quintessence surrounding field, get the following form
\begin{equation}\label{rqa}
(\frac{d\tilde{r}}{d\varphi})^2=\tilde{N}_q\epsilon\tilde{L}\tilde{r}^6+((E^2-\epsilon)\tilde{L}+\tilde{N}_q)r^4+2\epsilon \tilde{L}\tilde{r}^3-(1+\tilde{Q}^2\epsilon\tilde{L})\tilde{r}^2+2\tilde{r}-\tilde{Q}^2=R_q({\tilde{r}}),%=\sum_{i=0}^{6}a_{i}\tilde{r}^{i},
\end{equation}
\begin{itemize}
	\item For $\omega_{eff}\geq-\frac{1}{3}$ and $k\lambda=-2$, then 
\end{itemize}
\begin{eqnarray}\label{frqc2}
&ds^2=-f_q(\tilde{r})dt^2+\frac{d\tilde{r}^2}{f_q(\tilde{r})}+\tilde{r}^2d \Omega ^2,\\
&f_q(\tilde{r})=1-\frac{2+\tilde{N}_q}{\tilde{r}}+\frac{\tilde{Q}^2}{\tilde{r}^2},
\end{eqnarray}
and 
\begin{eqnarray}\label{veffqc2}
V_{eff}=(1-\frac{2+\tilde{N}_q}{\tilde{r}}+\frac{\tilde{Q}^2}{\tilde{r}^2})
(\epsilon+\frac{1}{\tilde{L}\tilde{r}^2}).
\end{eqnarray}
%the plots of $V_{eff}$ for the quintessence surrounding field are shown in Figs. \ref{pic:RQC1a} and \ref{pic:RQC2a}. 
The Eq. (\ref{r5}) can be written 
\begin{equation}\label{rqb}
(\frac{d\tilde{r}}{d\varphi})^2=(E^2-\epsilon)\tilde{L}r^4+(2+\tilde{N}_q)
\epsilon\tilde{L}\tilde{r}^3-(1+\tilde{Q}^2\epsilon\tilde{L}) \tilde{r}^2+
(2+\tilde{N}_q)\tilde{r}-\tilde{Q}^2=R_q({\tilde{r}}),%=\sum_{i=0}^{4}b_{i}\tilde{r}^{i},
\end{equation}
%\end{itemize}
\subsubsection{Analytical Solution of Geodesic Equations}\label{sec}
In this section, we present the analytical solution of the geodesic Eqs. (\ref{rqa}) and (\ref{rqb}). %by changing the variable $\tilde{r}=\frac{1}{u}$ for the test particle and light ray. 
%%%%%%%%%%%%%% seca
\paragraph{Null Geodesics}\label{seca}
For light ray $(\epsilon=0)$, Eqs. (\ref{rqa}) and (\ref{rqb}) are polynomials 
of degree four in the form $(\frac{d\tilde{r}}{d\varphi})=\sum_{i=0}^{4}a_{i}r^{i}$, 
which by substitution $\tilde{r}=\frac{1}{u}+\tilde{r}_R$, where $\tilde{r}_R $ 
is a zero of $R$, convert to a polynomial $R_3$ of degree $3$
\begin{equation}
(\frac{du}{d\phi})^2=R_3(u)=\sum_{j=1}^3 b_ju^j,\;\;\;\;\;u(\phi_0)=u_0,
\end{equation}
where
\begin{equation}
b_j=\frac{1}{(4-j)!}\frac{d^{(4-j)}R}{d\tilde{r}^{4-j}}(\tilde{r}_R),
\end{equation}
in which $b_j,\;(j=1,2,3) $ is an arbitrary constant of the relevant metric. Next, 
substitution $u=\frac{1}{b_3}(4y-\frac{b_2}{3}) $, transform $ R_3(u) $, to 
elliptical type differential equation as \cite{Soroushfar:2016esy}
\begin{equation}\label{wform}
(\frac{dy}{d\phi})^2=4y^3-g_2y-g_3=p_3(y).
\end{equation}
Equation (\ref{wform}) known as the Weierstrass form which 
\begin{equation}
g_2=\frac{1}{16}(\frac{4}{3}b_2^2-4b_1b_3),\;\;\;g_3=\frac{1}{16}(\frac{1}{3}b_1b_2b_3-\frac{2}{27}b_2^3b_3^2),
\end{equation}
are the Weierstrass invariants. So, the answer of Eq. (\ref{wform}), using the Weierstrass function, is as follows
\begin{equation}
y(\phi)=\wp(\phi-\phi_{in};g_2,g_3),
\end{equation}
in which
$\phi_{in}=\phi_0+\int_{y_0}^{\infty}\frac{dy}{\sqrt{4y^3-g_2y-g_3}}$ with $\phi_0=\frac{1}{4}(\frac{b_3}{\tilde{r}_0-\tilde{r}_R}+\frac{b_2}{3})$ depends only on the initial value $\phi_0$ 
and $\tilde{r}_0$. Eventually, the solution of polynomials of degree four is %Eqs. (\ref{rqa}) and (\ref{rqb}) become
\cite{Hackmann:2008zz}
\begin{equation}\label{eqrw}
\tilde{r}(\phi)=\frac{b_3}{4\wp(\phi-\phi_{in};g_2,g_3)-\frac{b_2}{3}}+\tilde{r}_R.
\end{equation}
This analytic solution is obtained for null geodesic in quintessence surrounding field in 
Rastall theory and is reliable in all regions of this space-times. The explanation and properties 
presented in this section are applied to solve all geodesic equations of elliptic type in this paper. %\cite{Hackmannbook}
\paragraph{Timelike Geodesics}\label{secb}
%%%%%%%%%%%%%%%%%%%%%%%%%
For the massive particle $(\epsilon=1)$ Eq. (\ref{rqa}) is a polynomial of order six and also of the hyperelliptic type.
By substitution $\tilde{r}=\frac{1}{u}+\tilde{r}_R$, where $\tilde{r}_R $ is a zero of $R$,
the equation of motion can be reduced to one of the two forms
%%%%%%%%%%%%%%%%%%
%The system (\ref{r5}) will be simplified and reform to two differential equations
%\begin{eqnarray}
%(y\frac{dy}{dx})^2=P(y),\\
%(\frac{dy}{dx})^2=P(y),
% \end{eqnarray}
%where high possible solvable degree for polynomial P is of order 6 \cite{Kagramanova2009}. 
%With polynomial of degree 3 or 4, geodesic equation is of elliptic type 
%and can be performed in terms of the Weierstrass $\wp$-functions and 
%with polynomial of degree 5 or 6, geodesic equation is of hyperelliptic type 
%and the solutions are given in terms of derivatives of Kleinian sigma functions 
%\cite{Soroushfar:2016yea,Hackmann:2008tu}.\\
%%%%%%%%%%%%%%%%
\begin{eqnarray}\label{twof}
\left(u\frac{du}{d\phi} \right) ^2=P_{5}(u),\\
\left(\frac{du}{d\phi} \right) ^2=P_{5}(u).
\end{eqnarray}
The analytic solution of above equations, which is extensively discussed in 
\cite{Enolski:2010if,Soroushfar:2016esy, Hackmann:2015ewa}, is given in form 
of derivatives of the Kleinian $\sigma$ function as
\begin{equation}
u(\varphi)=-\frac{\sigma_1(\varphi_{\infty})}{\sigma_2(\varphi_{\infty})}|_{\sigma(\varphi_{\infty})=0},
\end{equation}
with 
\begin{equation} 
\varphi_{\infty}=(\varphi_2,\varphi-\varphi^{'}_{in}),
\end{equation}
where $ \varphi^{'}_{in}=\varphi_{in}+\int_{\varphi_{in}}^{\infty}\frac{udu^{'}}{\sqrt{P_5(u^{'})}}$. 
The component $\varphi_2$ is determined by the condition $ \sigma(\varphi_{\infty})=0 $. 
Also, the function $\sigma_i$ is the ith derivative of Kleinian $\sigma$ function and $\sigma_z$ is
\begin{equation}
\sigma_z=Ce^{zt}kz\theta[g,\theta](2w^{-1}z;\tau),
\end{equation}
where $C$ is a constant, $\tau$ is the symmetric Riemann matrix, $\omega$ is the 
period matrix, $k=\eta(2)^{-1}$ which $\eta$ is the periodmatrix of the second kind 
and $ \theta $ is the Riemann function with characteristic $[g, h]$ which $2[g, h] =(0,1)^t+(1,1)^t\tau$ 
\cite{Hackmann:2008zza, Hackmann:2008tu, Soroushfar:2015wqa, Buchstaber:1997}, %tu-higher dimension
%, \cite{Buchstaber:1997:1997}.
%%%%%%%%%%%%%%%%%%%%%%%%%%%%%%
So the solution of Eq. (\ref{rqa}) becomes
\begin{eqnarray}\label{eqrsigma}
r(\varphi)=-\frac{\sigma_2(\varphi_{\infty})}{\sigma_1(\varphi_{\infty})} .
\end{eqnarray}
This analytic solution is obtained for timelike geodesic in the quintessence surrounding field 
in Rastall theory and is reliable in all regions of this space-times. The explanation and 
properties presented in this section are applied to solve all geodesic equations of hyperelliptic type in this paper.\\
%%%%%%%%%%%%%%%%%%%%
%%%%%%%%%%%%%%%%%%%%
We use of these analytical solutions to plot some example of possible orbits of test particles and light ray,
but before that we need to plot $\tilde{L}-E^2$ diagram for each cases. 
Solving $R_q(\tilde{r})=0$ and $\frac{dR_q(\tilde{r})}{d\tilde{r}}=0$ 
give us $E^2$ and $\tilde{L}$ equations. For massive particles $(\epsilon=1)$ 
in a black hole surrounded by the quintessence field with $k\lambda=\frac{1}{4}$, we have 
\begin{eqnarray}\label{leqc11}
\tilde{L}=-\frac{2\tilde{Q}^2+\tilde{r}^2-3\tilde{r}}{\tilde{r}^2(\tilde{N}_q\tilde{r}^4+\tilde{Q}^2-r)},\;\;\;E^2=\frac{(-\tilde{N}_q\tilde{r}^4+\tilde{Q}^2+\tilde{r}^2-2\tilde{r})^2}{(2\tilde{Q}^2+
	\tilde{r}^2-3\tilde{r})\tilde{r}^2},
\end{eqnarray}
and for massless particles $(\epsilon=0)$
\begin{eqnarray}\label{leqc12}
\tilde{L}=\frac{-\tilde{N}_q\tilde{r}^4+\tilde{Q}^2+\tilde{r}^2-2\tilde{r}}{E^2\tilde{r}^4}.
\end{eqnarray}
%%%%%%%%%%%%%%%%%%%%%%%%%%
With $k\lambda=-2$, for massive particle $(\epsilon=1)$ 
\begin{eqnarray}\label{leqc21}
\tilde{L}=-\frac{4\tilde{Q}^2-3(2+\tilde{N}_q)\tilde{r}+ 2\tilde{r}^2)^2}{(2\tilde{Q}^2-\tilde{N}_q\tilde{r}+2\tilde{r}^2-2\tilde{r})\tilde{r}^2},\;\;\;E^2=
\frac{2(\tilde{Q}^2-(2+\tilde{N}_q)\tilde{r}+ \tilde{r}^2)^2}{(4\tilde{Q}^2-
	3(2+\tilde{N}_q)\tilde{r}+2\tilde{r}^2)\tilde{r}^2},
\end{eqnarray}
and for massless particles $(\epsilon=0)$
\begin{eqnarray}\label{leqc22}
\tilde{L}=\frac{\tilde{Q}^2+\tilde{r}^2-(2+\tilde{N}_q)\tilde{r}}{E^2\tilde{r}^4}.
\end{eqnarray}
%%%%%%%%%%%%%%%%%%%%%%%%%%%%%
Figures of $\tilde{L}-E^2$ diagrams (Eqs. (\ref{leqc11})-(\ref{leqc22})) %for the case $k\lambda=\frac{1}{4}$ with the region of different types of geodesic motion, in the quintessence surrounding field, 
have been shown in Figs. \ref{pic:RQC1b} and \ref{pic:RQC2b}. % and plots of $\tilde{L}-E^2$ for the case $k\lambda=-2$ (Eqs. (\ref{frqc2}), (\ref{veffqc2})), is shown in Fig. \ref{pic:RQC1b}((c),(d)). 
Moreover, a summary of possible orbits type with numbers of zero points in each 
regions for both cases $k\lambda=\frac{1}{4}$ and $k\lambda=-2$ are shown 
in table \ref{tab:RQ1} and \ref{tab:RQ2} respectively. 
Also results of timelike and null geodesic effective potential (Eqs. (\ref{veffqc1}), 
(\ref{veffqc2})) are shown in Figs. \ref{pic:RQC1a} and \ref{pic:RQC2a}.
%%%%%%%%%%%%%%%%%%%%%%%%%%%%%%%%%%%%%%%%%%dust
\subsection{The black hole surrounded by dust field}
When the black hole is surrounded by the dust field, we put $\omega_s=\omega_d=0$ \cite{kieslev} 
and the metric (\ref{metrics}) writes as follows
\begin{eqnarray}\label{dust}
&ds^2=-f_d(\tilde{r})dt^2+\frac{d\tilde{r}^2}{f_d(\tilde{r})}+\tilde{r}^2d \Omega ^2,\\
&f_d(\tilde{r})=1-\frac{2}{\tilde{r}}+\frac{\tilde{Q}^2}{\tilde{r}^2}-
\frac{\tilde{N}_d}{\tilde{r}^{\frac{1-6k\lambda}{1-3k\lambda}}}.
\end{eqnarray}
%regarding the eq (\ref{rho}) and eq (\ref{w}), $\rho_d$ has effectively different from their GR such that %$\rho_d=\frac{3\lambda(1-4k\lambda)N_d}{(1-3k\lambda)^2}r^{-\frac{3-12k\lambda}{1-3k\lambda}}$, 
The equation of effective state parameter $\omega_{eff}$ can be obtained 
by comparing this metric with the original Kieslev metric (\ref{metrick}), as \cite{Heydarzade:2017wxu}
%By comparing this metric with the Kieslev metric (\ref{metrick}) in GR, one may obtain an effective equation of state parameter $w_{eff}$ for %the modification term resulting from the geometry of Rastall theory as \cite{Heydarzade:2017wxu}
\begin{equation}\label{weffdust}
\omega_{eff}=\frac{1}{3}(-1+\frac{1-6k\lambda}{1-3k\lambda}).
\end{equation}
By considering two values of $ \omega_{eff}\leq -\frac{1}{3}$ and $ w_{eff}\geq -\frac{1}{3}$ 
in Eq. (\ref{metrick}) \cite{Heydarzade:2017wxu},
the range values of $k\lambda$ in Eq. (\ref{weffdust}) are discernible as 
$\frac{1}{6} <k\lambda < \frac{1}{3}$ and $k\lambda < \frac{1}{6} \cup k\lambda >\frac{1}{3}$ respectively. 
%which leads to $ w_{eff}\leq -\frac{1}{3}$, and $k\lambda < \frac{1}{6} \cup k\lambda >\frac{1}{3}$ which leads to $ w_{eff}\geq -\frac{1}{3}$.\\
%%%%%%%%%%%%%%%%%%%%%
To solving the analytical solution of the equations of motion for surrounding dust field
\begin{itemize}
	\item we consider $k\lambda=\frac{2}{9}$ for $ \omega_{eff}\leq -\frac{1}{3}$,
\end{itemize} % in the first interval,
so the metric (\ref{dust}) can be written as
\begin{eqnarray}\label{frd1}
&ds^2=-f_d(\tilde{r})dt^2+\frac{d\tilde{r}^2}{f_d(\tilde{r})}+\tilde{r}^2d \Omega ^2,\\
&f_d(\tilde{r})=1-\frac{2}{\tilde{r}}+\frac{\tilde{Q}^2}{\tilde{r}^2}-\tilde{N}_d\tilde{r}
\end{eqnarray}
and 
\begin{eqnarray}\label{veffd1}
V_{eff}=(1-\frac{2}{\tilde{r}}+\frac{\tilde{Q}^2}{r^2}-\tilde{N}_d \tilde{r})(\epsilon+\frac{1}{\tilde{L}\tilde{r}^2}).
\end{eqnarray}
Then the metric (\ref{r5}) get the form
\begin{equation}\label{eqrd}
(\frac{d\tilde{r}}{d\varphi})^2=\tilde{N}_d\epsilon\tilde{L}\tilde{r}^5+(E^2-\epsilon)\tilde{L}r^4+
(2\tilde{L}\epsilon+\tilde{N}_d)\tilde{r}^3-(1+\tilde{Q}^2\epsilon\tilde{L})
\tilde{r}^2+2\tilde{r}-\tilde{Q}^2=R_d({\tilde{r}}),%=\sum_{i=0}^{5}a_{i}\tilde{r}^{i},
\end{equation}
For $\epsilon=0$, Eq. (\ref{eqrd}) is of elliptic type and %the polynomial $R_d(\tilde{r})$ can be reduced to forth order which by substituting  $r=\frac{1}{u}$, it reduces to third order and 
therefore has analytical solution same as Sec. \ref{sec}(a) %like (\ref{eqrw}).
\begin{equation}
\tilde{r}(\phi)=\frac{b_3}{4\wp(\phi-\phi_{in};g_2,g_3)-\frac{b_2}{3}}+\tilde{r}_R.
\end{equation}
But for $\epsilon=1$, Eq. (\ref{eqrd}) is of hyperelliptic type, 
therefore has analytical solution same as Sec. \ref{sec}(b) %Eq. (\ref{eqrsigma}).
\begin{eqnarray}
\tilde{r}(\varphi)=-\frac{\sigma_2(\varphi_{\infty})}{\sigma_1(\varphi_{\infty})}
\end{eqnarray}
%%%%%%%%%%%%%%%%%%%%%%%%%%%%%%%%%
%Figs. \ref{pic:RDb} and \ref{pic:RDa} show plots of $\tilde{L}-E^2$ and effective potentials, for surrounding dust field respectively corresponding to table \ref{tab:RD}.\\
%Plots of $\tilde{L}-E^2$ (Eqs. (\ref{}), (\ref{})) for the case $k\lambda=\frac{2}{9}$ with the region of different types of geodesic motion, in the dust surrounding field, has been shown in Fig. \ref{pic:RDb}. 
%%%%%%%%%%%%%%%%%%%%%%%%%%%%%%%%%%%
\begin{itemize}
	\item Our next possible choice is $k\lambda=\frac{1}{4}$, with $\omega_{eff}\geq -\frac{1}{3}$, 
\end{itemize}
so the metric (\ref{dust}) can be written as
\begin{eqnarray}\label{frd2}
& ds^2=-f_d(\tilde{r})dt^2+\frac{d\tilde{r}^2}{f_d(\tilde{r})}+\tilde{r}^2d \Omega ^2,\\
& f_d(\tilde{r})=1-\frac{2}{\tilde{r}}+\frac{\tilde{Q}^2}{\tilde{r}^2}-\tilde{N}_d \tilde{r}^2,
\end{eqnarray}
with the effective potential
\begin{eqnarray}\label{veffd2}
V_{eff}=(1-\frac{2}{\tilde{r}}+\frac{\tilde{Q}^2}{r^2}-\tilde{N}_d \tilde{r}^2)(\epsilon+\frac{1}{\tilde{L}\tilde{r}^2}).
\end{eqnarray}
then
\begin{equation}\label{eqrdust}
(\frac{d\tilde{r}}{d\varphi})^2=\tilde{N}_d\epsilon\tilde{L}\tilde{r}^6+((E^2-\epsilon)\tilde{L}+\tilde{N}_d)r^4+
2\tilde{L}\epsilon\tilde{r}^3-(1+\tilde{Q}^2\epsilon\tilde{L})\tilde{r}^2+2\tilde{r}-\tilde{Q}^2=R_d(\tilde{r}),%=\sum_{i=0}^{6}a_{i}\tilde{r}^{i},
\end{equation}
In order to obtain the analytical solution, for $\epsilon=0$, Eq. (\ref{eqrdust}) is of elliptic type 
which has analytical solution same as Sec. \ref{sec}(a) 
\begin{equation}
\tilde{r}(\phi)=\frac{b_3}{4\wp(\phi-\phi_{in};g_2,g_3)-\frac{b_2}{3}}+\tilde{r}_R.
\end{equation}
But for $\epsilon=1$, %, we substitute $r=\frac{1}{u}$, and transform (\ref{eqrdust}) to $(u\frac{du}{d\varphi})^2=P_5(u)$. Then
the analytical solution of Eq. (\ref{eqrdust}) is similar to Sec. \ref{sec}(b) 
\begin{eqnarray}
\tilde{r}(\varphi)=-\frac{\sigma_2(\varphi_{\infty})}{\sigma_1(\varphi_{\infty})}
\end{eqnarray}
%Note that for this part our, so 
%%%%%%%%%%%%%%%%%%%%%%%%%%%%%
In next, as discussed in the previous section, we need to plot $\tilde{L}-E^2$ diagram. So 
by solving $R_d(\tilde{r})=0$ and $\frac{dR_d(\tilde{r})}{d\tilde{r}}=0$ 
for massive particles $(\epsilon=1)$ with $k\lambda=\frac{2}{9}$
%we have for massive particles $(\epsilon=1)$ in the black hole surrounded by dust field with $k\lambda=\frac{2}{9}$, we have 
\begin{eqnarray}\label{led1}
\tilde{L}=-\frac{-\tilde{N}_d\tilde{r}^3+4\tilde{Q}^2+2\tilde{r}^2-6\tilde{r}}
{\tilde{r}^2(\tilde{N}_d\tilde{r}^3+2\tilde{Q}^2-2\tilde{r})},\;\;\;E^2=\frac{2(-\tilde{N}_d\tilde{r}^3+\tilde{Q}^2+\tilde{r}^2-2\tilde{r})^2}{(-\tilde{N}_d\tilde{r}^3+4\tilde{Q}^2+2\tilde{r}^2-6\tilde{r})\tilde{r}^2},
\end{eqnarray}
and for massless particles $(\epsilon=0)$
\begin{eqnarray}\label{led2}
\tilde{L}=-\frac{-\tilde{N}_d\tilde{r}^3+\tilde{Q}^2+\tilde{r}^2-2r}{E^2\tilde{r}^4}.
\end{eqnarray}
%%%%%%%%%%%%%%%%%%%%%%%
The figures of $\tilde{L}-E^2$ diagram (Eqs. (\ref{led1})-(\ref{led2})) 
with the region of different types of geodesic motion in the dust surrounding field 
with the case $k\lambda=\frac{2}{9}$, has been illustrated in Fig. \ref{pic:RDb} while $\tilde{L}-E^2$ diagram
for the case $k\lambda=\frac{1}{4}$ are identical with the same case 
in the quintessence surrounding field (Figs. \ref{pic:RQC1b}, \ref{pic:RQC1a}).  
Moreover, a summary of possible orbits type with numbers of 
zero points in each region is shown in table \ref{tab:RD}. 
Also results of timelike and null geodesic effective potential (Eqs. (\ref{veffd1}), (\ref{veffd2})) 
has been shown in Fig. \ref{pic:RDa}.
%%%%%%%%%%%%%%%%%%%%%%%%%%%%%%%%%radiation
\subsection{The black hole surrounded by the Radiation field}
When the radiation is surrounding field, we put $w_r=\frac{1}{3}$ \cite{kieslev}, 
so the metric (\ref{metrics}) can be written as
\begin{eqnarray}\label{frr}
&ds^2=-f_r(\tilde{r})dt^2+\frac{d\tilde{r}^2}{f_r(\tilde{r})}+\tilde{r}^2d \Omega ^2,\\
&f_r(\tilde{r})=1-\frac{2}{\tilde{r}}+\frac{\tilde{Q}^2-\tilde{N}_r}{\tilde{r}^2},
\end{eqnarray}
with the effective potential
\begin{eqnarray}\label{veffr}
V_{eff}=(1-\frac{2}{\tilde{r}}+\frac{\tilde{Q}^2-\tilde{N}_r}{\tilde{r}^2})(\epsilon+\frac{1}{\tilde{L}\tilde{r}^2}).
\end{eqnarray}
The metric (\ref{frr}) is the 
Reissner-Nordstr\"om metric of a black hole 
with an effective charge $Q_{eff}=\sqrt{\tilde{Q}^2-\tilde{N}_r}$. 
Due to the similarity of the metric of a black hole surrounded by the radiation field (Eq. (\ref{frr})) with GR, 
the geometric effects of the Rastall parameters 
do not observe. %appear seem. 
Therefore the Eq. (\ref{r5}) 
for the black hole surrounded by the Radiation field becomes 
\begin{equation}\label{drradiation}
(\frac{d\tilde{r}}{d\varphi})^2=(E^2-\epsilon)\tilde{L}\tilde{r}^4+2\epsilon\tilde{L}\tilde{r}^3
-((\tilde{Q}^2-\tilde{N}_r)\epsilon\tilde{L}+1)\tilde{r}^2+2\tilde{r}-(\tilde{Q}^2-\tilde{N}_r)=R_r({\tilde{r}}),
\end{equation}
%In order to obtain the analytical solution, for $\epsilon=0$, Eq. (\ref{drradiation}) is of elliptic type 
%which has analytical solution same as Sec. \ref{sec}(a) 
which for both massive and massless particles has analytical solution similar to Sec. \ref{sec}(a) 
\begin{equation}
\tilde{r}(\phi)=\frac{b_3}{4\wp(\phi-\phi_{in};g_2,g_3)-\frac{b_2}{3}}.
\end{equation}
%with an analysis similar to Eq. (\ref{rqb}), for radiation surrounding field with $\epsilon=0$
%%%%%%%%%%%%%%%%%%%%%%%%%%%%
To plot $\tilde{L}-E^2$ diagram for a black hole surrounded by the radiation field, 
by solving $R_r(\tilde{r})=0$ and $\frac{dR_r(\tilde{r})}{d\tilde{r}}=0$, 
for massive particles $(\epsilon=1)$ we have
%For massive particles $(\epsilon=1)$ in the black hole surrounded by the radiation field, we have 
\begin{eqnarray}\label{ler1}
\tilde{L}=-\frac{2(\tilde{Q}^2-\tilde{N}_r)+\tilde{r}^2-3\tilde{r}}{\tilde{r}^2
	((\tilde{Q}^2-\tilde{N}_r)-\tilde{r})},\;\;\;E^2=\frac{((\tilde{Q}^2-\tilde{N}_r)
	+\tilde{r}^2-2\tilde{r})^2}{(2(\tilde{Q}^2-\tilde{N}_r)+\tilde{r}^2-3\tilde{r})\tilde{r}^2},
\end{eqnarray}
and for massless particles $(\epsilon=0)$
\begin{eqnarray}\label{ler2}
\tilde{L}=\frac{(\tilde{Q}^2-\tilde{N}_r)+\tilde{r}^2-2\tilde{r}}{E^2\tilde{r}^4}.
\end{eqnarray}
The figures of $\tilde{L}-E^2$ diagrams (Eqs. (\ref{ler1})-(\ref{ler2})) 
with the region of different types of geodesic motion, in the radiation surrounding field, 
have been shown in Fig. \ref{pic:RRb}. Also plots of effective potential %for radiation surrounding field 
(Eq. (\ref{veffr})) is shown in Fig. \ref{pic:RRa}. %corresponding to table \ref{tab:RR}.
% which are identical with the case $k\lambda=-2$ in quintessence and surrounding field as our expected.
Moreover, a summary of possible orbits type with numbers of zero points in each regions 
for both massive and massless particles is shown in table \ref{tab:RR}. 
%%%%%%%%%%%%%%%%%%%%%%%% cosmological constant %%%%%%%%%
\subsection{The black hole surrounded by the cosmological constant field}
For the cosmological constant surrounding field, we put $w_c=-1$ \cite{kieslev}, 
so the metric (\ref{metrics}) can be written as 
\begin{eqnarray}\label{frcc}
&ds^2=-f_c(\tilde{r})dt^2+\frac{d\tilde{r}^2}{f_c(\tilde{r})}+\tilde{r}^2d \Omega ^2,\\
&f_c(\tilde{r})=1-\frac{2}{\tilde{r}}+\frac{\tilde{Q}^2}{\tilde{r}^2}-\tilde{N}_c \tilde{r}^2
\end{eqnarray}
with effective potential
\begin{eqnarray}\label{veffcc}
V_{eff}=(1-\frac{2}{\tilde{r}}+\frac{\tilde{Q}^2}{\tilde{r}^2}-\tilde{N}_c \tilde{r}^2)
(\epsilon+\frac{1}{\tilde{L}\tilde{r}^2}).
\end{eqnarray}
so the Eq. (\ref{r5}) gets the form
\begin{equation}\label{drcosmological}
(\frac{d\tilde{r}}{d\varphi})^2=\tilde{N}_c\epsilon\tilde{L}\tilde{r}^6+((E^2-\epsilon)\tilde{L}+\tilde{N}_c)r^4+
2\tilde{L}\epsilon\tilde{r}^3-(1+\tilde{Q}^2\epsilon\tilde{L})\tilde{r}^2+2\tilde{r}-\tilde{Q}^2=R_c({\tilde{r}}),%=\sum_{i=0}^{6}a_{i}\tilde{r}^{i},
\end{equation}
%In order to obtain the analytical solution, for $\epsilon=0$, Eq. (\ref{eqrdust}) is of elliptic type which has analytical solution same as Sec. \ref{sec}(a) %like (\ref{eqrw}).
The analytical solution of (\ref{drcosmological}) for massive particles $(\epsilon=1)$,  %as same as (\ref{rqa}) and (\ref{eqrdust}), 
is same as Sec. \ref{sec}(b) 
\begin{eqnarray}
r(\varphi)=-\frac{\sigma_2(\varphi_{\infty})}{\sigma_1(\varphi_{\infty})}
\end{eqnarray}
and to solve null geodesics $(\epsilon=0)$, Eq. (\ref{drcosmological}) is a polynomial of degree
four, which has analytical solution like Sec. \ref{sec}(a)  
\begin{equation}
\tilde{r}(\phi)=\frac{b_3}{4\wp(\phi-\phi_{in};g_2,g_3)-\frac{b_2}{3}}+\tilde{r}_R.
\end{equation}
Due to exact similarity of Eq. (\ref{drcosmological}) with Eq. (\ref{rqa}) in the quintessence 
surrounding field, except with difference between coefficient, 
solving $R_c(\tilde{r})=0$ and $\frac{dR_c(\tilde{r})}{d\tilde{r}}=0$, 
give us $E^2$ and $\tilde{L}$ diagram similar to the quintessence 
surrounding field (Sec. \ref{sec}(a), \ref{sec}(b)). 
Thus for massive particles $(\epsilon=1)$ in the black hole surrounded 
by the cosmological constant surrounding field, we have
\begin{eqnarray}\label{lecc1}
\tilde{L}=-\frac{2\tilde{Q}^2+\tilde{r}^2-3\tilde{r}}{\tilde{r}^2
	(\tilde{N}_c\tilde{r}^4+\tilde{Q}^2-\tilde{r})},\;\;\;E^2=\frac{(-\tilde{N}_c\tilde{r}^4
	+\tilde{Q}^2+ \tilde{r}^2-2\tilde{r})^2}{(2\tilde{Q}^2
	+\tilde{r}^2-3\tilde{r}) \tilde{r}^2},
\end{eqnarray}
and for massless particles $(\epsilon=0)$
\begin{eqnarray}\label{lecc2}
\tilde{L}=\frac{-\tilde{N}_c\tilde{r}^4+\tilde{Q}^2+\tilde{r}^2-2\tilde{r}}{E^2\tilde{r}^4}.
\end{eqnarray}
Therefore figures of $\tilde{L}-E^2$ diagram (Eqs. (\ref{lecc1}), (\ref{lecc2})) 
for the black hole surrounded by the cosmological constant background, are like 
the quintessence and also the dust surrounding field (with $k\lambda=\frac{1}{4}$),  
%for the case $k\lambda=\frac{1}{4}$ with the region of different types of geodesic motion, in the quintessence surrounding field, 
which has been shown in Figs. \ref{pic:RQC1b} and \ref{pic:RDb}. % and plots of $\tilde{L}-E^2$ for the case $k\lambda=-2$ (Eqs. (\ref{frqc2}), (\ref{veffqc2})), is shown in Fig. \ref{pic:RQC1b}((c),(d)). 
Also results of timelike and null geodesic effective potential Eq. (\ref{veffcc}) is same as 
the figure of the quintessence surrounding field, which is shown in Fig. \ref{pic:RQC1a}. % and \ref{pic:RQC2a}.
%%%%%%%%%%%%%%%%%%%%%%%%   Phantom Field
\subsection{The black hole surrounded by the phantom field}
For the phantom surrounding field, we put $\omega_c=-\frac{4}{3}$ \cite{kieslev}, 
so the metric (\ref{metrics}) can be obtained as
\begin{eqnarray}\label{frp1}
&ds^2=-f_p(\tilde{r})dt^2+\frac{d\tilde{r}^2}{f_p(\tilde{r})}+\tilde{r}^2d \Omega ^2,\\
&f_p(\tilde{r})=1-\frac{2}{\tilde{r}}+\frac{\tilde{Q}^2}{\tilde{r}^2}-\frac{\tilde{N}_p}
{\tilde{r}^{\frac{-3+2k\lambda}{1+k\lambda}}}.
\end{eqnarray}
The equation of effective state parameter $\omega_{eff}$ can be obtained by 
comparing the Eq. (\ref{frp1}) with the Kieslev metric (\ref{metrick}) \cite{Heydarzade:2017wxu} 
\begin{equation}\label{weffp1}
\omega_{eff}=\frac{1}{3}(-1-\frac{3-2k\lambda}{1+k\lambda}).
\end{equation}
Now by considering two values of $ \omega_{eff}\leq -\frac{1}{3}$ and $\omega_{eff}\geq -\frac{1}{3}$ 
in Eq. (\ref{metrick}) \cite{Heydarzade:2017wxu},
the range values of $k\lambda$ in Eq. (\ref{weffp1}) are discernible as 
$-1 \leq k\lambda < \frac{3}{2}$ and $k\lambda \leq -1 \cup k\lambda \geq \frac{3}{2}$ respectively.  
In following, we consider these range values of $k\lambda$ to solving the analytical solution of the equations of motion,  
\begin{itemize}
	\item In the range $\omega_{eff}\leq -\frac{1}{3}$, for $k\lambda=\frac{2}{3}$, we have
\end{itemize}
\begin{eqnarray}\label{frp2}
&ds^2=-f_p(\tilde{r})dt^2+\frac{d\tilde{r}^2}{f_p(\tilde{r})}+\tilde{r}^2d \Omega ^2,\\
&f_p(\tilde{r})=1-\frac{2}{\tilde{r}}+\frac{\tilde{Q}^2}{\tilde{r}^2}-\tilde{N}_p \tilde{r},
\end{eqnarray}
with effective potential
\begin{eqnarray}\label{veffp2}
V_{eff}=(1-\frac{2}{\tilde{r}}+\frac{\tilde{Q}^2}{\tilde{r}^2}-\tilde{N}_p \tilde{r})(\epsilon+\frac{1}{\tilde{L}\tilde{r}^2}),
\end{eqnarray}
so the Eq. (\ref{r5}) get the form
\begin{equation}\label{drphantom}
(\frac{d\tilde{r}}{d\varphi})^2=\tilde{N}_p\epsilon\tilde{L}\tilde{r}^5+(E^2-\epsilon)\tilde{L}\tilde{r}^4+
(2\tilde{L}\epsilon+\tilde{N}_p)\tilde{r}^3-(1+\tilde{Q}^2\epsilon\tilde{L})\tilde{r}^2+2\tilde{r}-\tilde{Q}^2=R_p({\tilde{r}}),
%=\sum_{i=0}^{5}a_{i}\tilde{r}^{i},
\end{equation}
Eq. (\ref{drphantom}) is %a polynomials of degree five and 
exactely similar to massive case $k\lambda=\frac{2}{9}$ in the dust surrounding field 
(Eq. (\ref{eqrd})), except with difference between metric coefficient. 
Thus the analytical solution for $\epsilon=0$ 
is same as Sec. \ref{sec}(a) Weierstrass form Eq. (\ref{eqrw})
\begin{equation}
\tilde{r}(\phi)=\frac{b_3}{4\wp(\phi-\phi_{in};g_2,g_3)-\frac{b_2}{3}}+\tilde{r}_R .
\end{equation}
and for $\epsilon=1$, Eq. (\ref{drphantom}) is of order five which 
answer is the same as Sec. \ref{sec}(b) %given by
%Kleinian sigma function %Eq. (\ref{eqrsigma})
\begin{eqnarray}
\tilde{r}(\varphi)=-\frac{\sigma_2(\varphi_{\infty})}{\sigma_1(\varphi_{\infty})} .
\end{eqnarray}
%%%%%%%%%%%%%%%%%%%%%%%%%%%%%
\begin{itemize}
	\item In the range of $ \omega_{eff}\geq -\frac{1}{3}$, for $k\lambda=\frac{1}{4}$,
\end{itemize}
\begin{eqnarray}\label{frp3}
&ds^2=-(f_p(\tilde{r}))dt^2+\frac{d\tilde{r}^2}{f_p(\tilde{r})}+\tilde{r}^2d \Omega ^2,\\
&f_p(\tilde{r})=1-\frac{2}{\tilde{r}}+\frac{\tilde{Q}^2}{\tilde{r}^2}-\tilde{N}_p\tilde{r}^2,
\end{eqnarray}
with effective potential
\begin{eqnarray}\label{veffp3}
V_{eff}=(1-\frac{2}{\tilde{r}}+\frac{\tilde{Q}^2}{\tilde{r}^2}-\tilde{N}_p\tilde{r}^2)
(\epsilon+\frac{1}{\tilde{L}\tilde{r}^2}),
\end{eqnarray}
and
\begin{equation}\label{drp3}
(\frac{d\tilde{r}}{d\varphi})^2=\tilde{N}_p\epsilon\tilde{L}\tilde{r}^6+((E^2-\epsilon)\tilde{L}+\tilde{N}_p)r^4+
(2\tilde{L}\epsilon)\tilde{r}^3-(1+\tilde{Q}^2\epsilon\tilde{L})\tilde{r}^2+2\tilde{r}-\tilde{Q}^2=R_p({\tilde{r}}),%=\sum_{i=0}^{6}a_{i}\tilde{r}^{i},
\end{equation}
Eq. (\ref{drp3}) is similar to Eq. (\ref{rqa}) in quintessence and also like Eq. (\ref{eqrdust}) 
in the dust surrounding field. 
Thus the analytical solutions % as same as the answers in the same case in both quintessence and dust fields,  
are same as %given by Weierstrass form 
Sec. \ref{sec}(a) 
for massless and %Kleinian sigma function 
Sec. \ref{sec}(b) for massive geodesic. 
%%%%%%%%%%%%%%%%%%%case 3
\begin{itemize}
	\item The next possible value in the range $\omega_{eff}\geq -\frac{1}{3}$ is $k\lambda=4$, 
\end{itemize}
So, the metric (\ref{r5}) will be as the form 
\begin{eqnarray}\label{frp4}
&ds^2=-f_p(\tilde{r})dt^2+\frac{d\tilde{r}^2}{f_p(\tilde{r})}+\tilde{r}^2d \Omega ^2,\\
&f_p(\tilde{r})=1-\frac{2+\tilde{N}_p}{\tilde{r}}+\frac{\tilde{Q}^2}{\tilde{r}^2},
\end{eqnarray}
with effective potential
\begin{eqnarray}\label{veffp4}
V_{eff}=(1-\frac{2+\tilde{N}_p}{\tilde{r}}+\frac{\tilde{Q}^2}{\tilde{r}^2})(\epsilon+\frac{1}{\tilde{L}\tilde{r}^2}),
\end{eqnarray}
then
\begin{equation}\label{drp4}
(\frac{d\tilde{r}}{d\varphi})^2=(E^2-\epsilon)\tilde{L}r^4+(\tilde{N}_p\epsilon+2\epsilon)
\tilde{L}\tilde{r}^3-(1+\tilde{Q}^2\epsilon\tilde{L})\tilde{r}^2+2\tilde{r}
+\tilde{N}_p\tilde{r}-\tilde{Q}^2=R_p(\tilde{r}),
%=\sum_{i=0}^{4}a_{i}\tilde{r}^{i},
\end{equation}
Analytical solution of the Eq.~(\ref{drp4}) for both massive and massless geodesic is given by 
Weierstrass form same as Sec \ref{sec}(a). \\
%%%%%%%%%%%%%%%%%%%%%%%%%%%%%
For a phantom surrounding field with $k\lambda=4$, 
solving $R_p(\tilde{r})=0$ and $\frac{dR_p(\tilde{r})}{d\tilde{r}}=0$ 
give us $\tilde{L}-E^2$ diagram 
similar to the case $k\lambda=-2$ in quintessence surrounding field. 
So for massive particles $(\epsilon=1)$ % in a black hole surrounded 
%by the phantom field $(k\lambda=4)$, we have 
\begin{eqnarray}\label{lep1}
\tilde{L}=-\frac{4\tilde{Q}^2+2\tilde{r}^2-3(2+\tilde{N}_p)\tilde{r}}{\tilde{r}^2
	(2Q^2-(2+\tilde{N}_p)\tilde{r})},\;\;\;E^2=\frac{2(\tilde{Q}^2
	+\tilde{r}^2-(2+\tilde{N}_p)\tilde{r})^2}{(4\tilde{Q}^2+2\tilde{r}^2
	-3(2+\tilde{N}_p)\tilde{r})\tilde{r}^2},
\end{eqnarray}
and for massless particles $(\epsilon=0)$
\begin{eqnarray}\label{lep2}
\tilde{L}=\frac{(\tilde{Q}^2+\tilde{r}^2-(2+\tilde{N}_p)\tilde{r}}{E^2\tilde{r}^4}.
\end{eqnarray}
Therefore the figures of $\tilde{L}-E^2$ diagram (Eqs. (\ref{lep1}), (\ref{lep2})) 
with the region of different types of geodesic motion in the 
phantom surrounding field, has been shown in Fig. \ref{pic:RPb}. Also plots of 
effective potential (Eq. (\ref{veffp4})) is shown in Fig. \ref{pic:RP2}. 
Moreover, a summary of possible orbits type with numbers of zero 
points in each region %for both massive and massless 
are shown in table \ref{tab:RP}. 
%%%%%%%%%%%%%%%%%%%%%%%%%%%%%%%%%%
\section{Orbits}\label{sec4}
\begin{figure}[!ht]
	\centering
	\subfigure[]{
		\includegraphics[width=0.26\textwidth]{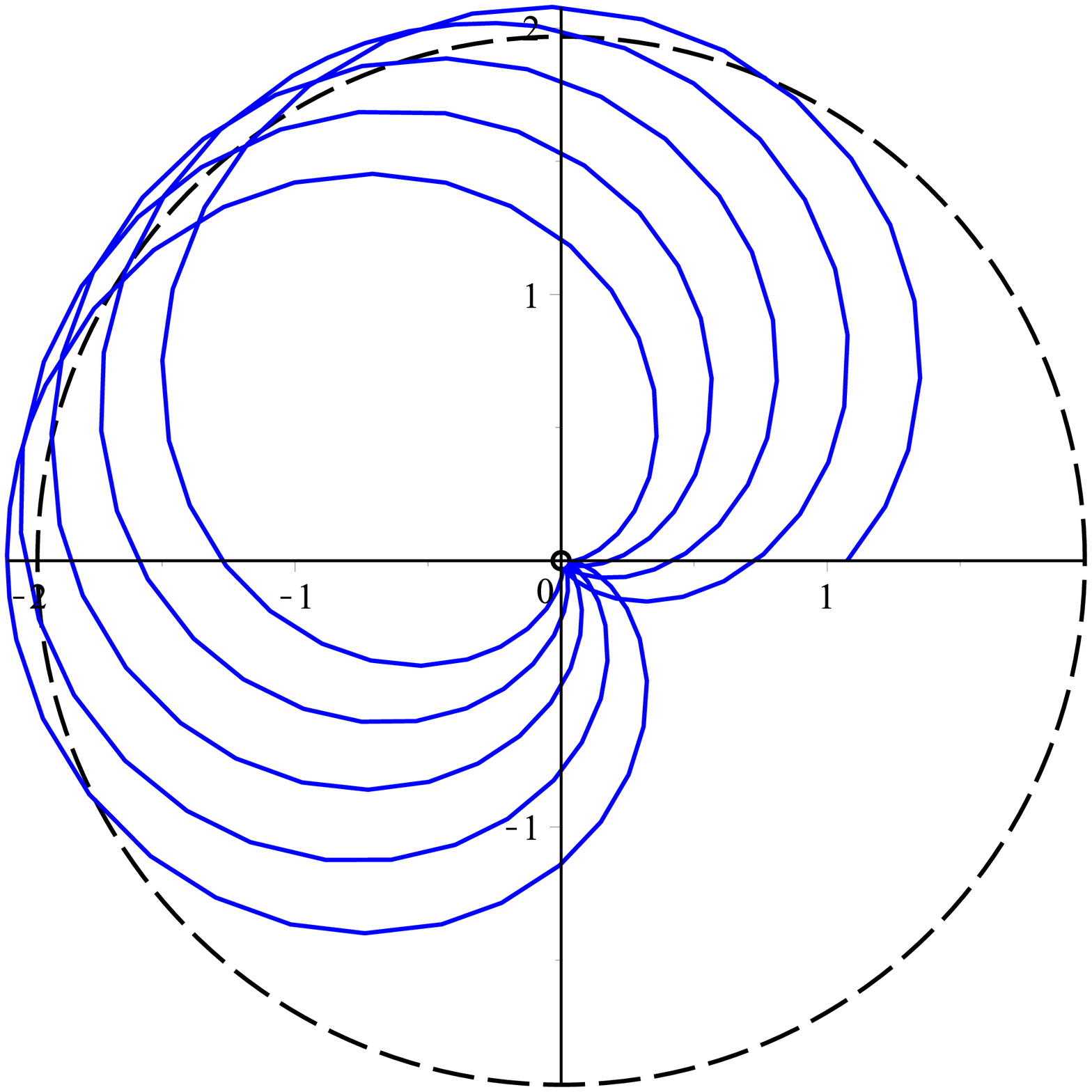}
	}
	\subfigure[]{
		\includegraphics[width=0.26\textwidth]{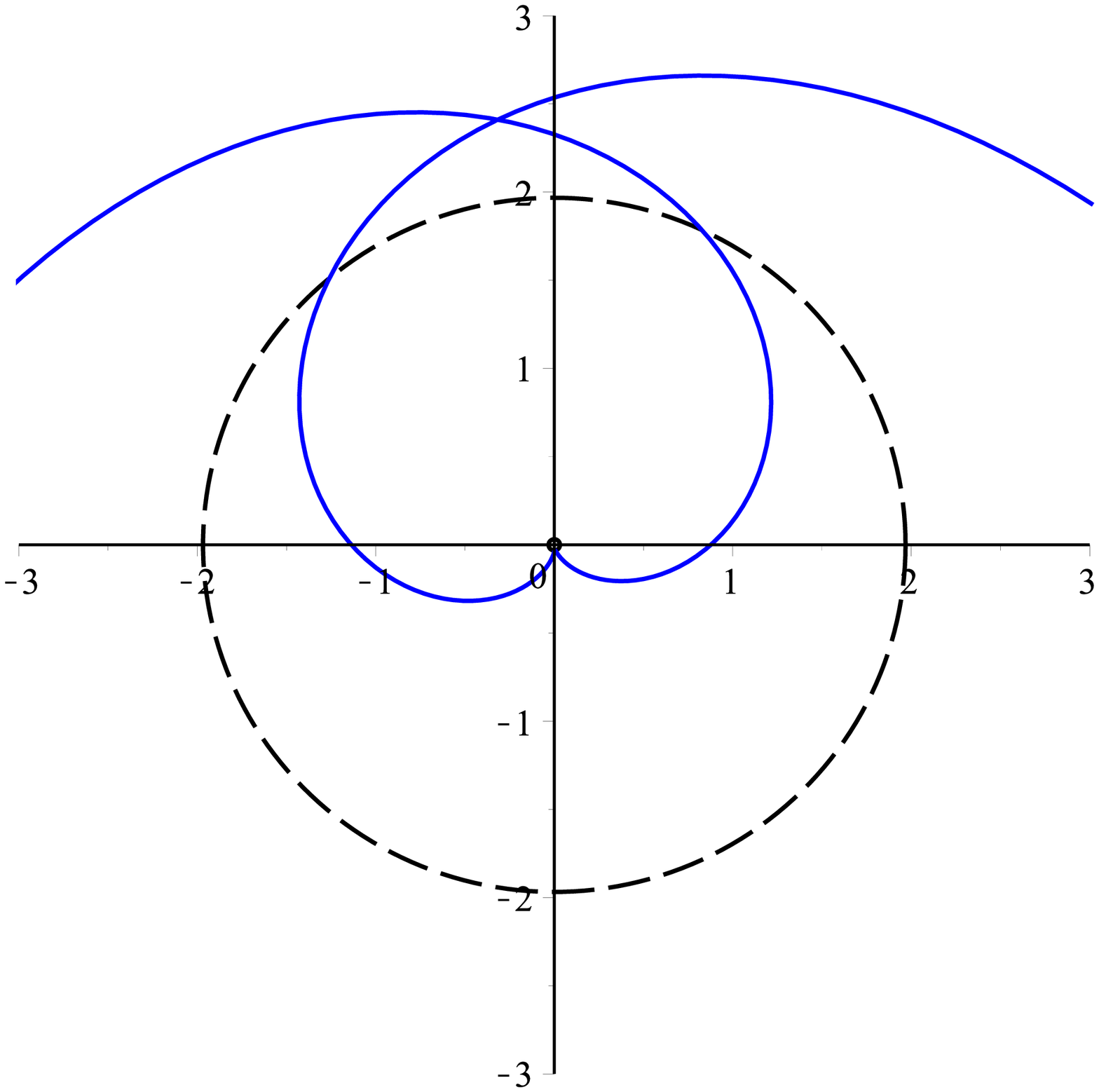}
	}
	\subfigure[]{
		\includegraphics[width=0.26\textwidth]{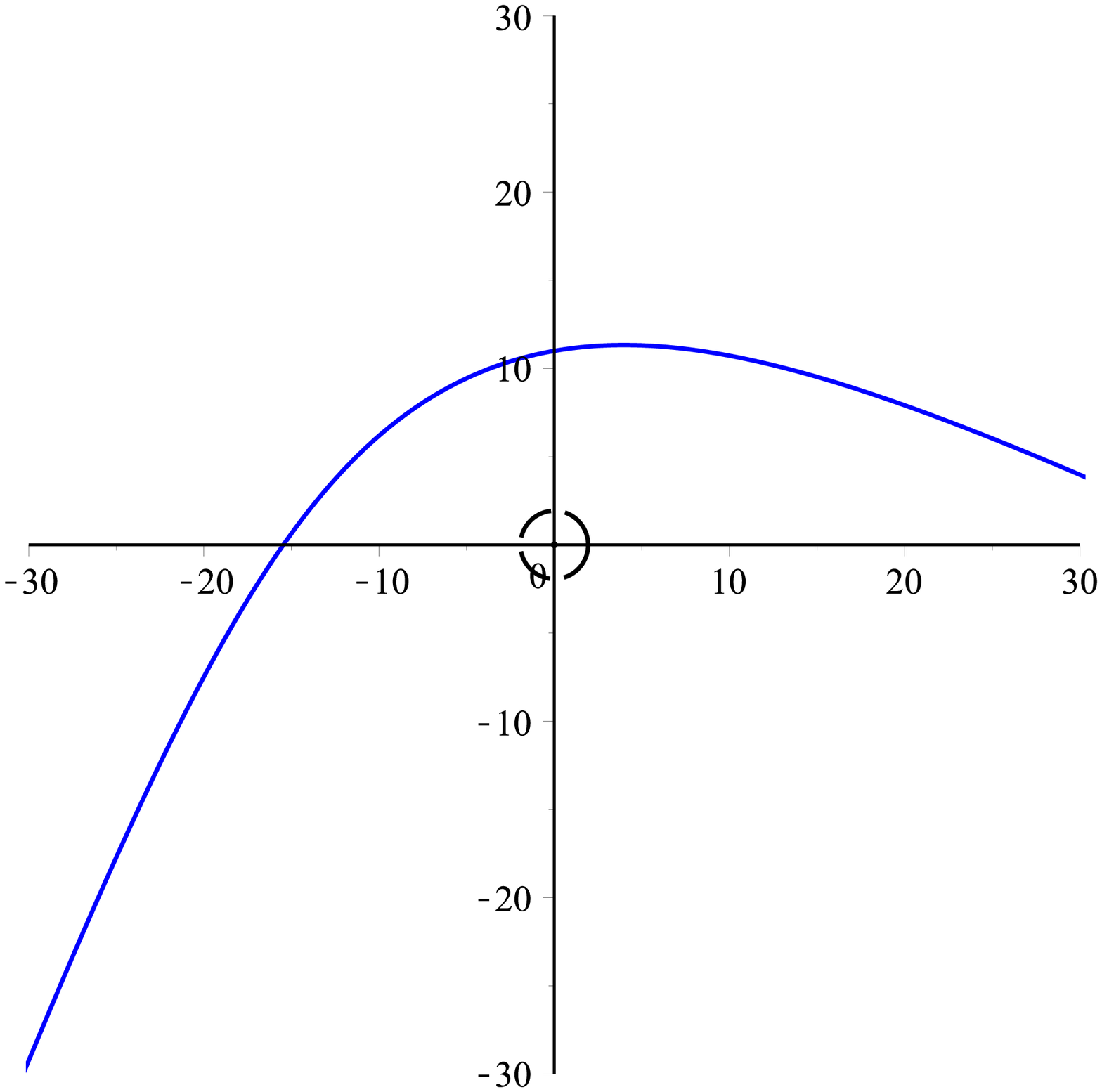}
	}
	\subfigure[]{
		\includegraphics[width=0.26\textwidth]{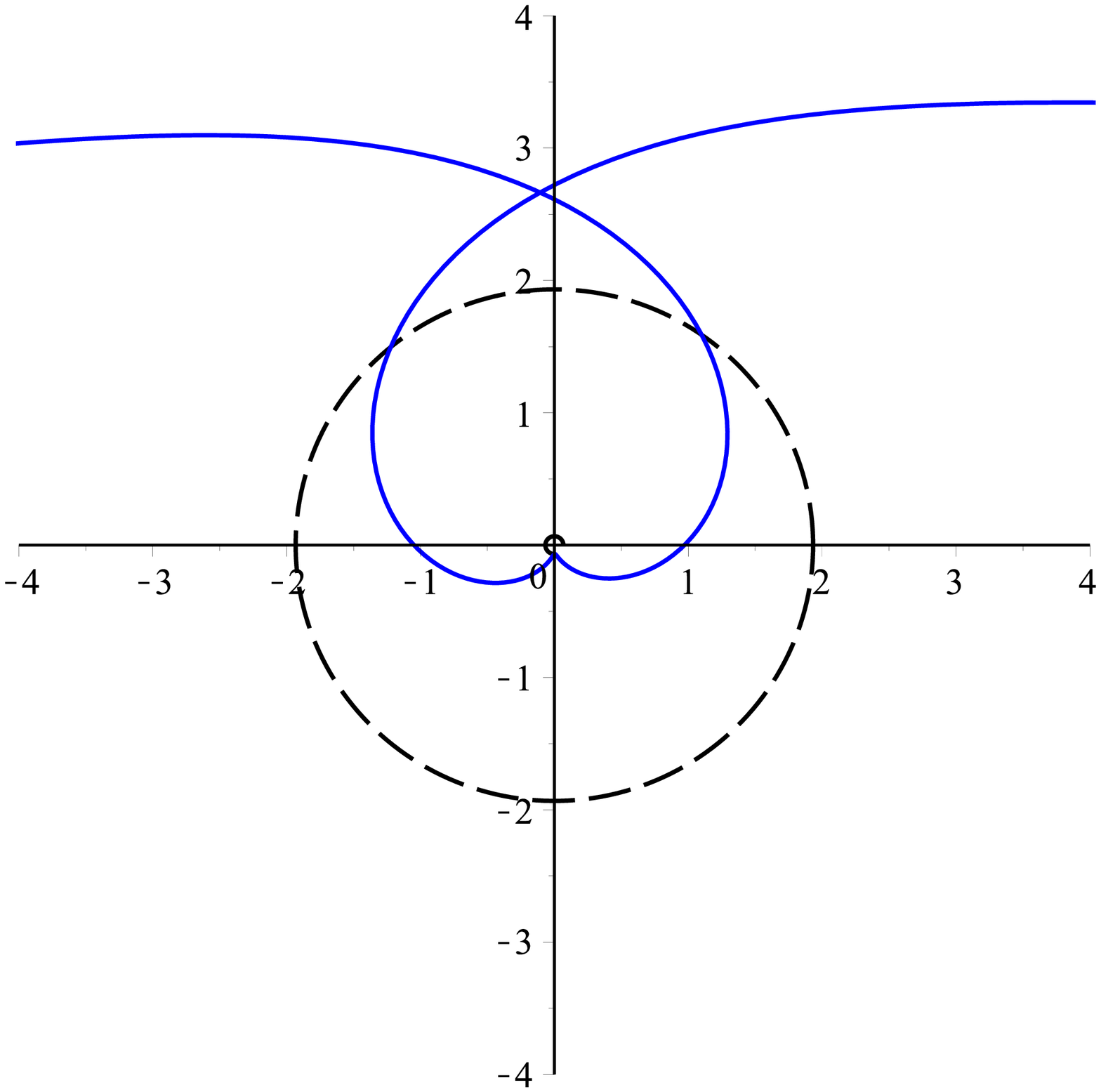}
	}
	\subfigure[]{
		\includegraphics[width=0.26\textwidth]{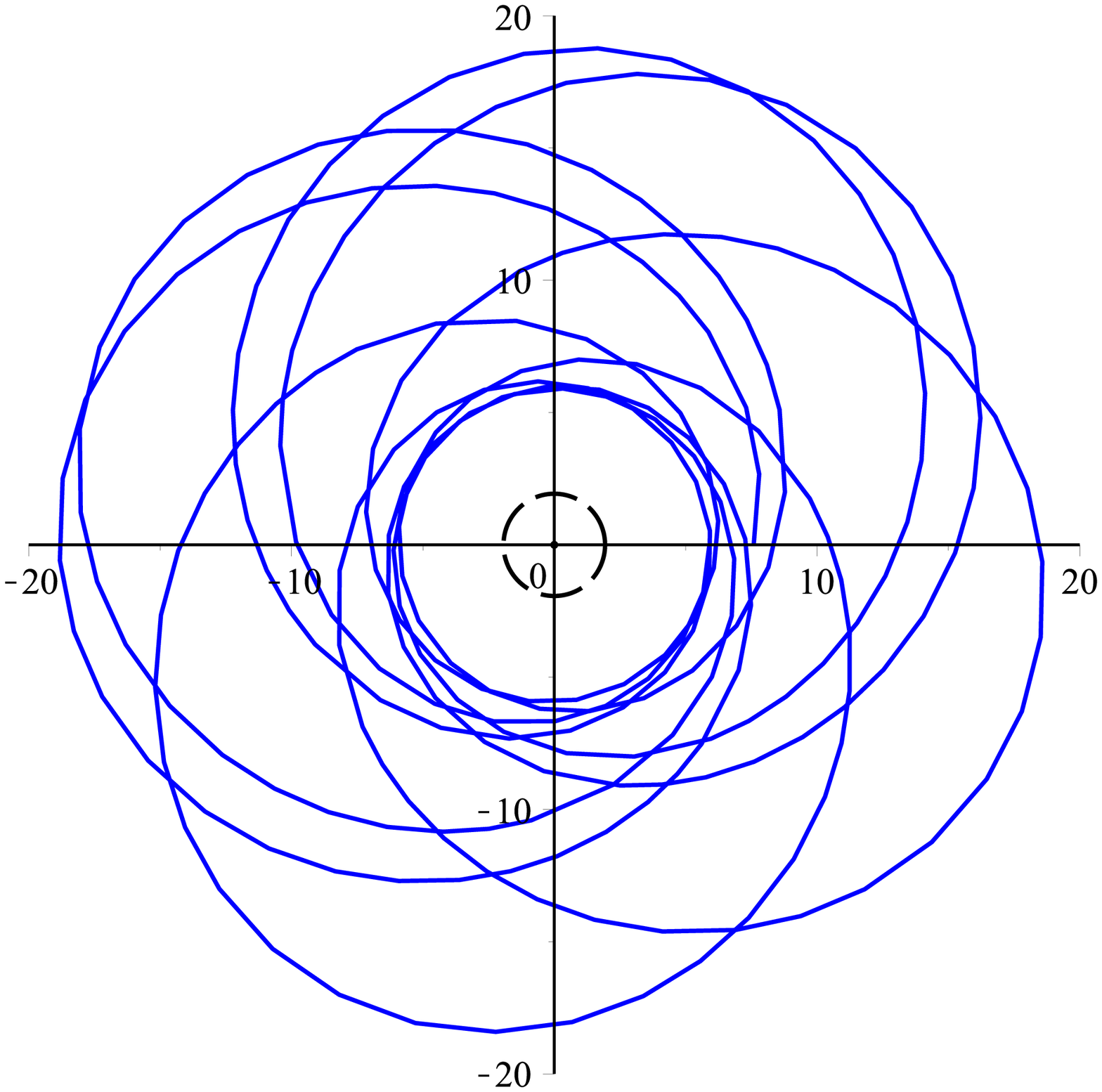}
	}
	\subfigure[]{
		\includegraphics[width=0.26\textwidth]{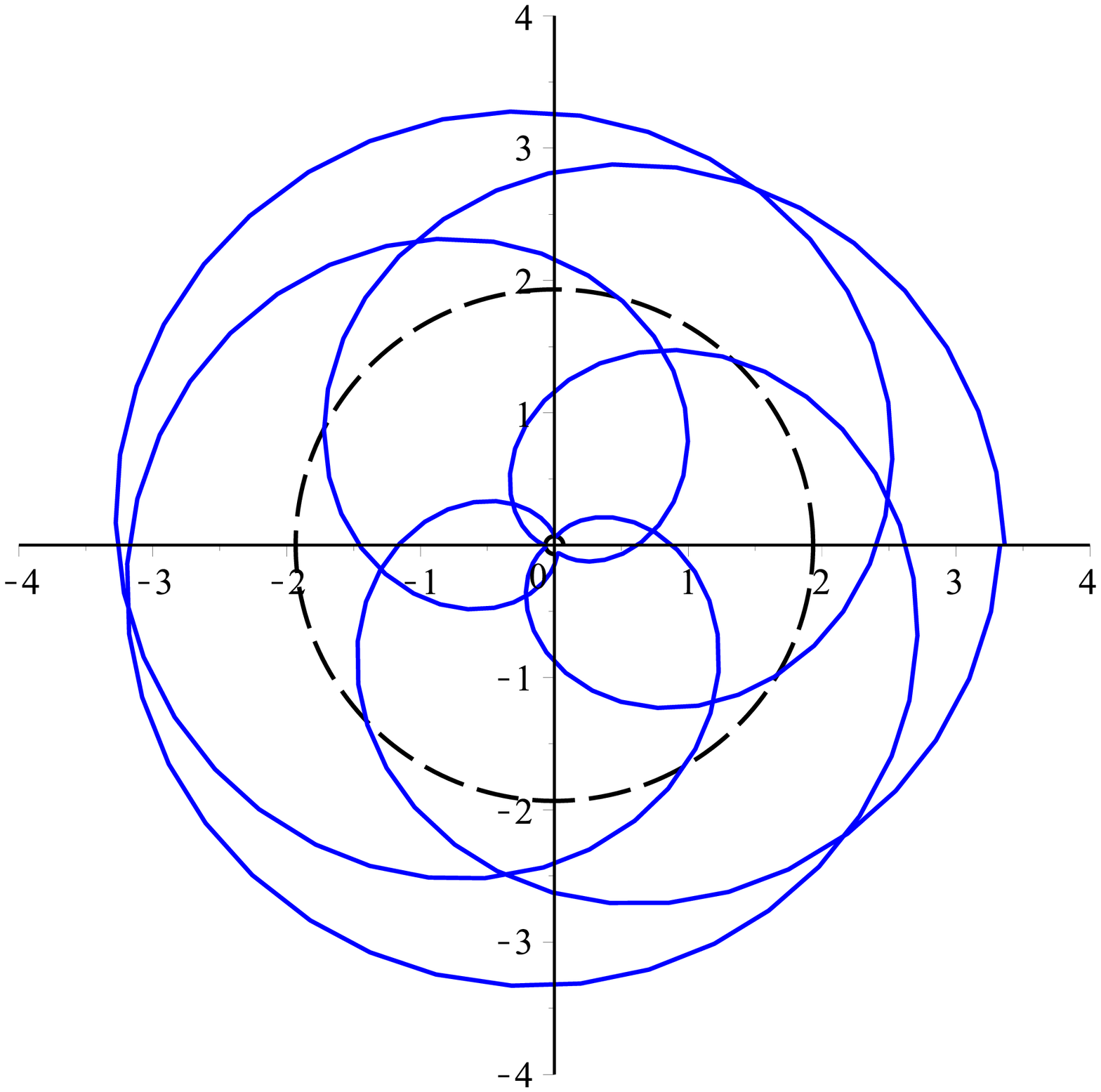}
	}
	\subfigure[]{
		\includegraphics[width=0.26\textwidth]{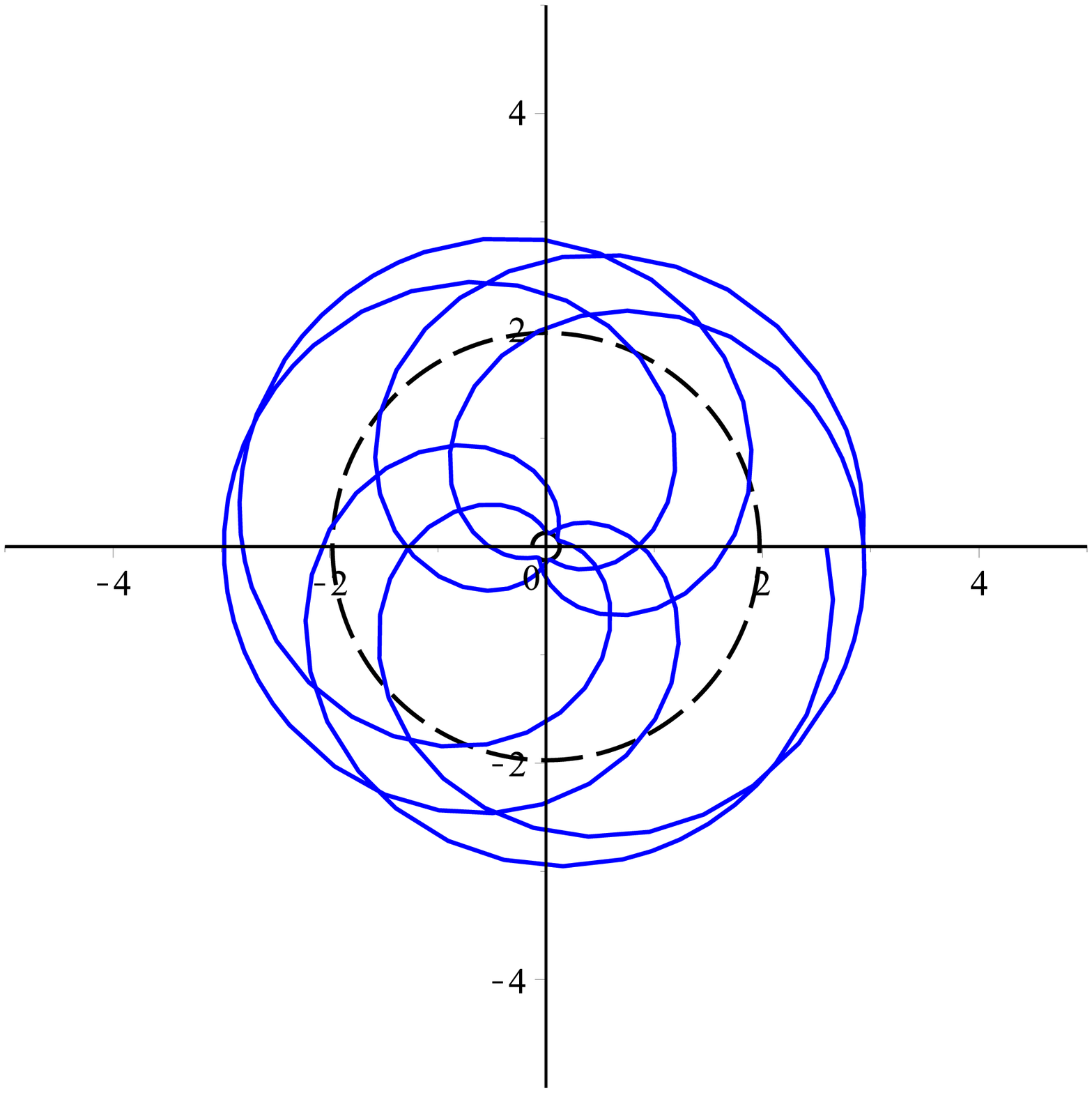}
	}
	\subfigure[]{
		\includegraphics[width=0.26\textwidth]{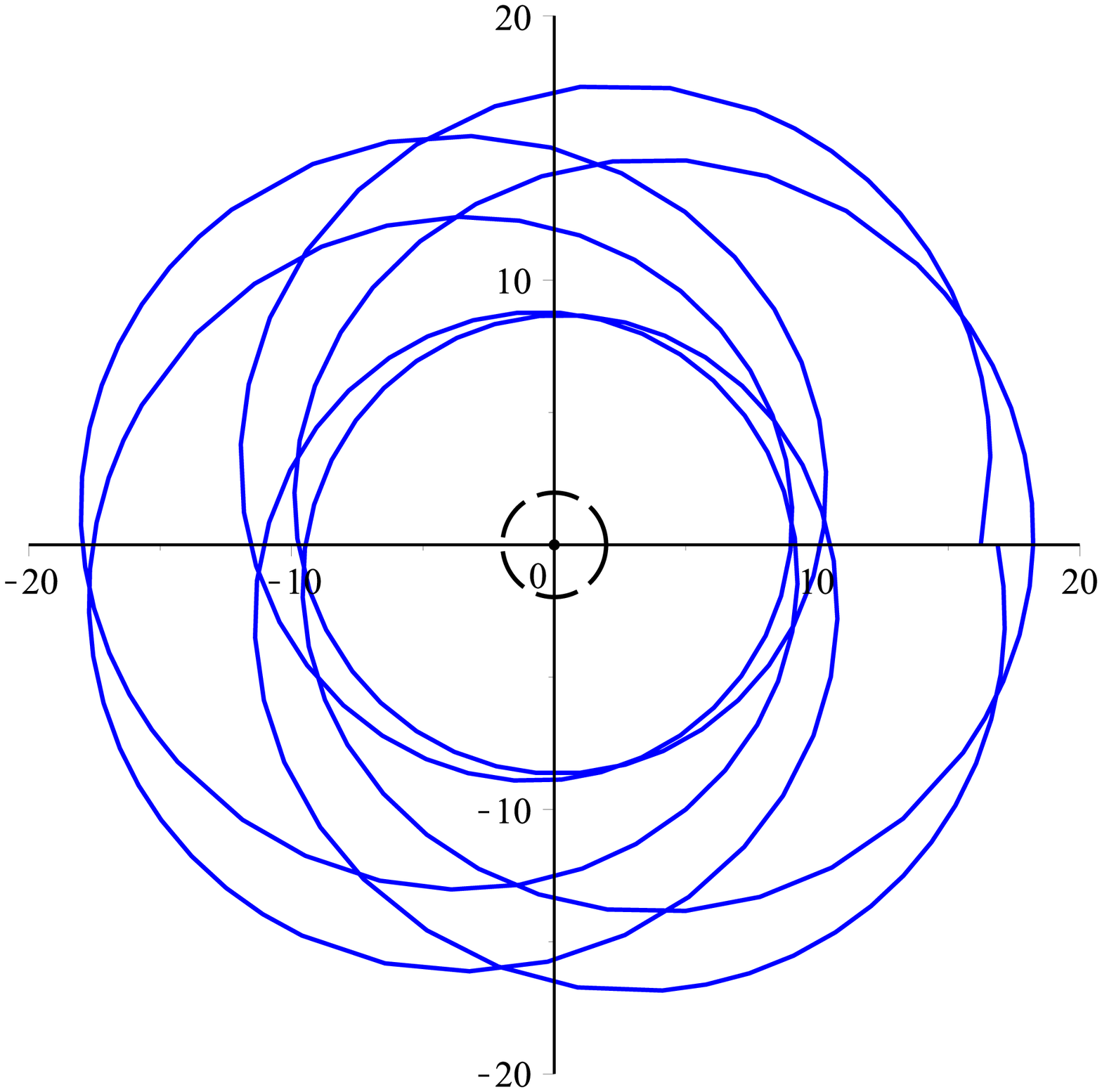}
	}
	\subfigure[]{
		\includegraphics[width=0.26\textwidth]{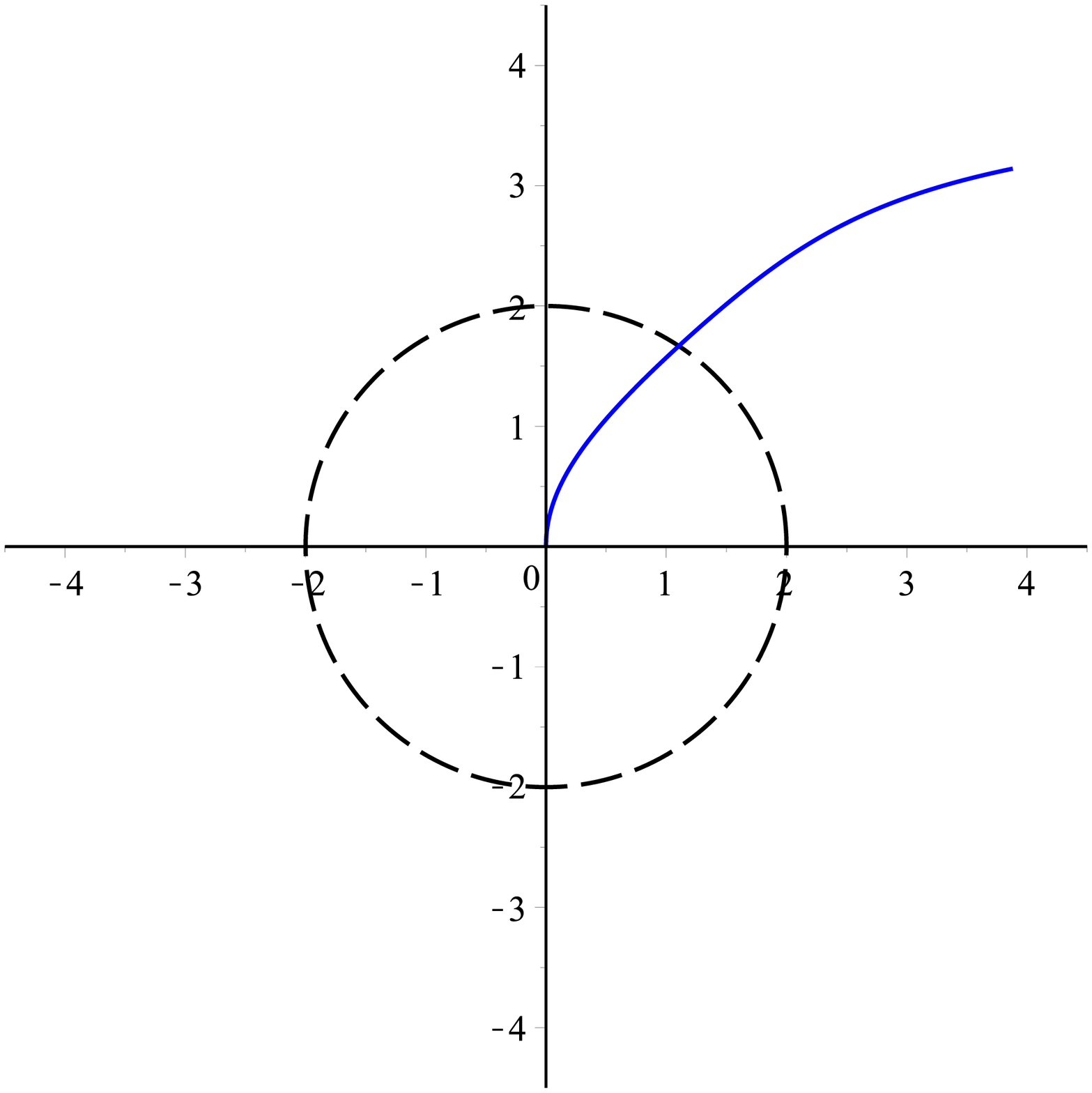}
	}
	\caption{\footnotesize Possible orbits for a black hole surrounded by perfect fluid: Possible orbits for the quintessence surrounding field for $k\lambda=\frac{1}{4}$ 
		corresponding to the  table \ref{tab:RQ1}. (a): Many-world Bound Orbit, with parameters $\tilde{Q}=0.45$, $E=\sqrt{0.93}$ and 
		(b): Two-world Escape Orbit with parameters $\tilde{Q}=\sqrt{0.25}$, $E=\sqrt{0.93}$, $\tilde{N}=0.12$, $L=0.06$. 
	Possible orbits for the radiation field %for $k\lambda=\frac{1}{4}$ 
	corresponding to table \ref{tab:RR}. (c): Escape Orbit, with parameters 
	$\tilde{Q}=0.45$, $E=\sqrt{0.93}$, (d): Two-world Escape Orbit with parameters 
	$\tilde{Q}=\sqrt{0.25}$, $E=\sqrt{0.93}$, $\tilde{N}=0.12$, $L=0.06$, (e): Bound Orbit, 
	with parameters $\tilde{Q}=0.45$, $E=\sqrt{0.93}$ and (f): Many-world Bound Orbit 
	with parameters $E=\sqrt{0.93}$, $\tilde{N}=0.12$ and $\tilde{Q}=\sqrt{0.25}$. 
	Possible orbits for the phantom surrounding field for 
	$k\lambda=\frac{2}{3}$ corresponding to table \ref{tab:RP}. (g): Many-world Bound Orbit, with parameters $\tilde{Q}=0.45$,
	$E=\sqrt{0.93}$ and (h): Bound Orbit with parameters $E=\sqrt{0.93}$, $L=0.06$, $\tilde{N}=0.12$. 
	(i): Terminating orbits for a Schwarzschild black hole corresponding to table \ref{tab:Sh}
	with parameters $E=\sqrt{0.93}$, $L=0.06$, $\tilde{N}=0$ and $\tilde{Q}=0$.
	The blue line shows the path of the orbit and the circle represents the horizon.
}
	\label{pic:RQC1O}
\end{figure}

%%%%%%%%%%%%%%%%%%%%%%%%%
%%%%%%%%%%%%%%%%%%%%%%%%%%%
In this section, we use the analysis provided in the previous sections for 
geodesic equations as well as their analytical solutions,
$\tilde{L}-E^2$ diagrams and also effective potential to plot some examples of possible orbit. 
So we begin with introducing different types of possible orbits. Suppose $\tilde{r}_-$ be the inner horizon
and $\tilde{r}_+$ be the outer event horizon.
\begin{enumerate}	
	\item \textit{Terminating orbit} (TO) with ranges either $\tilde{r} \in [0, \infty)$ or $\tilde{r} \in [0, r_1)$
	with 	$r_1  \geq r_+$.
	\item \textit{Escape orbit} (EO) with range $\tilde{r} \in [r_1, \infty)$ with $r_1>\tilde{r}_+$. 
	%or with range $\tilde{r} \in (-\infty, r_1]$ with  $r_1<0$.
	\item \textit{Bound orbit} (BO) with range $\tilde{r} \in [r_1, r_2]$ with
	\begin{enumerate}
		\item $r_1, r_2  > r_+$, or 
		\item $ 0 < r_1, r_2 < r_-$.
	\end{enumerate}
	\item \textit{Two-world escape orbit} (TEO) with range $[r_1, \infty)$ where $0<r_1 < r_-$.	
	\item \textit{Many-world bound orbit} (MBO) with range $\tilde{r} \in [r_1, r_2]$ 
	where $0<r_1 \leq r_-$ and $r_2 \geq r_+$.
\end{enumerate}
For a certain parameters (E, L, Q, N), different type of orbit is dependent on primary location  
of the test particle or light ray.
In the following, we explain the possible orbit types and examples of effective potentials.
\begin{enumerate}
	\item In Region O, there is no real positive zero,
	so the possible type of orbit is TO [see fig. \ref{pic:RQC1O}(i)].
	
	\item In region I, there is one real positive zero and the kind of possible orbit is 
	TEO [see figs. \ref{pic:RQC1O} (b) and \ref{pic:RQC1O} (d)]. 
	
	\item In region II, there are two real positive zeros, therefore the kind of possible orbit is 
	MBO [see figs. \ref{pic:RQC1O} (a), \ref{pic:RQC1O} (f) and \ref{pic:RQC1O} (g)]. 
	
	\item In region III, there are three real and positive zeros, therefore the kind of two possible orbits are 
	EO [see fig. \ref{pic:RQC1O} (c)] and MBO 
	[see figs. \ref{pic:RQC1O} (a), \ref{pic:RQC1O} (f) and \ref{pic:RQC1O} (g)].
	
	\item In region IV, there are four real and positive zeros, so the kind of two 
	possible orbits are BO [see figs. \ref{pic:RQC1O} (e), \ref{pic:RQC1O} (h)] and 
	MBO [see figs. \ref{pic:RQC1O} (a), \ref{pic:RQC1O} (f) and \ref{pic:RQC1O} (g)].
	
	\item In region V, there are five real and positive zeros, so the kind of three possible orbits are 
	EO [see fig. \ref{pic:RQC1O} (c)], BO [see figs. \ref{pic:RQC1O} (e), \ref{pic:RQC1O} (h)] and 
	MBO [see figs. \ref{pic:RQC1O} (a), \ref{pic:RQC1O} (f) and \ref{pic:RQC1O} (g)].
\end{enumerate}
%\clearpage
%%%%%%%%%%%%% Quin-C1(LE2-Table-potential)   %%%%%%%%%%%%
%%%%%%%%%%%%%%%% QC1
%\paragraph{Quintessence surrounding field (with $k\lambda=\frac{1}{4}$)}
\begin{enumerate}
	\item Quintessence surrounding field (with $k\lambda=\frac{1}{4}$)
	
		\begin{table}[!ht]
		\begin{center}
			\begin{tabular}{|l|l|c|l|}
				%{|lccll|}
				\hline
				region & pos.zeros & range of $\tilde{r}$ &  orbit \\
				\hline\hline
				I & 1 &
				$|$$--$$\lVert$$\bullet$$\textbf{--------}$$\lVert$$\textbf{-----------------------}$%$\dashrightarrow$
				
				& TEO
				\\  \hline
				III & 3 &
				$|$$--$$\lVert$$\bullet$$\textbf{--------}$$\lVert$$\textbf{---}$$\bullet$$--$$\bullet$$\textbf{--------------}$%$\dashrightarrow$
				& MBO, EO
				\\ \hline
				V & 5 &
				$|$$--$$\lVert$$\bullet$$\textbf{--------}$$\lVert$$\textbf{---}$$\bullet$$--$$\bullet$$\textbf{------}$$\bullet$$-$$\bullet$$\textbf{---}$%$\dashrightarrow$
				& MBO, BO, EO
				\\ \hline
			\end{tabular}
			\caption{Types of orbits of the quintessence surrounding field for $k\lambda=\frac{1}{4}$. The lines represent the range of the orbits. The dots
				show the turning points of the orbits. The positions of the two horizons are marked by a vertical double line. The
				single vertical line indicates the singularity at $\tilde{r}=0$.}
			\label{tab:RQ1}
		\end{center}
	\end{table}
	
	\begin{figure}[!ht]
		\centering
		\subfigure[]{
			\includegraphics[width=0.2\textwidth]{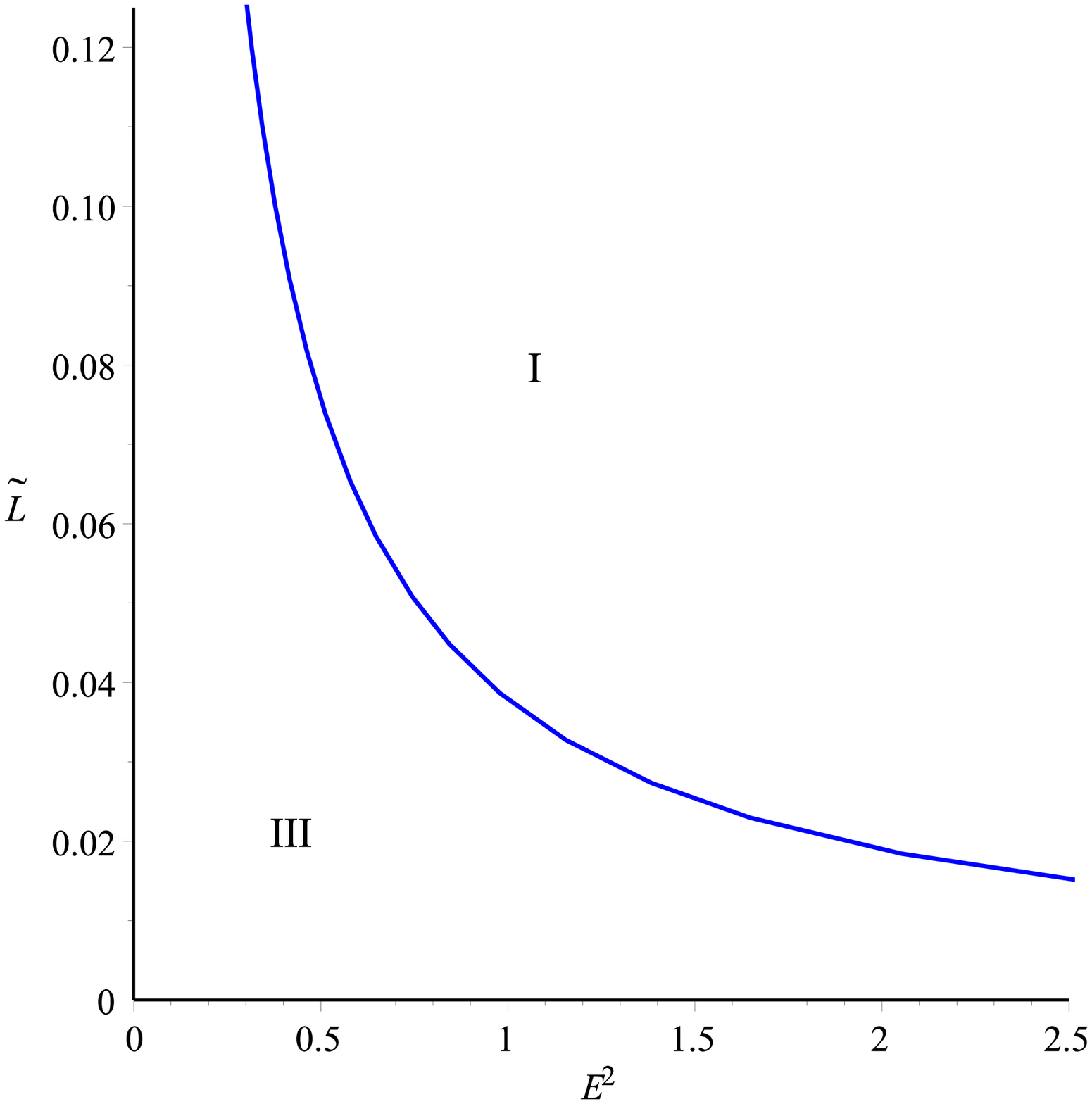}
		}
		\subfigure[]{
			\includegraphics[width=0.2\textwidth]{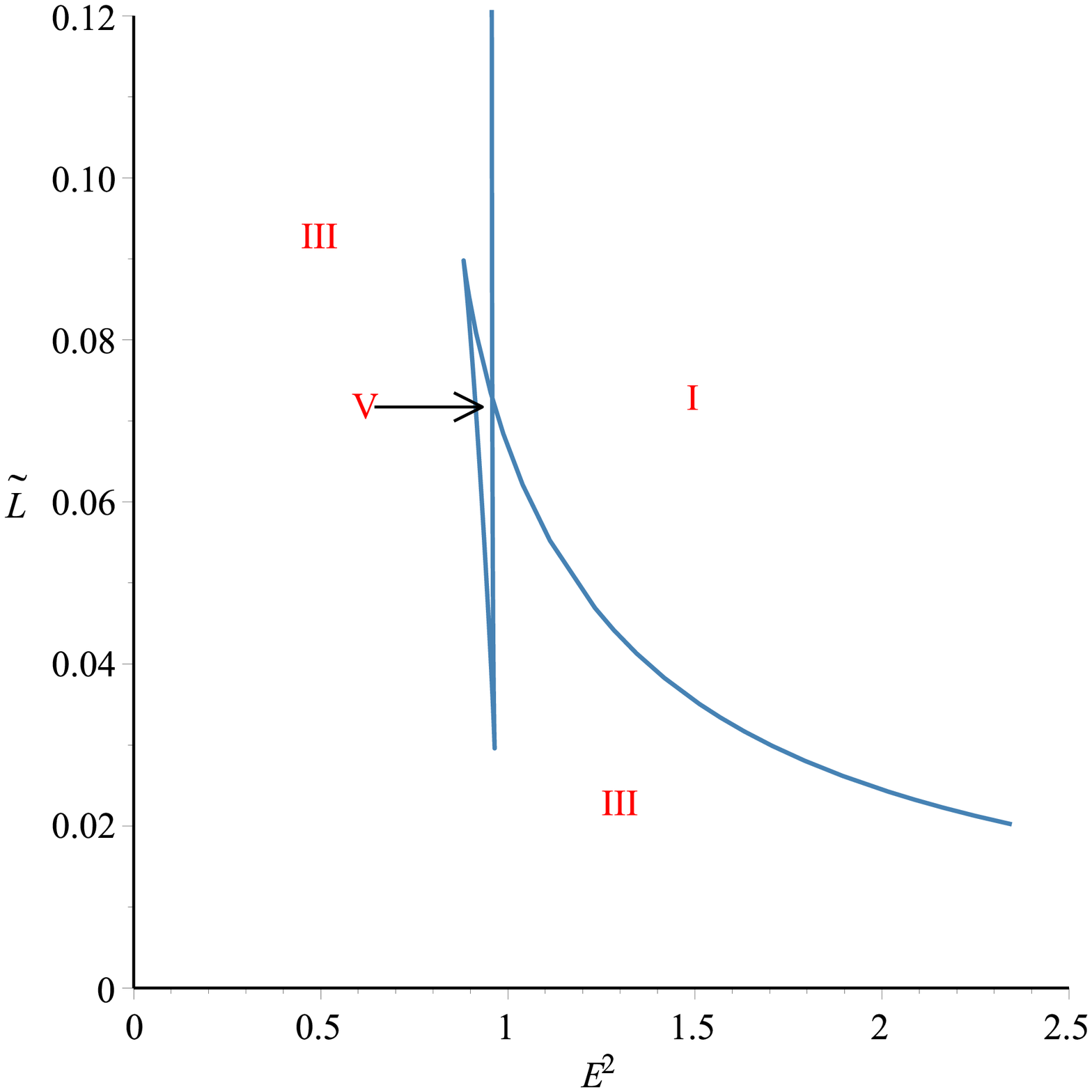}
		}
		\caption{\footnotesize Plots of $L-E^2$ diagram and regions of different types of geodesic motion in 
			quintessence surrounding field with the parameters $\tilde{N}=\frac{1}{3*10^5}, \tilde{Q}=\sqrt{0.25}$ 
			for (a): Null geodesic ($\epsilon=0$) and (b): Timelike geodesic ($\epsilon=1$). 
			The numbers of positive real zeros in these regions are: I=1, III=3, V=5.}
		\label{pic:RQC1b}
	\end{figure}
	%%%%%%%%%%%%%%%%%%% QC1-potential
	\begin{figure}[!ht]
		\centering
		\subfigure[$\epsilon=1$, $\tilde{N}=\frac{1}{3*10^5}$, $\tilde{Q}=\sqrt{0.25}$, $\tilde{L}=0.07$, $E=1$]{
			\includegraphics[width=0.17\textwidth]{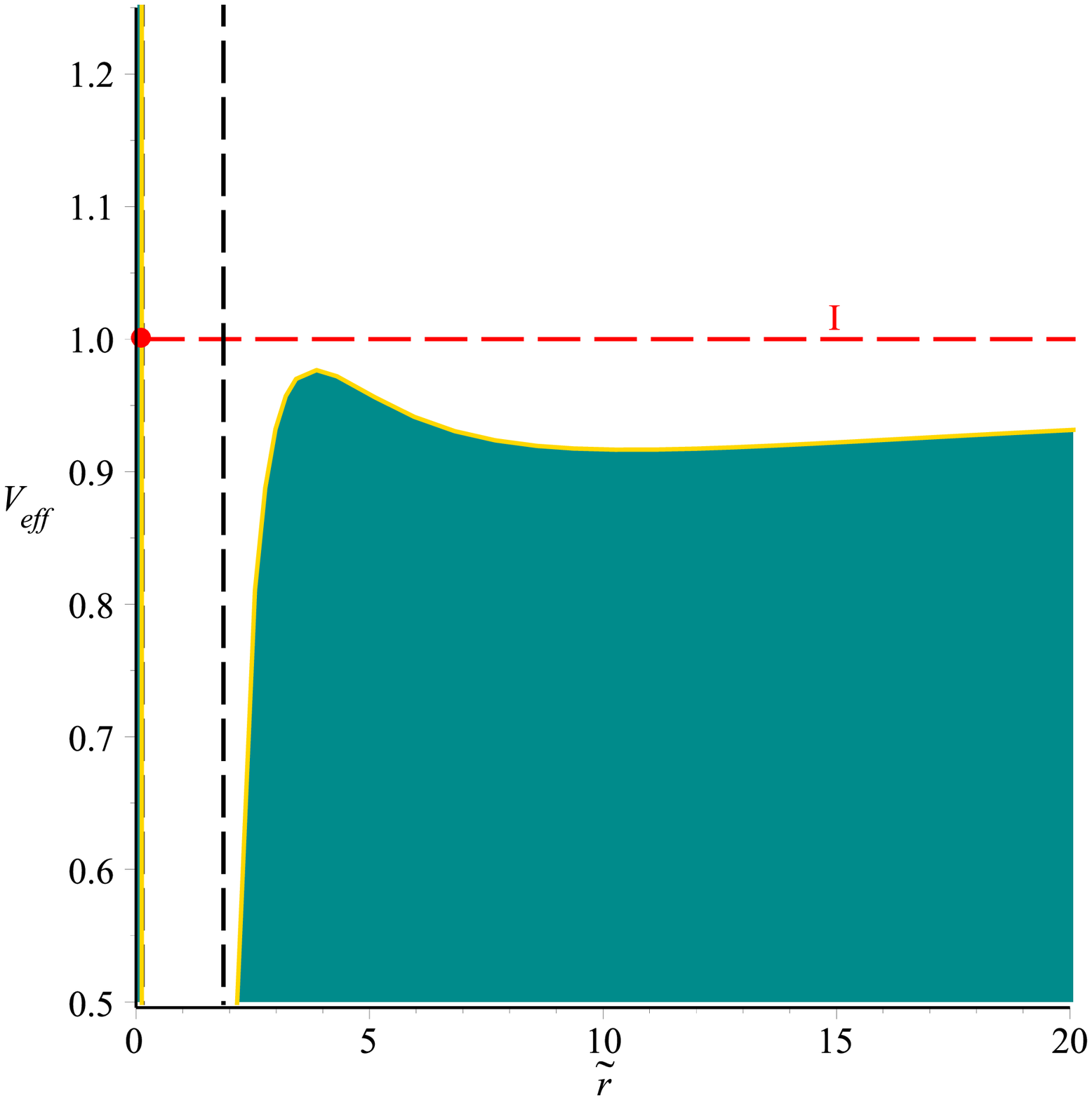}
		}
		\subfigure[$\epsilon=1$, $\tilde{N}=\frac{1}{3*10^5}$, $\tilde{Q}=\sqrt{0.25}$, $\tilde{L}=0.07$, $E=\sqrt{0.97}$]{
			\includegraphics[width=0.17\textwidth]{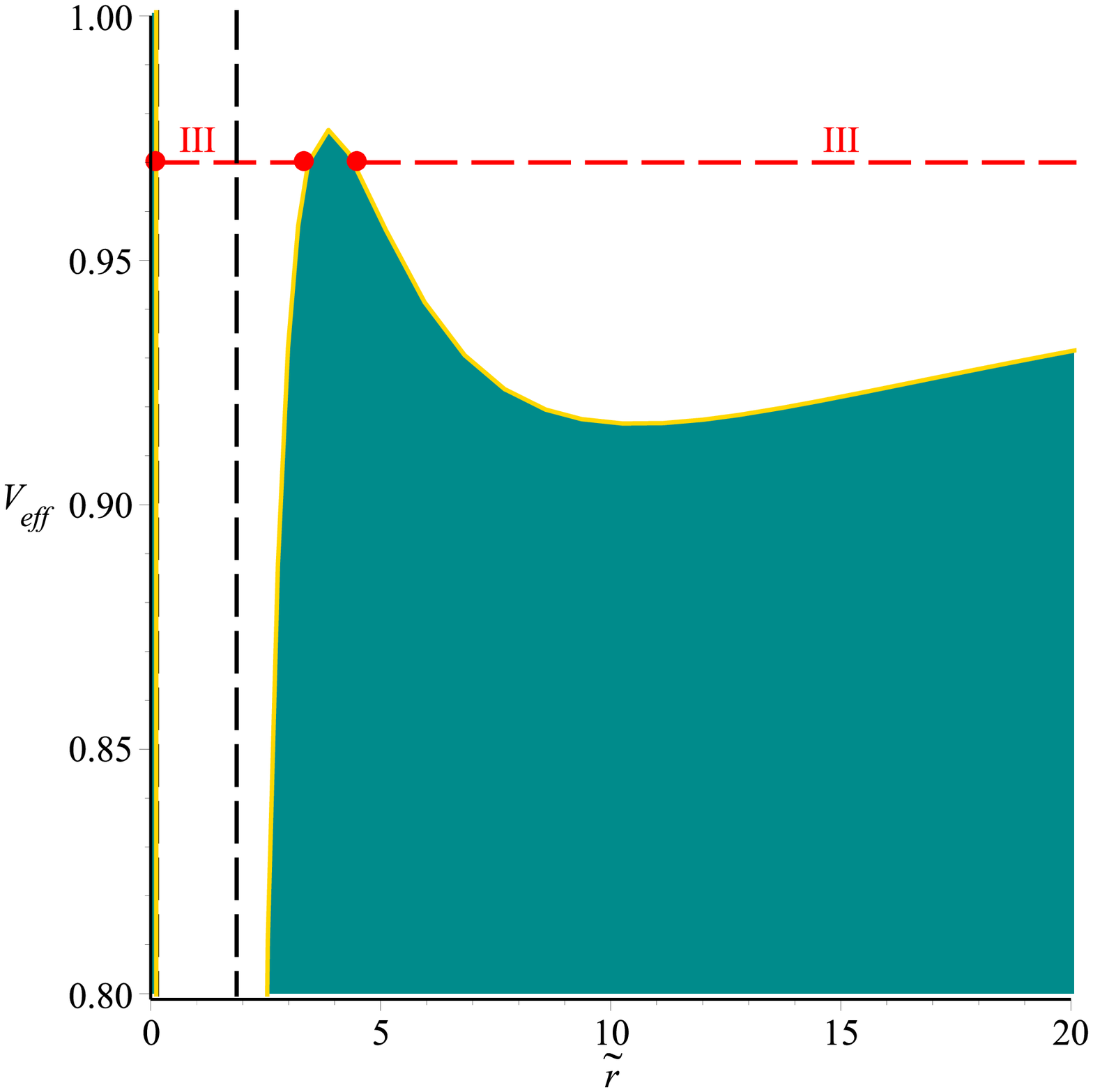}
		}
		\subfigure[$\epsilon=1$, $\tilde{N}=\frac{1}{3*10^5}$, $\tilde{Q}=\sqrt{0.25}$, $\tilde{L}=0.07$, $E=\sqrt{0.95}$]{
			\includegraphics[width=0.17\textwidth]{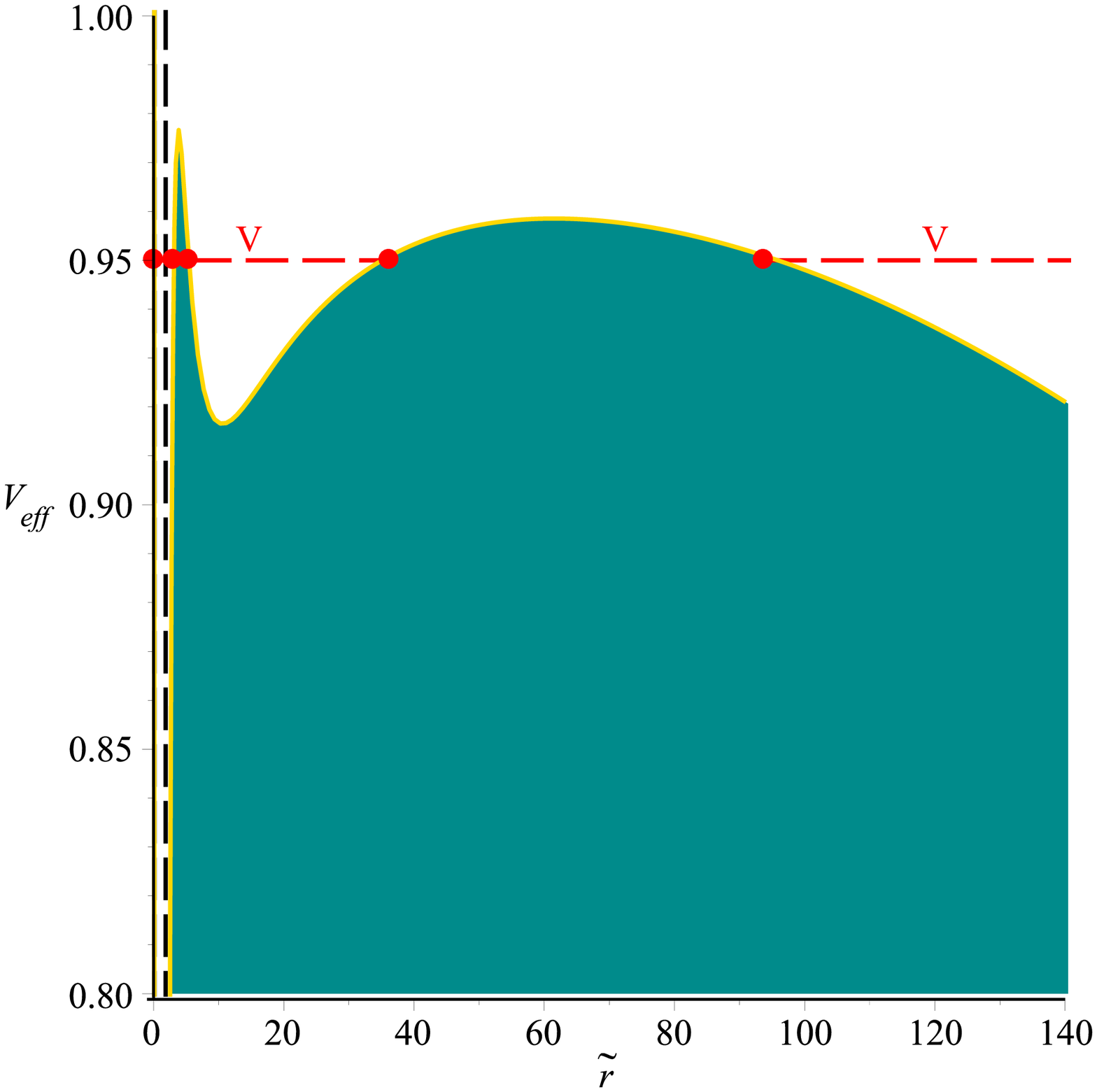}
		}
		\subfigure[Clouseup of figure (c)]{
			\includegraphics[width=0.17\textwidth]{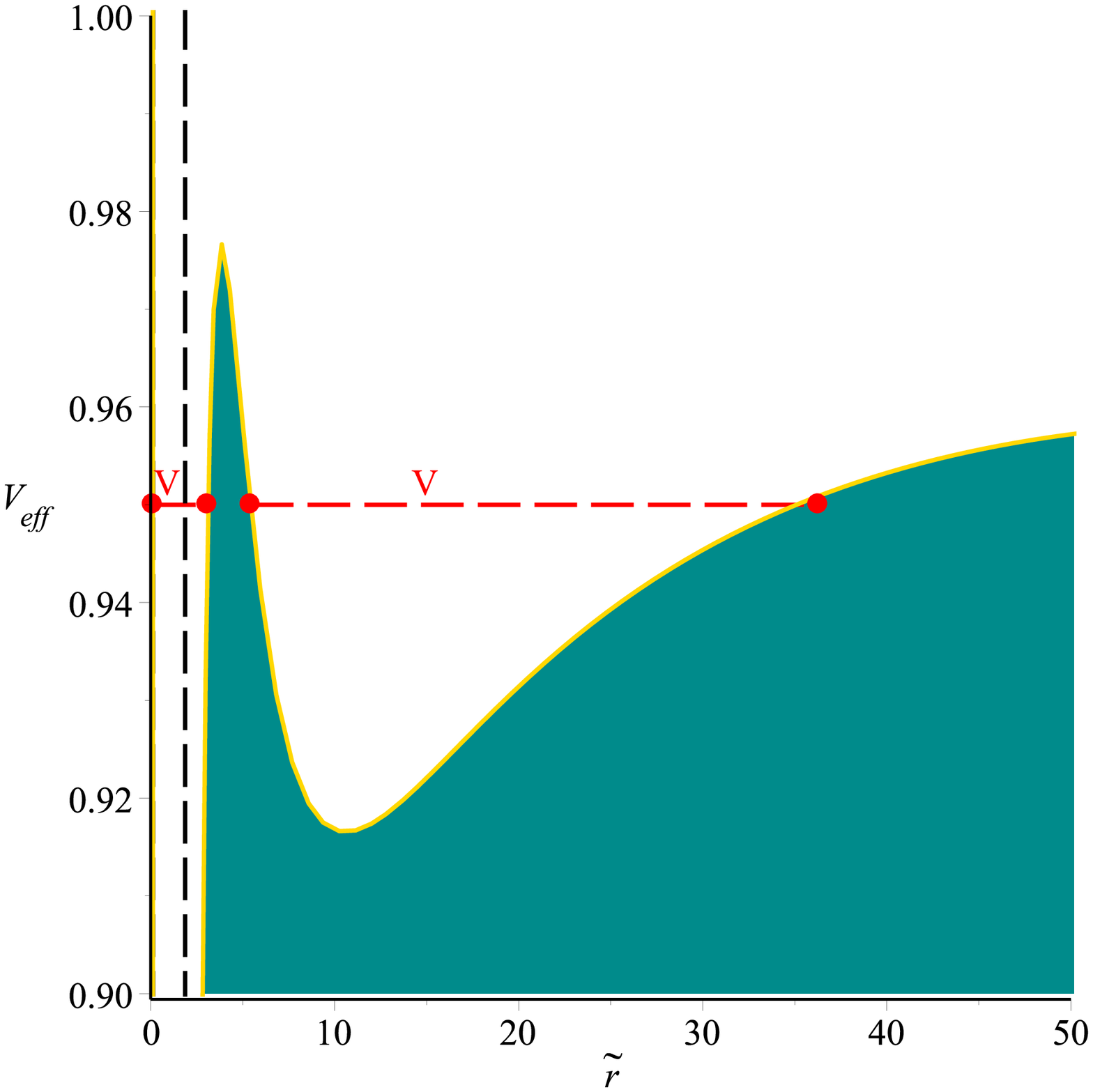}
		}
		\subfigure[$\epsilon=0$, $\tilde{N}=\frac{1}{3*10^5}$, $\tilde{Q}=0.25$, $\tilde{L}=0.05$]{
			\includegraphics[width=0.17\textwidth]{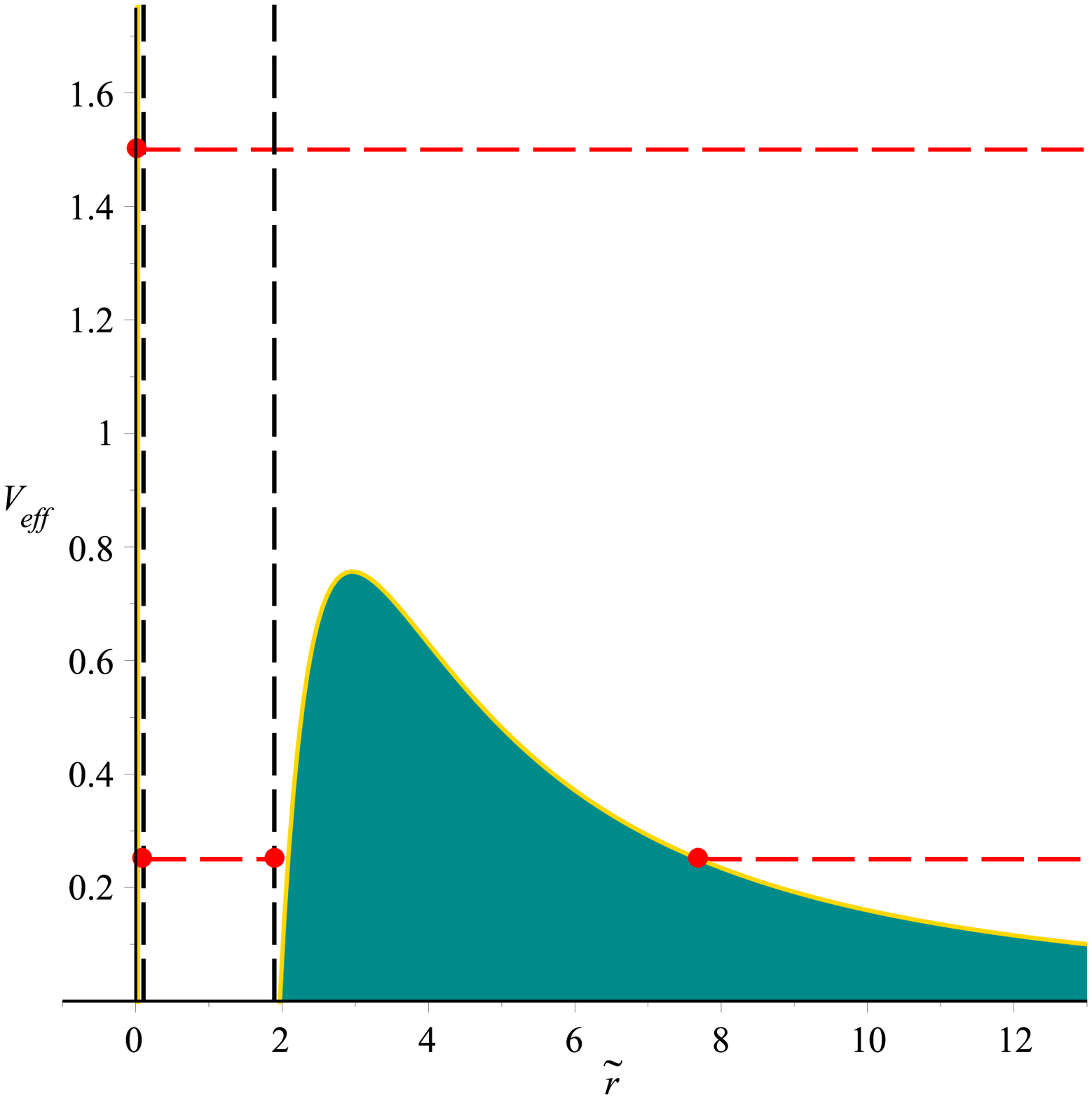}
		}
		\caption{\footnotesize Plots of the effective potential for the different orbit types of Table \ref{tab:RQ1}, 
			for the case of quintessence surrounding field with the parameters $k\lambda=\frac{1}{4}$. 
			The horizontal red-dashed lines denotes the squared energy parameter $E^2$. 
			The vertical black dashed lines show the position of the horizons. 
			The red dots marks denote the zeros of the polynomial $R$, which are the 
			turning points of the orbits. In the cyan area, no motion is possible since $\tilde{R} < 0$ .}
		\label{pic:RQC1a}
	\end{figure}
	% \clearpage
	\clearpage
	
	%%%%%%%%%%%%%%%%%%%%%%%%%%%%%% QC2-potential
	% \paragraph{Quintessence surrounding field (with $k\lambda=-2$) }
	\item Quintessence surrounding field (with $k\lambda=-2$)
	
	\begin{table}[!ht]
		\begin{center}
			\begin{tabular}{|l|l|c|l|}
				%{|lccll|}
				\hline
				region & pos.zeros & range of $\tilde{r}$ &  orbit \\
				\hline\hline
				I & 1 &
				$|$$--$$\lVert$$\bullet$$\textbf{--------}$$\lVert$$\textbf{-----------------------}$%$\dashrightarrow$
				
				& TEO
				\\  \hline
				II & 2 &
				$|$$--$$\lVert$$\bullet$$\textbf{--------}$$\lVert$$\textbf{---}$$\bullet$$-------$%$\dashrightarrow$
				& MBO
				\\ \hline
				III & 3 &
				$|$$--$$\lVert$$\bullet$$\textbf{--------}$$\lVert$$\textbf{---}$$\bullet$$--$$\bullet$$\textbf{--------------}$%$\dashrightarrow$
				& MBO, EO
				\\ \hline
				IV & 4 &
				$|$$--$$\lVert$$\bullet$$\textbf{--------}$$\lVert$$\textbf{---}$$\bullet$$--$$\bullet$$\textbf{-----}$$\bullet$ $---$
				& MBO, BO
				\\ \hline
			\end{tabular}
			\caption{Types of orbits of the quintessence surrounding field for $k\lambda=-2$. The lines represent the range of the orbits. The dots
				show the turning points of the orbits. The positions of the two horizons are marked by a vertical double line. The
				single vertical line indicates the singularity at $\tilde{r}=0$.}
			\label{tab:RQ2}
		\end{center}
	\end{table}
	
	\begin{figure}[!ht]
		\centering
		\subfigure[]{
			\includegraphics[width=0.2\textwidth]{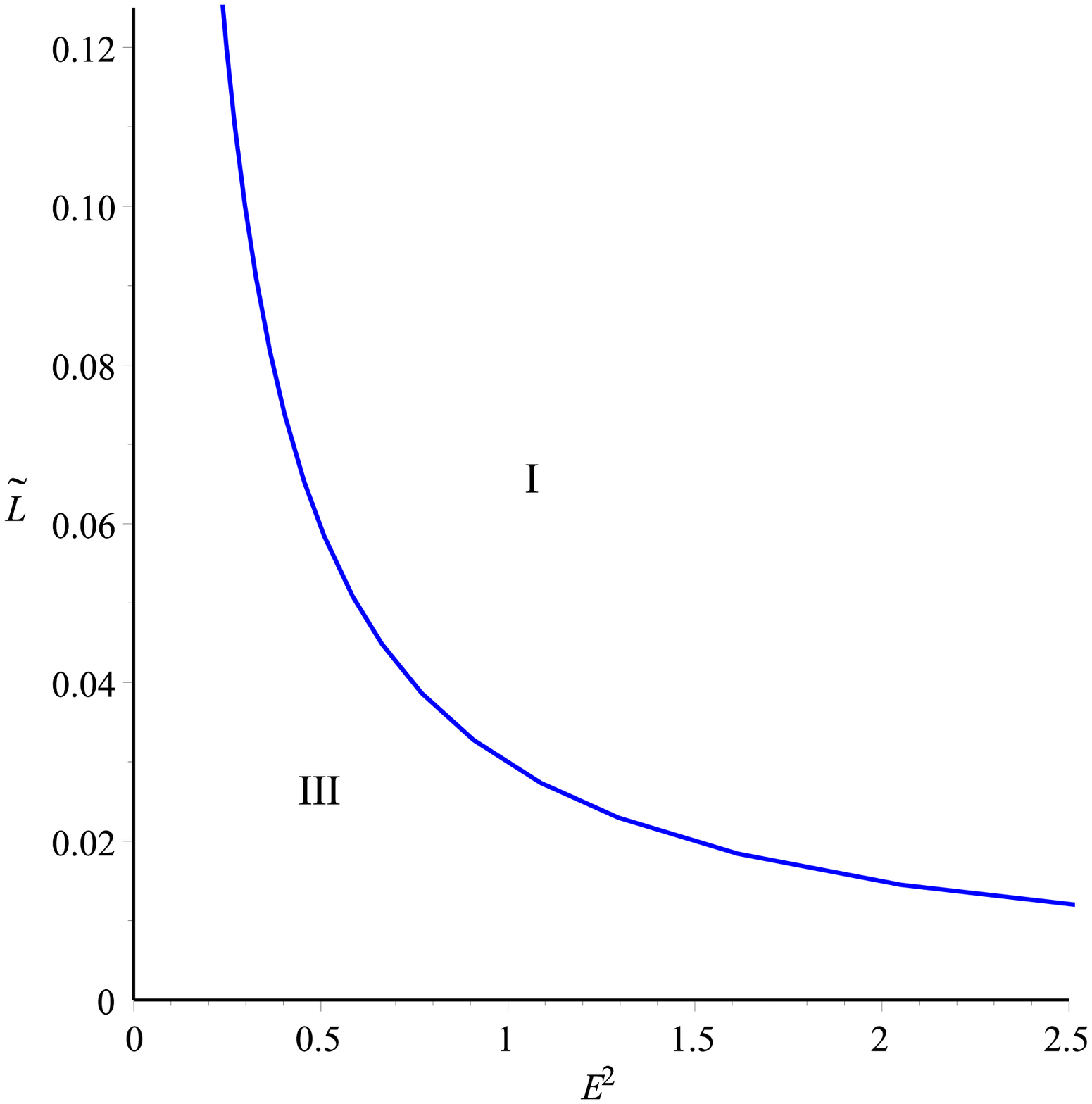}
		}
		\subfigure[]{
			\includegraphics[width=0.2\textwidth]{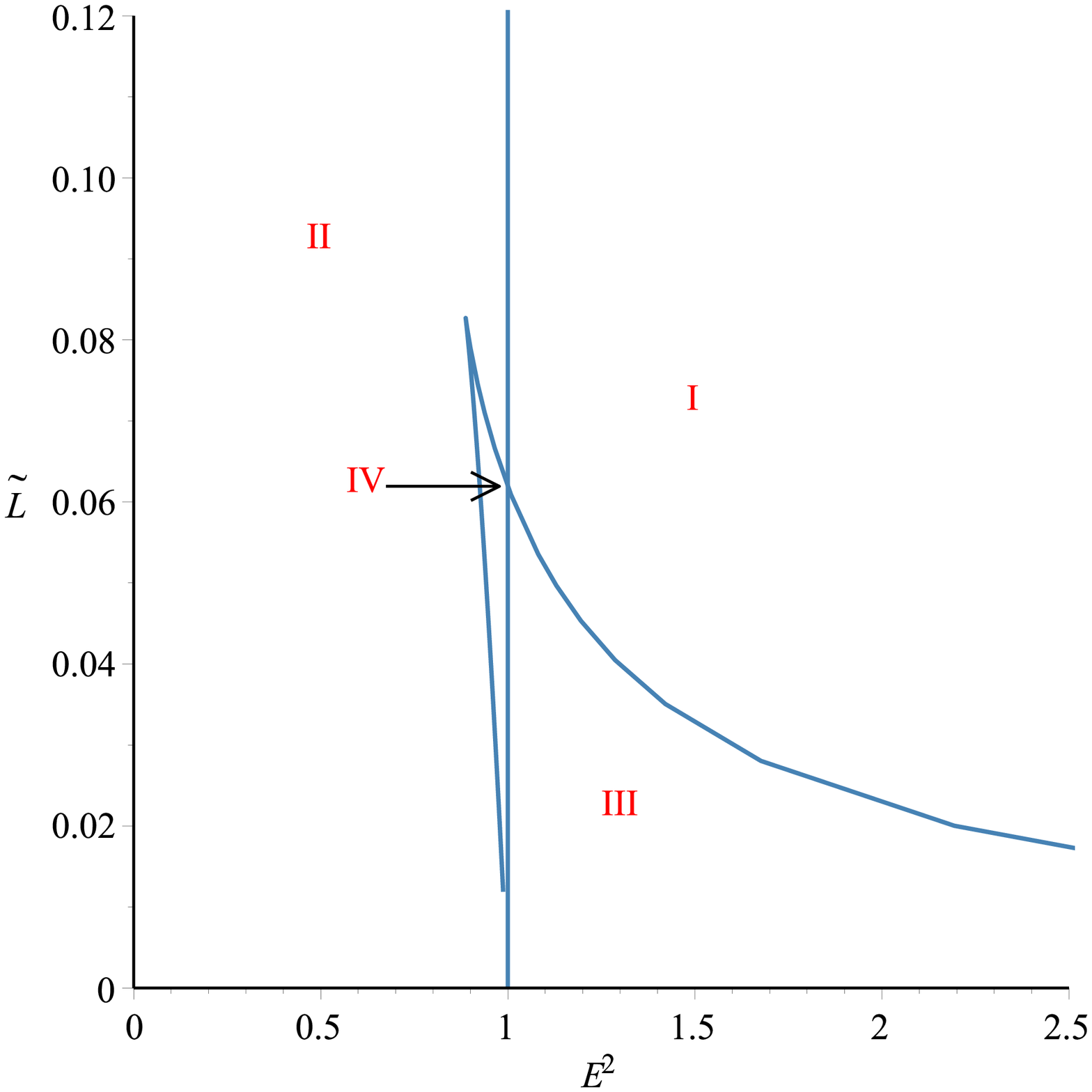}
		}
		\caption{\footnotesize Plots of $L-E^2$ diagram and region of different types of geodesic motion in a 
			quintessence surrounding field with the parameters $k\lambda=-2$, $\tilde{N}=0.025$ and $\tilde{Q}=\sqrt{0.25}$ 
			corresponding to table \ref{tab:RQ2} for (a): Null and (b): Timelike geodesic. The numbers 
			of positive real zeros in these regions are: I=1, II=2, III=3, IV=4.}
		\label{pic:RQC2b}
	\end{figure}
	%\clearpage 
	%%%%%%%%%%%%%%%%%%%%%%%%%
	\begin{figure}[!ht]
		\centering
		\subfigure[$\epsilon=1$, $\tilde{N}=0.025$, $\tilde{Q}=\sqrt{0.25}$, $\tilde{L}=0.05$, $E=\sqrt{1.75}$]{
			\includegraphics[width=0.17\textwidth]{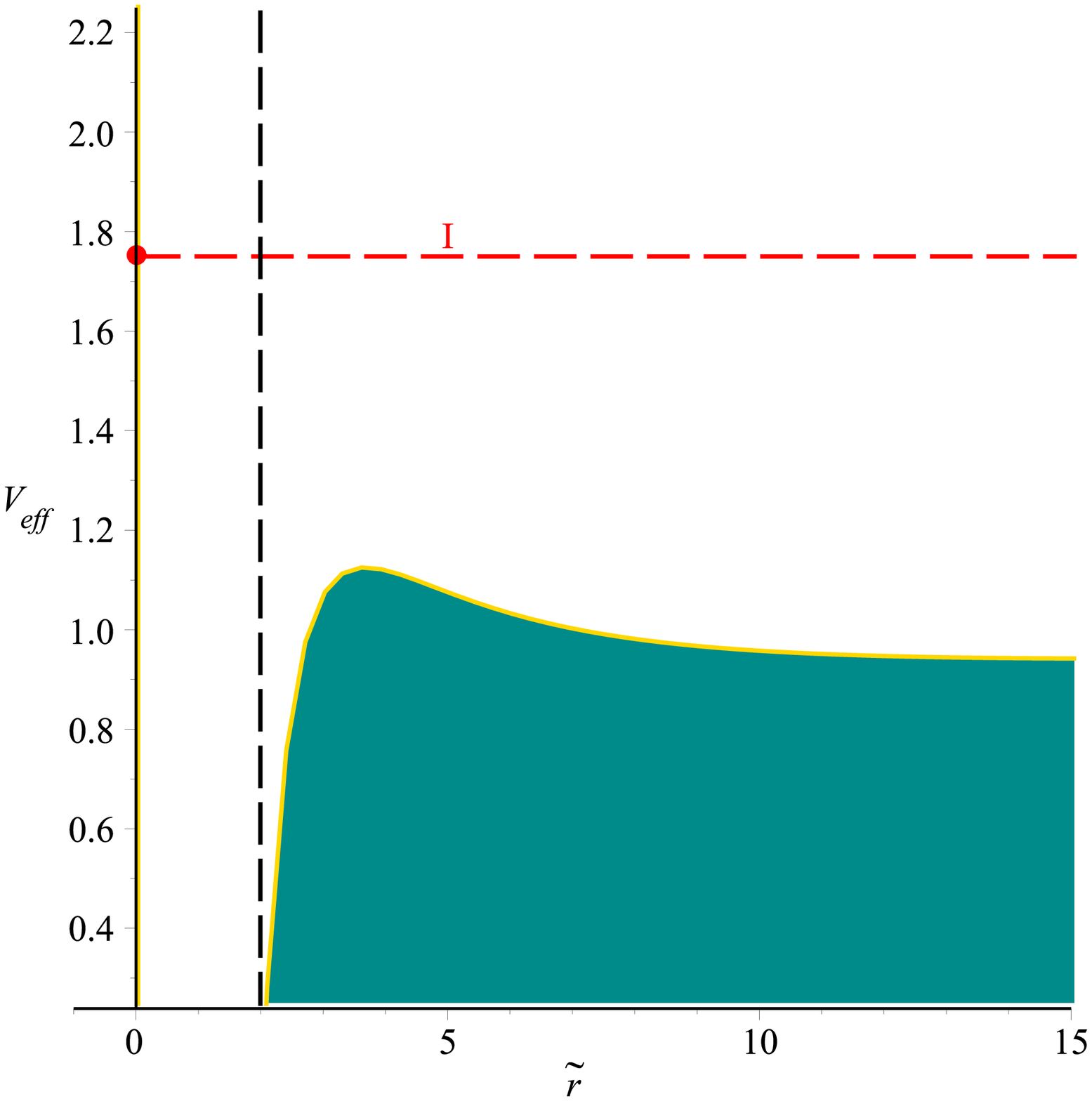}
		}
		\subfigure[$\epsilon=1$, $\tilde{N}=0.025$, $\tilde{Q}=\sqrt{0.25}$, $\tilde{L}=0.08$, $E=\sqrt{0.5}$]{
			\includegraphics[width=0.17\textwidth]{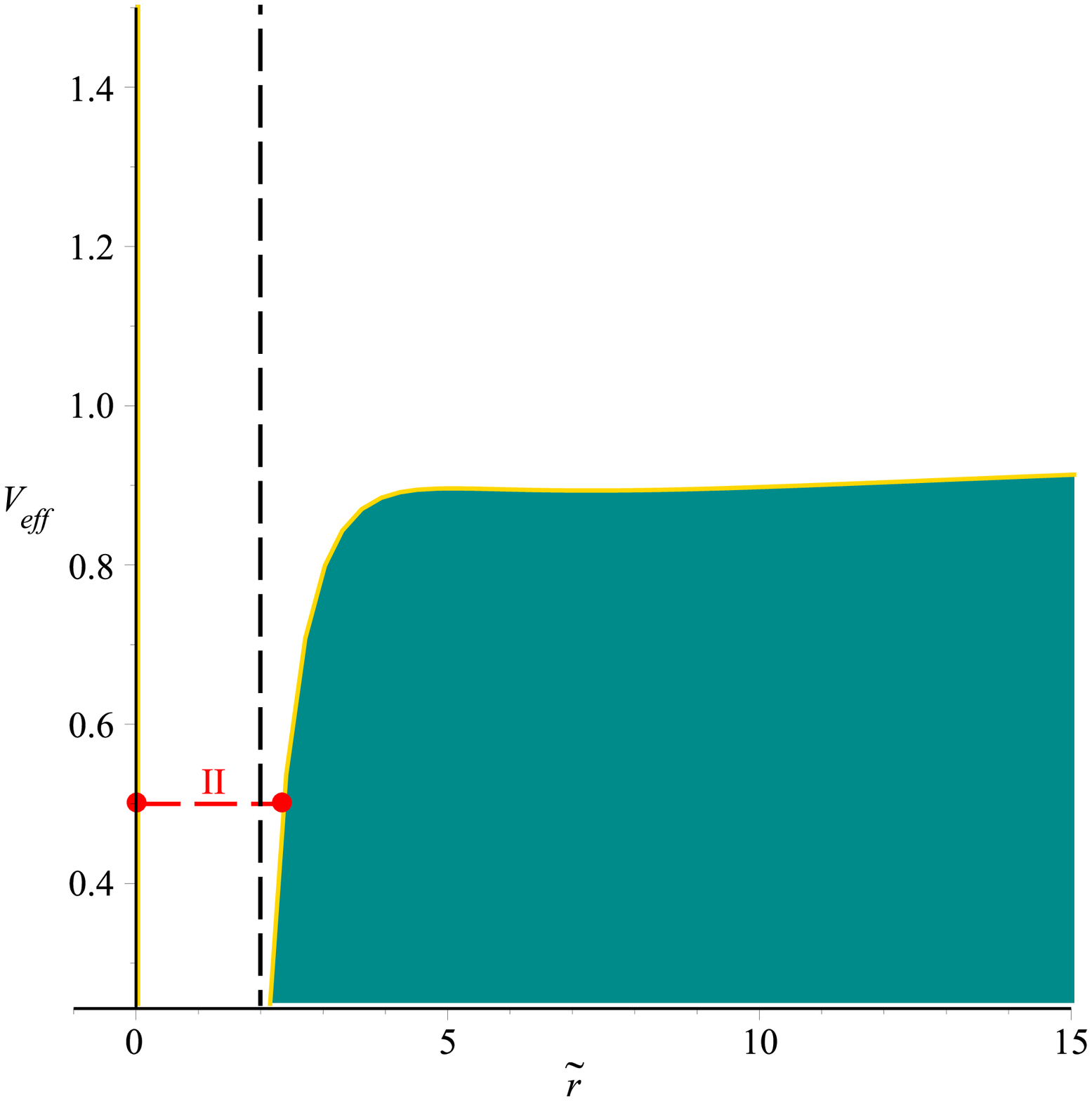}
		}
		\subfigure[$\epsilon=1$, $\tilde{N}=0.025$, $\tilde{Q}=\sqrt{0.25}$, $\tilde{L}=0.02$, $E=\sqrt{1.75}$]{
			
			\includegraphics[width=0.17\textwidth]{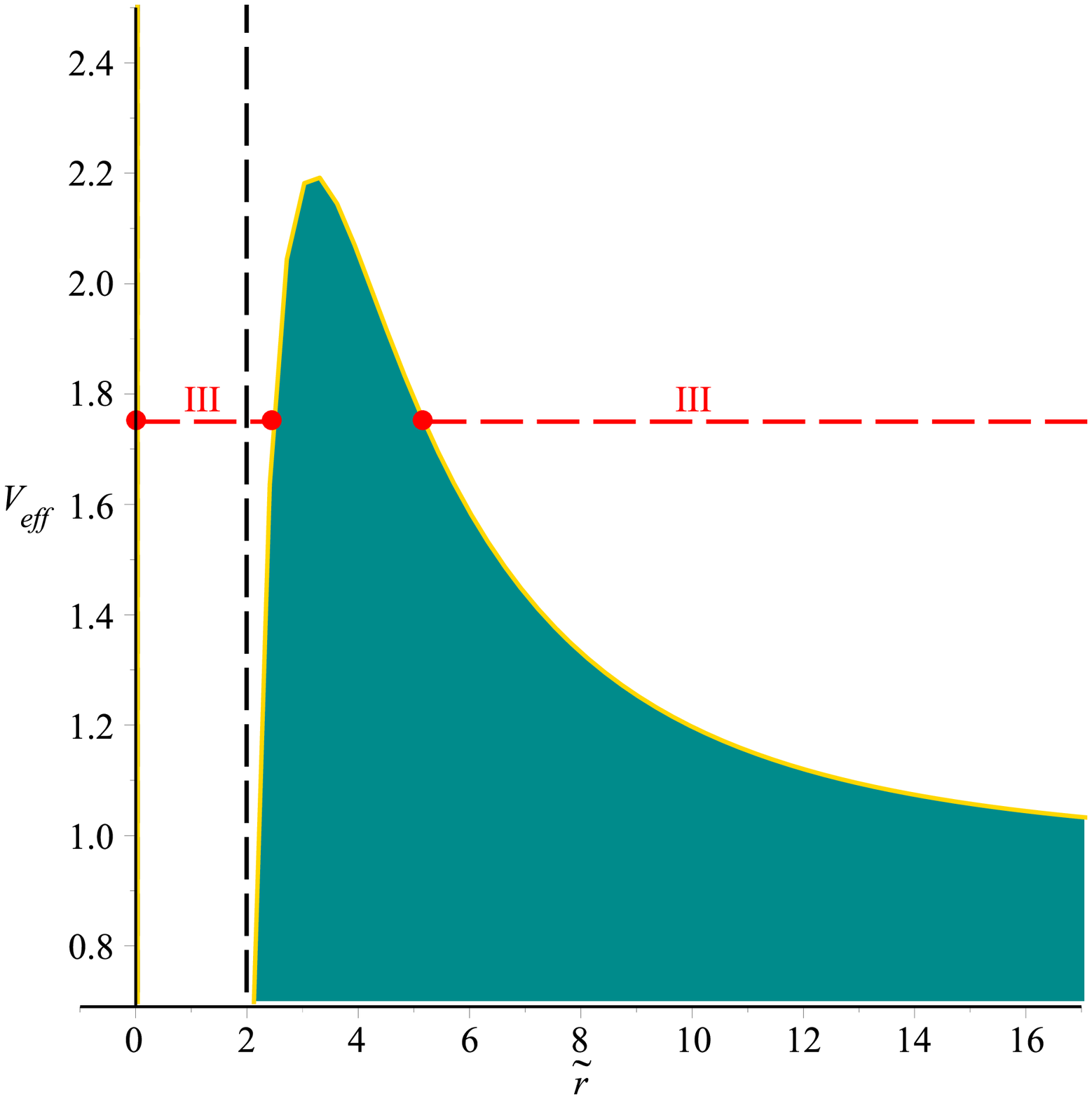}
		}
		\subfigure[$\epsilon=1$, $\tilde{N}=0.025$, $\tilde{Q}=\sqrt{0.25}$, $\tilde{L}=0.07$, $E=\sqrt{0.93}$]{
			\includegraphics[width=0.17\textwidth]{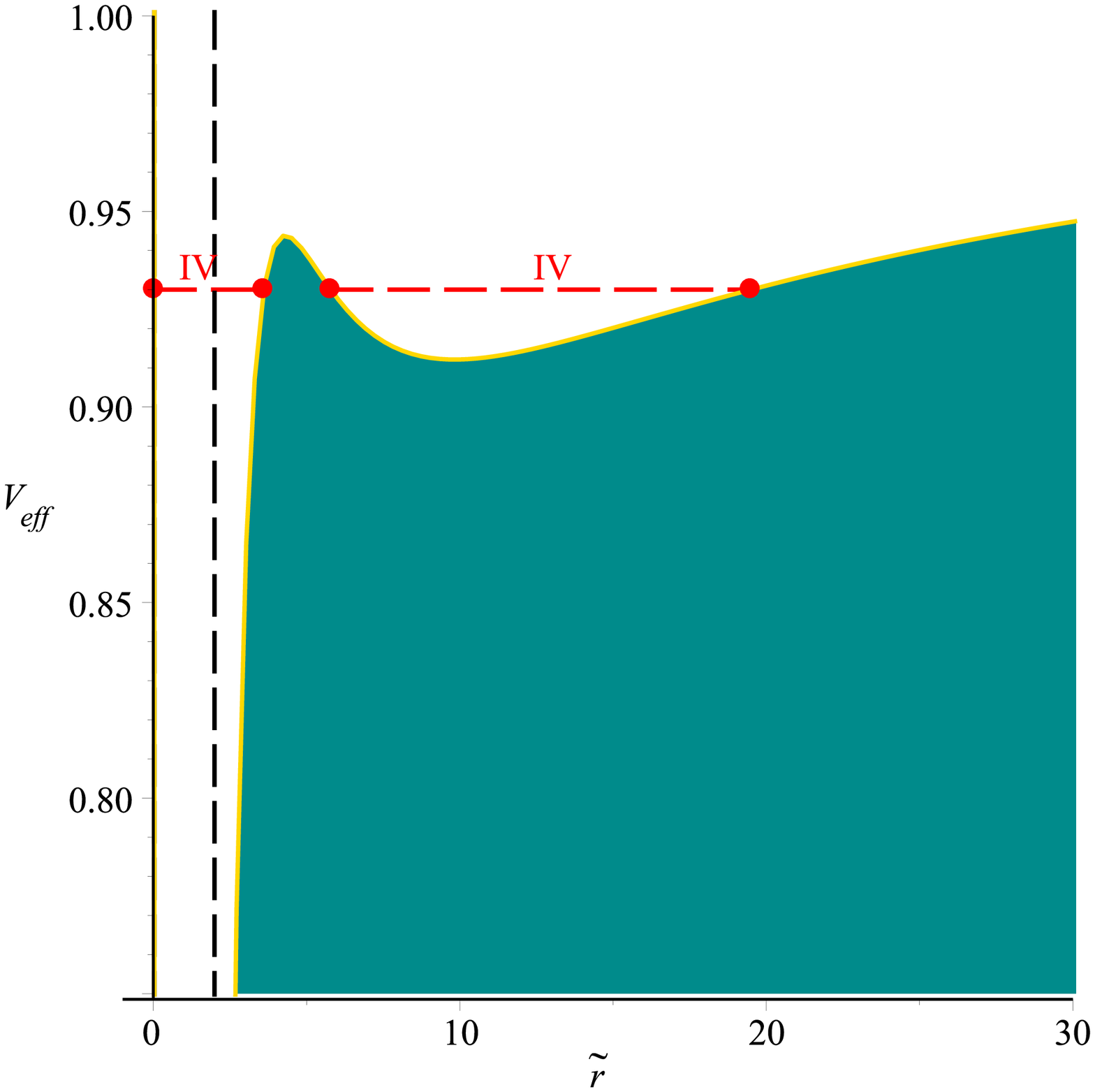}
		}
		\subfigure[$\epsilon=0$, $\tilde{N}=0.025$, $\tilde{Q}=\sqrt{0.25}$, $\tilde{L}=0.05$]{
			\includegraphics[width=0.17\textwidth]{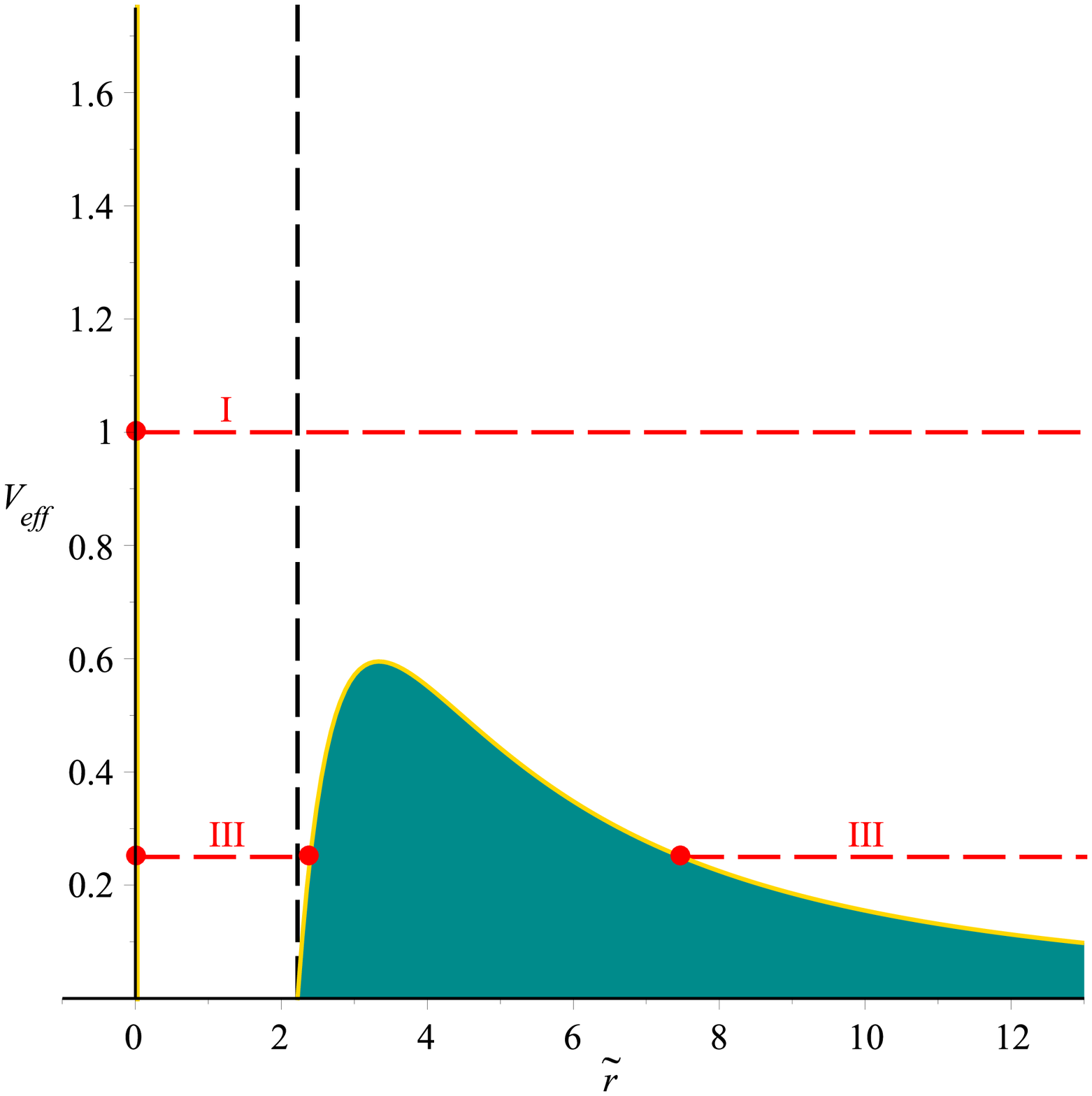}
		}
		\caption{\footnotesize Plots of the effective potential for the different orbit types of Table \ref{tab:RQ2}, 
			for the case of quintessence surrounding field with the parameters $k\lambda=-2$. 
			The horizontal red-dashed lines denotes the squared energy parameter $E^2$. 
			The vertical black dashed lines show the position of the horizons. 
			The red dots marks denote the zeros of the polynomial $R$, which are the 
			turning points of the orbits. In the cyan area, no motion is possible since $\tilde{R} < 0$ . }
		\label{pic:RQC2a}
	\end{figure}
	% \clearpage
	\clearpage
	%%%%%%%%%%%%%%%%%%%   Dust(LE2-Table-potential)   %%%%%%%%%%%%%%%%%%%%
	%%%%%%%%%%%%%%%%%%%%%%%%%%%%%
	%\paragraph{Dust surrounding field (with $k\lambda=\frac{2}{9}$) }
	\item Dust surrounding field (with $k\lambda=\frac{2}{9}$) 

\begin{table}[!ht]
	\begin{center}
		\begin{tabular}{|l|l|c|l|}
			%{|lccll|}
			\hline
			region & pos.zeros & range of $\tilde{r}$ &  orbit \\
			\hline\hline
			I & 1 &
			$|$$--$$\lVert$$\bullet$$\textbf{--------}$$\lVert$$\textbf{-----------------------}$%$\dashrightarrow$
			
			& TEO
			\\  \hline
			III & 3 &
			$|$$--$$\lVert$$\bullet$$\textbf{--------}$$\lVert$$\textbf{---}$$\bullet$$--$$\bullet$$\textbf{--------------}$%$\dashrightarrow$
			& MBO, EO
			\\ \hline
			V & 5 &
			$|$$--$$\lVert$$\bullet$$\textbf{--------}$$\lVert$$\textbf{---}$$\bullet$$--$$\bullet$$\textbf{------}$$\bullet$$-$$\bullet$$\textbf{---}$%$\dashrightarrow$
			& MBO, BO, EO
			\\ \hline
		\end{tabular}
		\caption{Types of orbits of the quintessence surrounding field for $k\lambda=\frac{2}{9}$. The lines represent the range of the orbits. The dots
			show the turning points of the orbits. The positions of the two horizons are marked by a vertical double line. The
			single vertical line indicates the singularity at $\tilde{r}=0$.}
		\label{tab:RD}
	\end{center}
\end{table}

	\begin{figure}[!ht]
		\centering
		\subfigure[]{
			\includegraphics[width=0.2\textwidth]{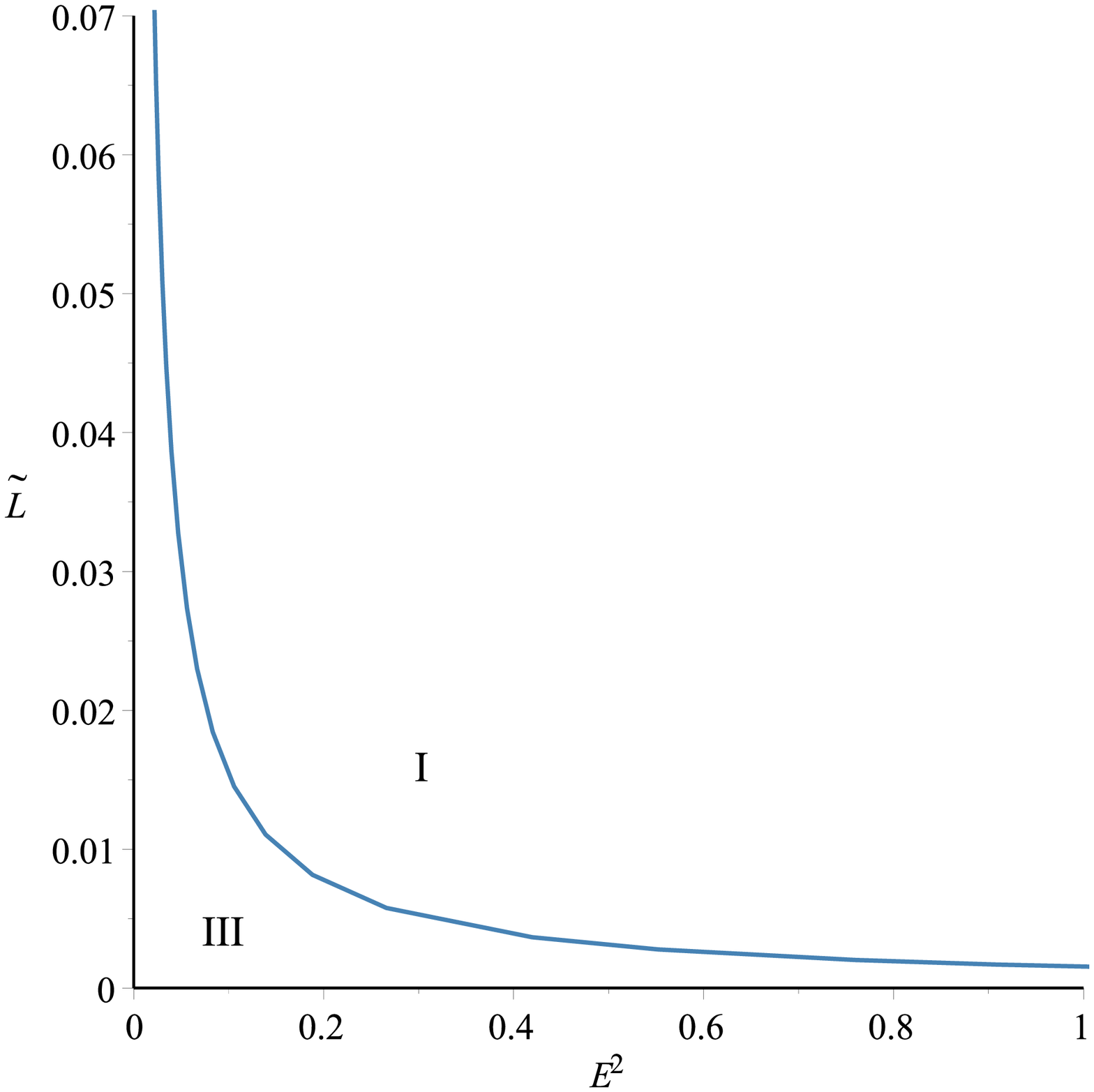}
		}
		\subfigure[]{
			\includegraphics[width=0.2\textwidth]{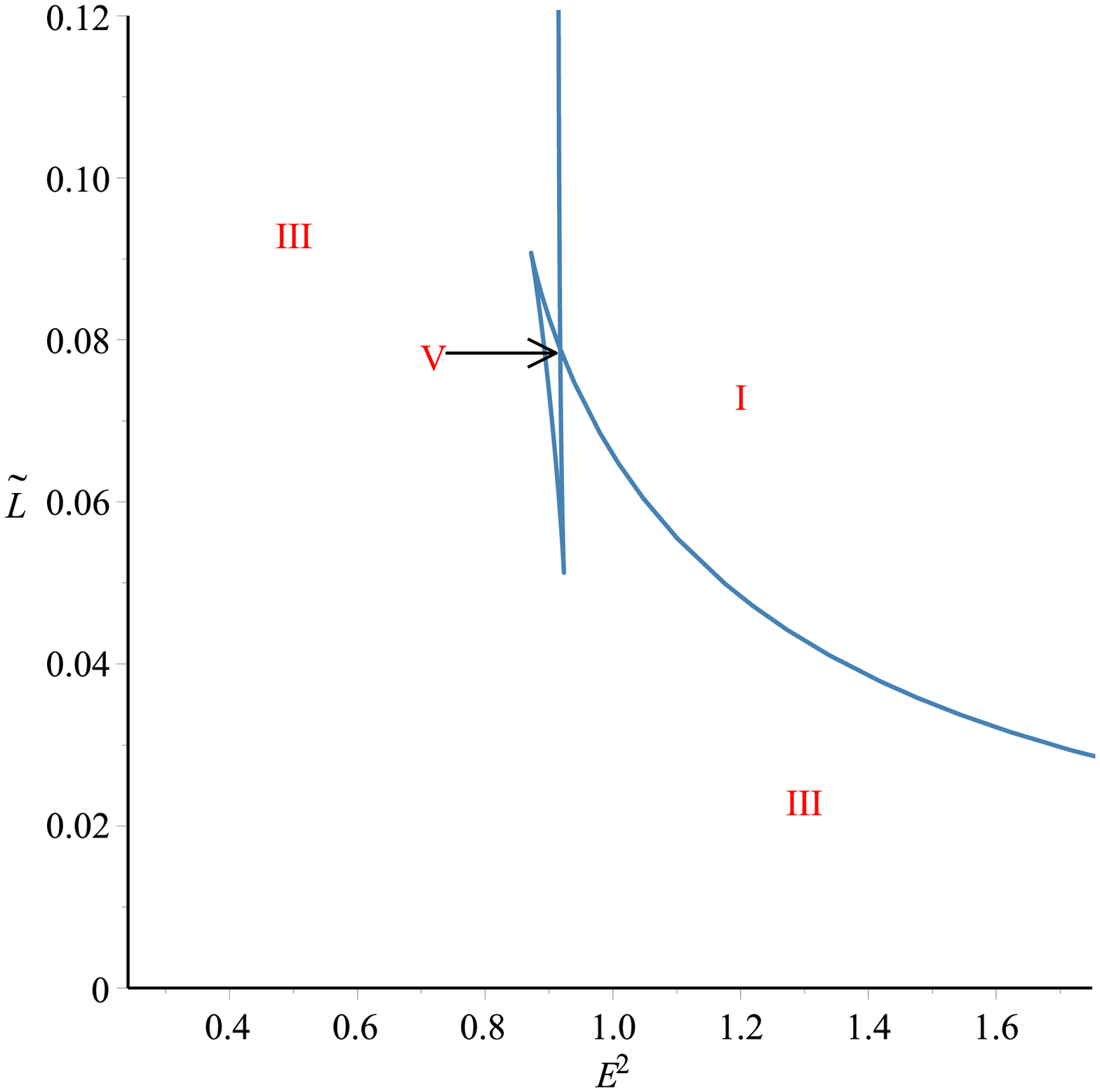}
		}
		\caption{\footnotesize Plots of $L-E^2$ diagram and region of different types of geodesic motion 
			for the dust surrounding field with the parameters $k\lambda=\frac{2}{9}$, $\tilde{N}=10^{-3}$, 
			$\tilde{Q}=0.2$ corresponding to table \ref{tab:RD}
			for (a): Null geodesic and (b): Timelike geodesic. 
			The numbers of positive real zeros in these regions are: I=1, III=3, V=5.}
		\label{pic:RDb}
	\end{figure}
	%\clearpage
	%%%%%%%%%%%%%%%%%%%%%%%%%%%%%%%%%
	\begin{figure}[!ht]
		\centering
		\subfigure[$\epsilon=1$, $\tilde{N}=10^-3$, $\tilde{Q}=\sqrt{0.25}$, $\tilde{L}=0.05$, $E=\sqrt{1.5}$]{
			\includegraphics[width=0.17\textwidth]{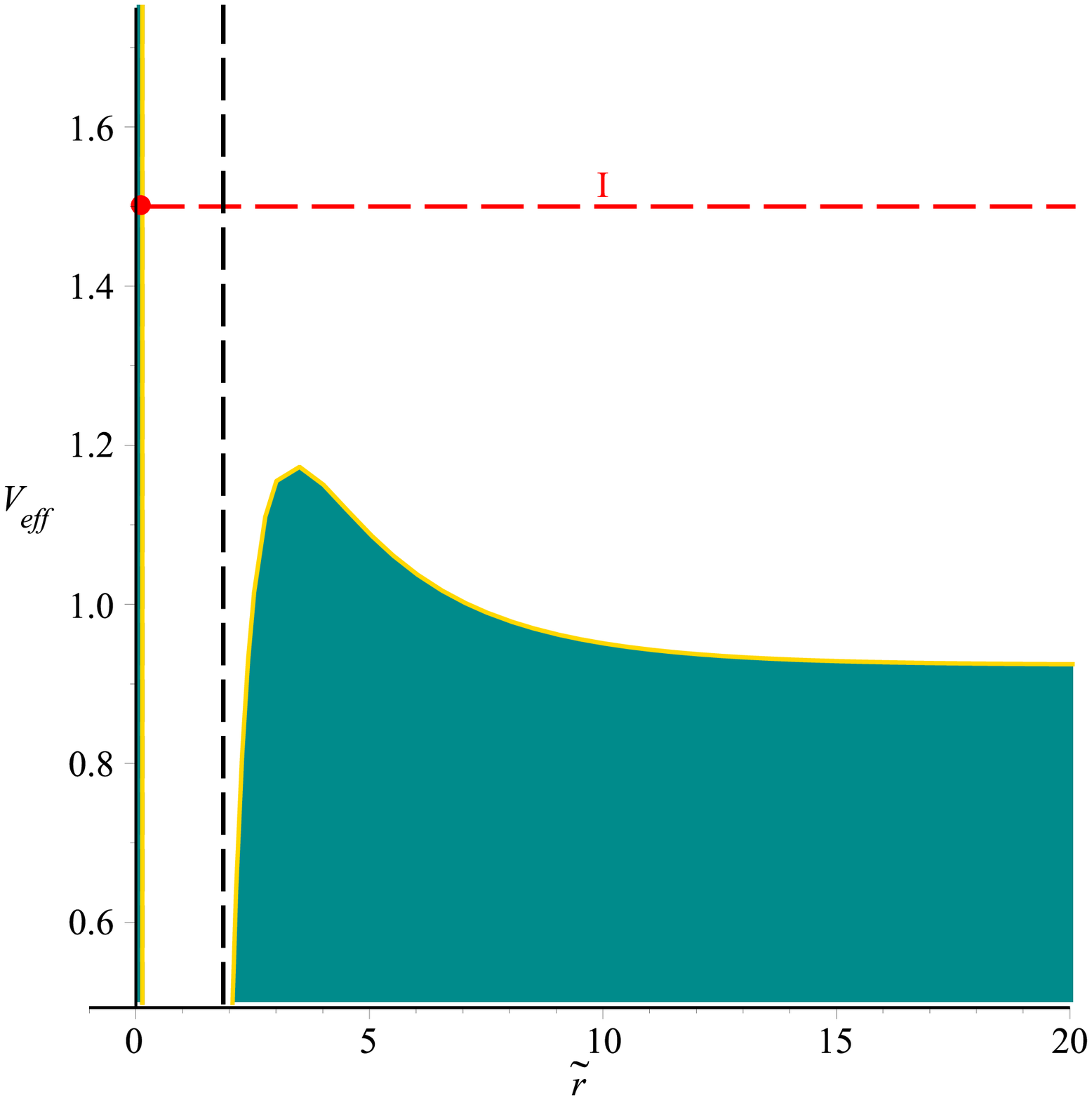}
		}
		\subfigure[$\epsilon=1$, $\tilde{N}=10^-3$, $\tilde{Q}=\sqrt{0.25}$, $\tilde{L}=0.025$, $E=\sqrt{1.25}$]{
			\includegraphics[width=0.17\textwidth]{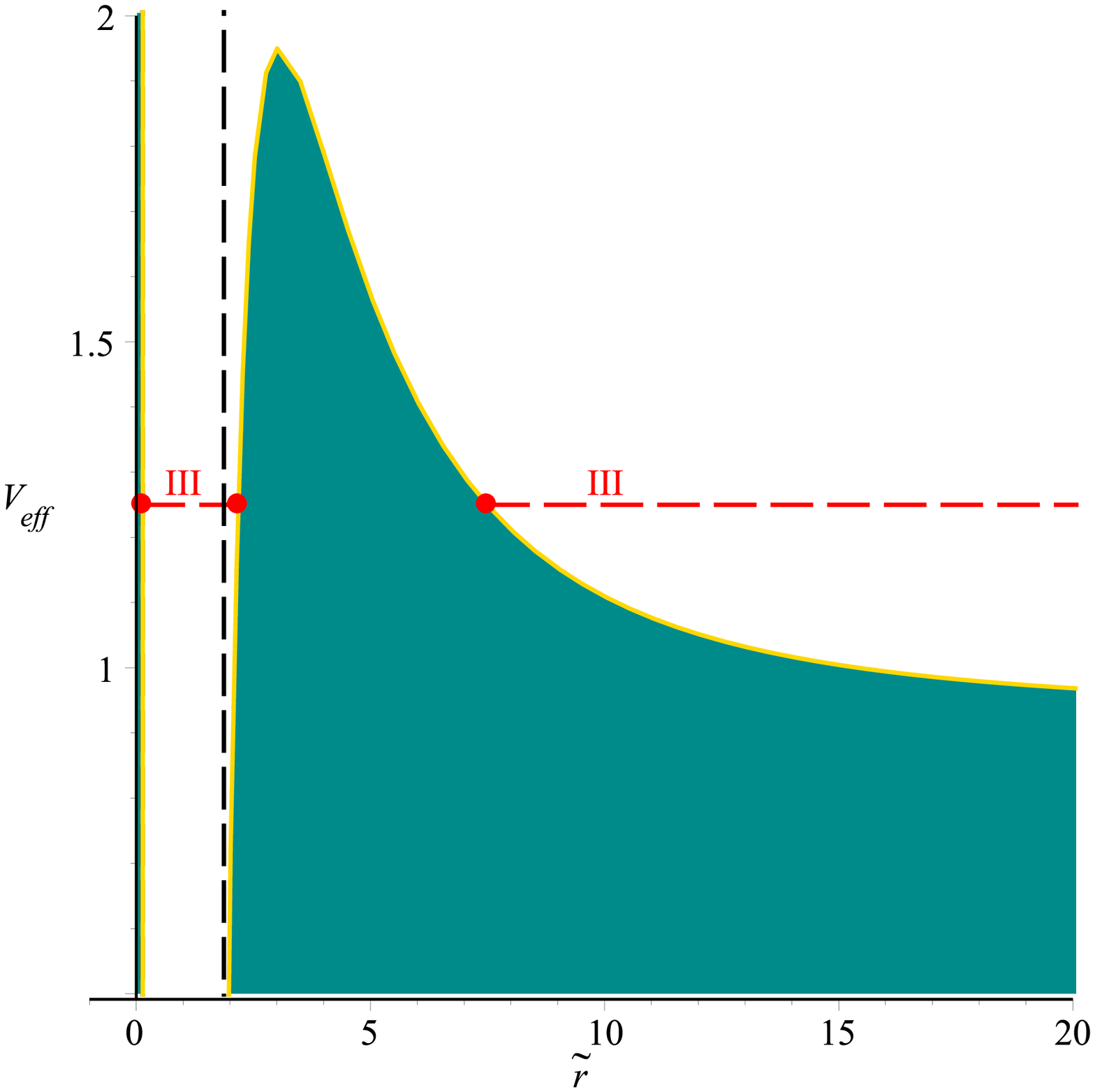}
		}
		\subfigure[$\epsilon=1$, $\tilde{N}=10^-3$, $\tilde{Q}=\sqrt{0.25}$, $\tilde{L}=0.075$, $E=\sqrt{0.9}$]{
			\includegraphics[width=0.17\textwidth]{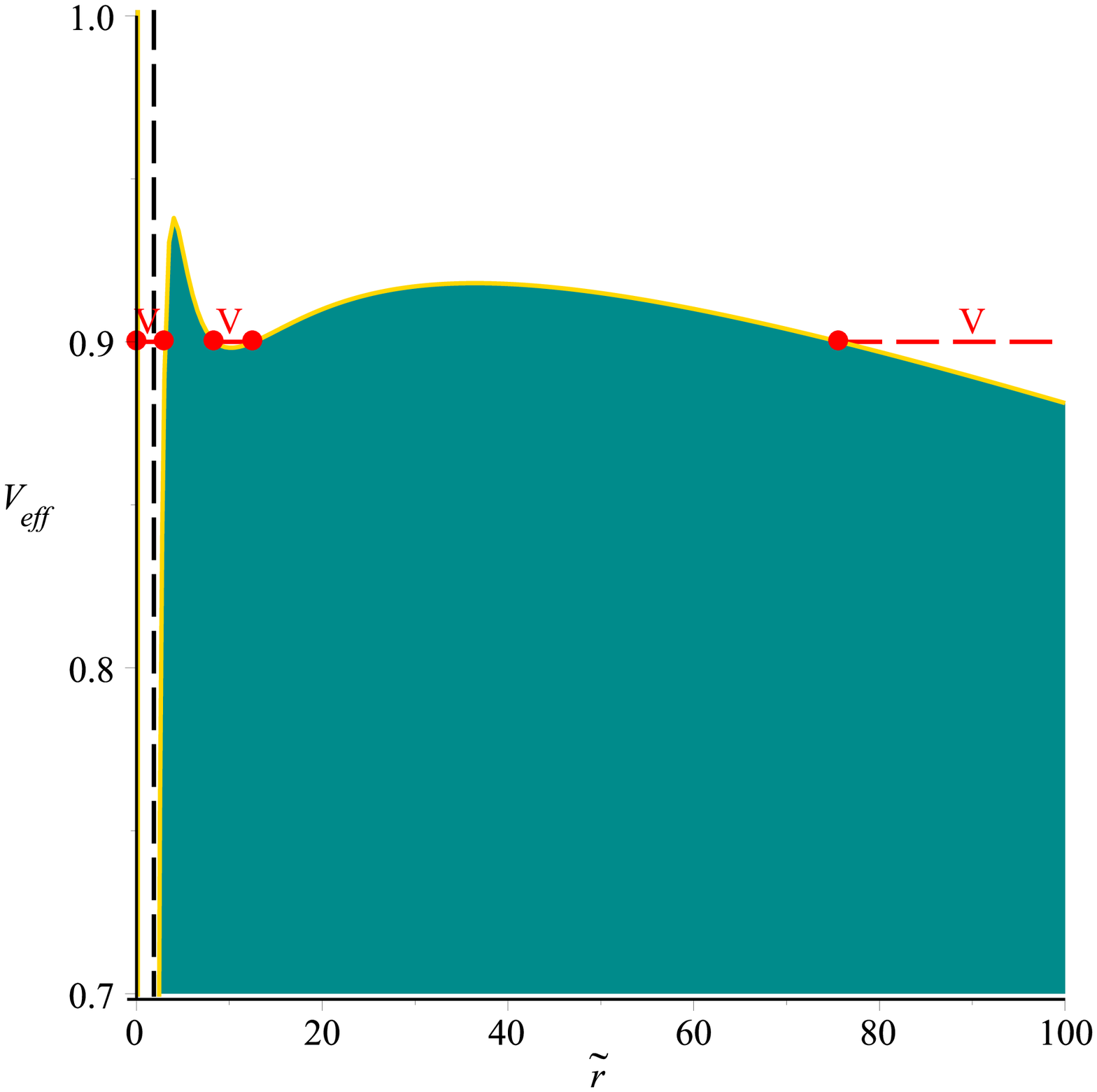}
		}
		-            \subfigure[Cluseup of figure (c)]{
			\includegraphics[width=0.17\textwidth]{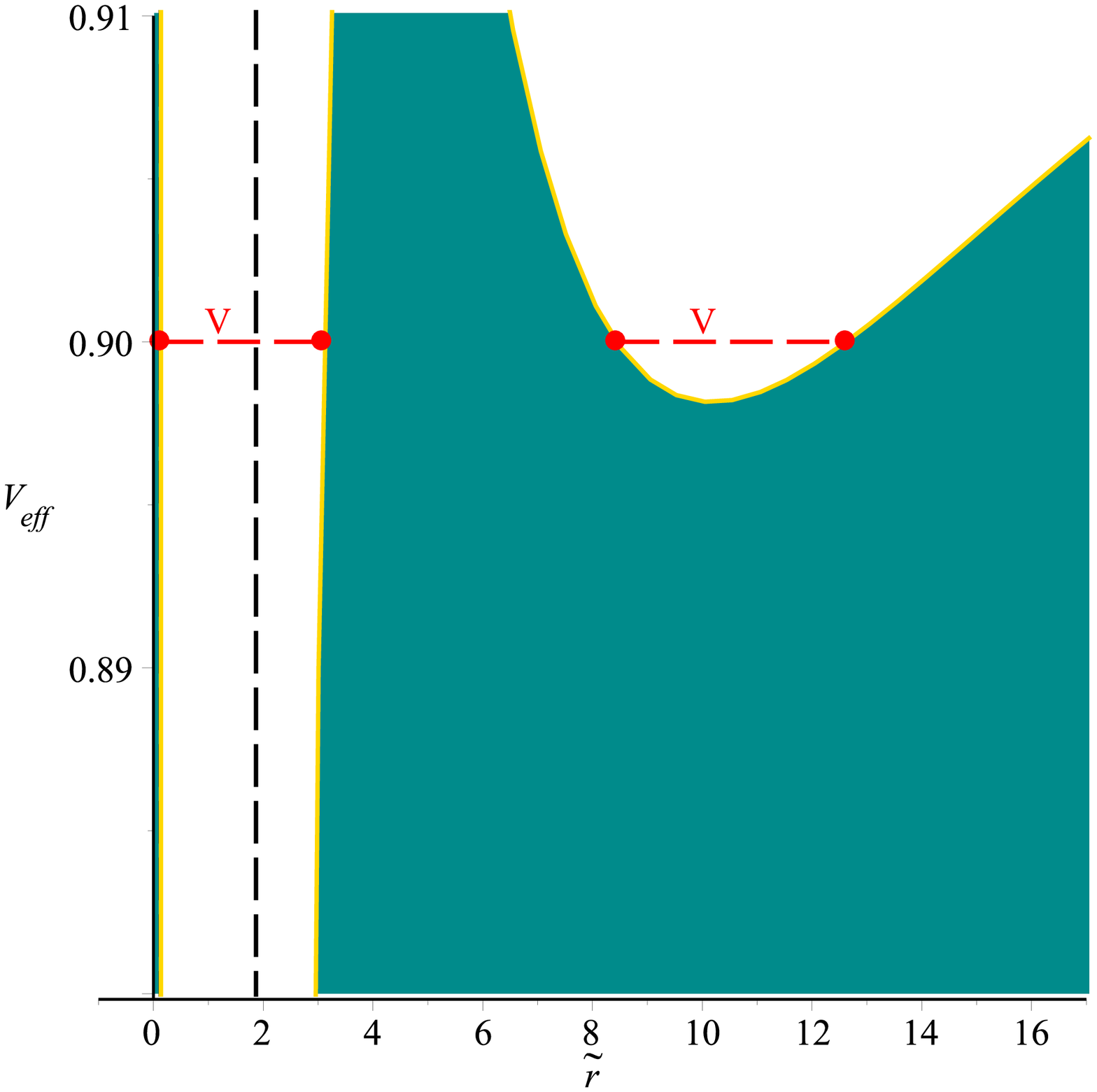}
		} 
		\subfigure[$\epsilon=0$, $\tilde{N}=0.075$, $\tilde{Q}=\sqrt{0.25}$, $\tilde{L}=0.02$]{
			\includegraphics[width=0.17\textwidth]{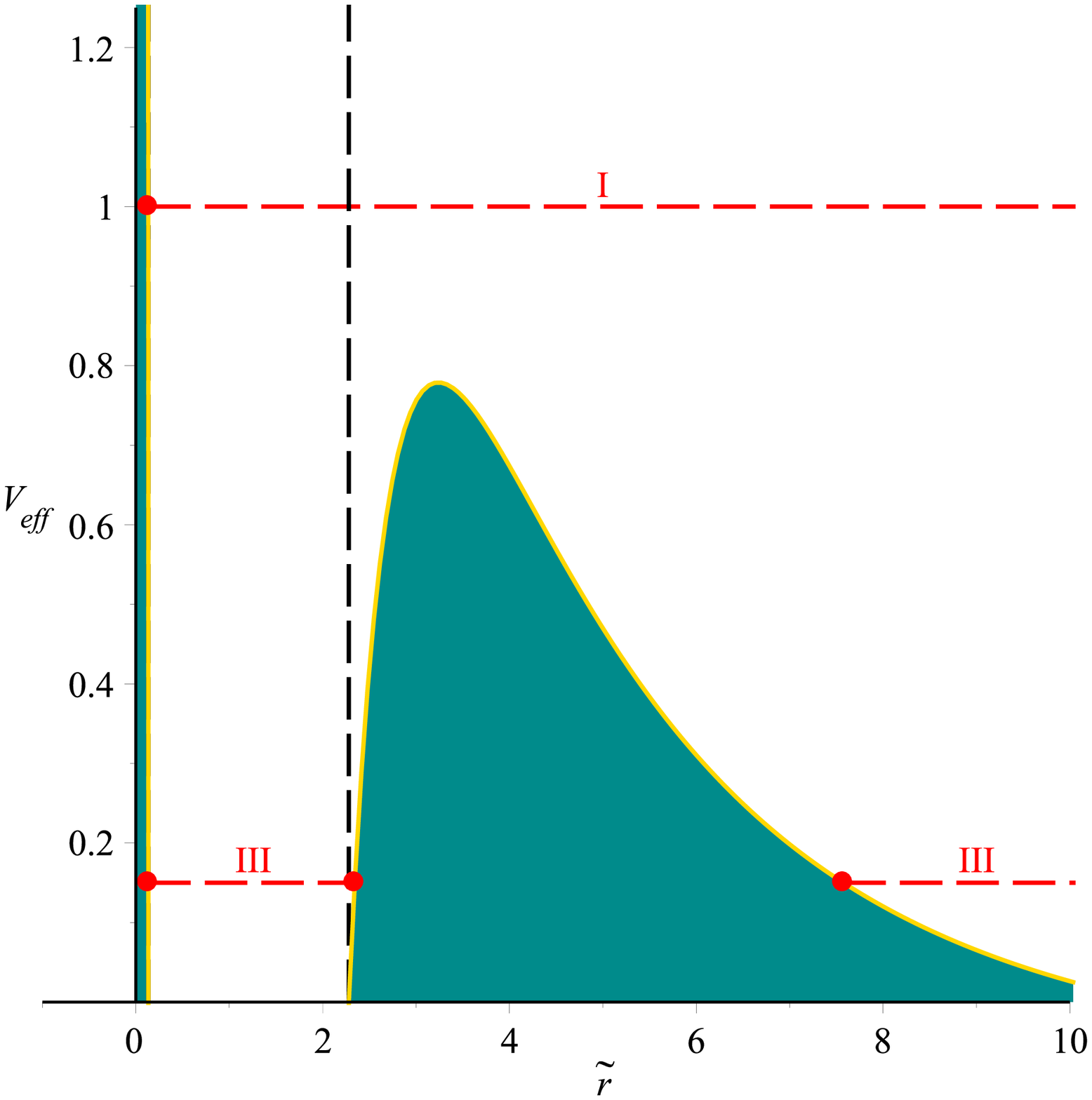}
		}
		\caption{\footnotesize Plots of the effective potential for the different orbit types of Table \ref{tab:RD}, 
			for the case of dust surrounding field with the parameters $k\lambda=\frac{2}{9}$. 
			The horizontal red-dashed lines denotes the squared energy parameter $E^2$. 
			The vertical black dashed lines show the position of the horizons. 
			The red dots marks denote the zeros of the polynomial $R$, which are the 
			turning points of the orbits. In the cyan area, no motion is possible since $\tilde{R} < 0$ .}
		\label{pic:RDa}
	\end{figure}
	%\clearpage	
	\clearpage
	%%%%%%%%%%%%%%%%%%%%%%%%%%%%%
	%%%%%%%%%%%%%%%%%%%   Radiation(LE2-Table-potential)   %%%%%%%%%%%%%%%%%%%%
	%%%%%%%%%%%%%%%%%%%%%%%%%%%%%%
	%  \paragraph{Radiation surrounding field}
	\item Radiation surrounding field
	
		%%%%%%%%%%%%%%%%%%%%%%%%
	\begin{table}[!ht]
		\begin{center}
			\begin{tabular}{|l|l|c|l|}
				%{|lccll|}
				\hline
				region & pos.zeros & range of $\tilde{r}$ &  orbit \\
				\hline\hline
				I & 1 &
				$|$$--$$\lVert$$\bullet$$\textbf{--------}$$\lVert$$\textbf{-----------------------}$%$\dashrightarrow$
				
				& TEO
				\\  \hline
				II & 2 &
				$|$$--$$\lVert$$\bullet$$\textbf{--------}$$\lVert$$\textbf{---}$$\bullet$$-------$%$\dashrightarrow$
				& MBO
				\\ \hline
				III & 3 &
				$|$$--$$\lVert$$\bullet$$\textbf{--------}$$\lVert$$\textbf{---}$$\bullet$$--$$\bullet$$\textbf{--------------}$%$\dashrightarrow$
				& MBO, EO
				\\ \hline
				IV & 4 &
				$|$$--$$\lVert$$\bullet$$\textbf{--------}$$\lVert$$\textbf{---}$$\bullet$$--$$\bullet$$\textbf{-----}$$\bullet$ $---$
				& MBO, BO
				\\ \hline
			\end{tabular}
			\caption{Types of orbits of the radiation surrounding field. The lines represent the range of the orbits. The dots
				show the turning points of the orbits. The positions of the two horizons are marked by a vertical double line. The
				single vertical line indicates the singularity at $\tilde{r}=0$.}
			\label{tab:RR}
		\end{center}
	\end{table}
	
	\begin{figure}[!ht]
		\centering
		\subfigure[]{
			\includegraphics[width=0.2\textwidth]{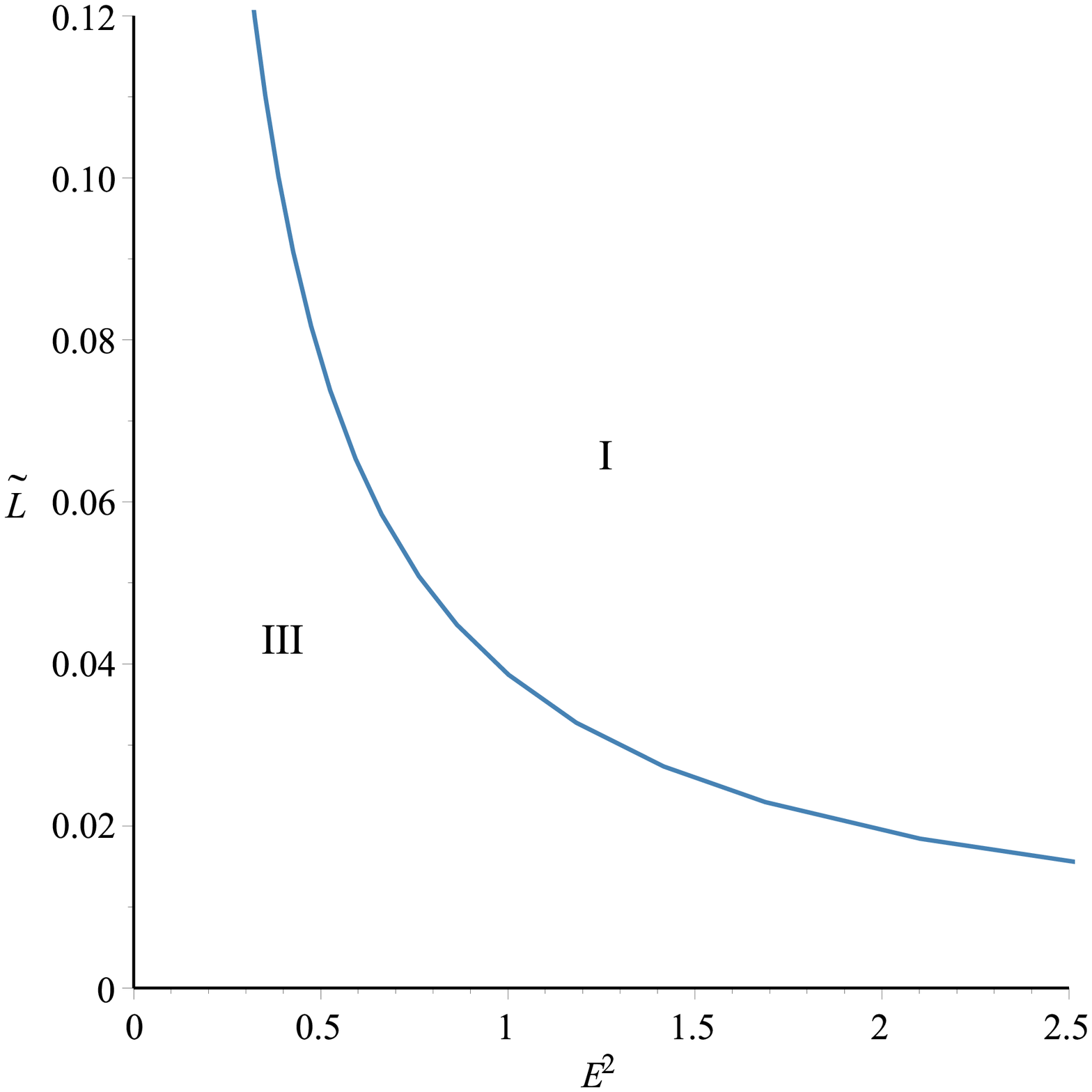}
		}
		\subfigure[]{
			\includegraphics[width=0.2\textwidth]{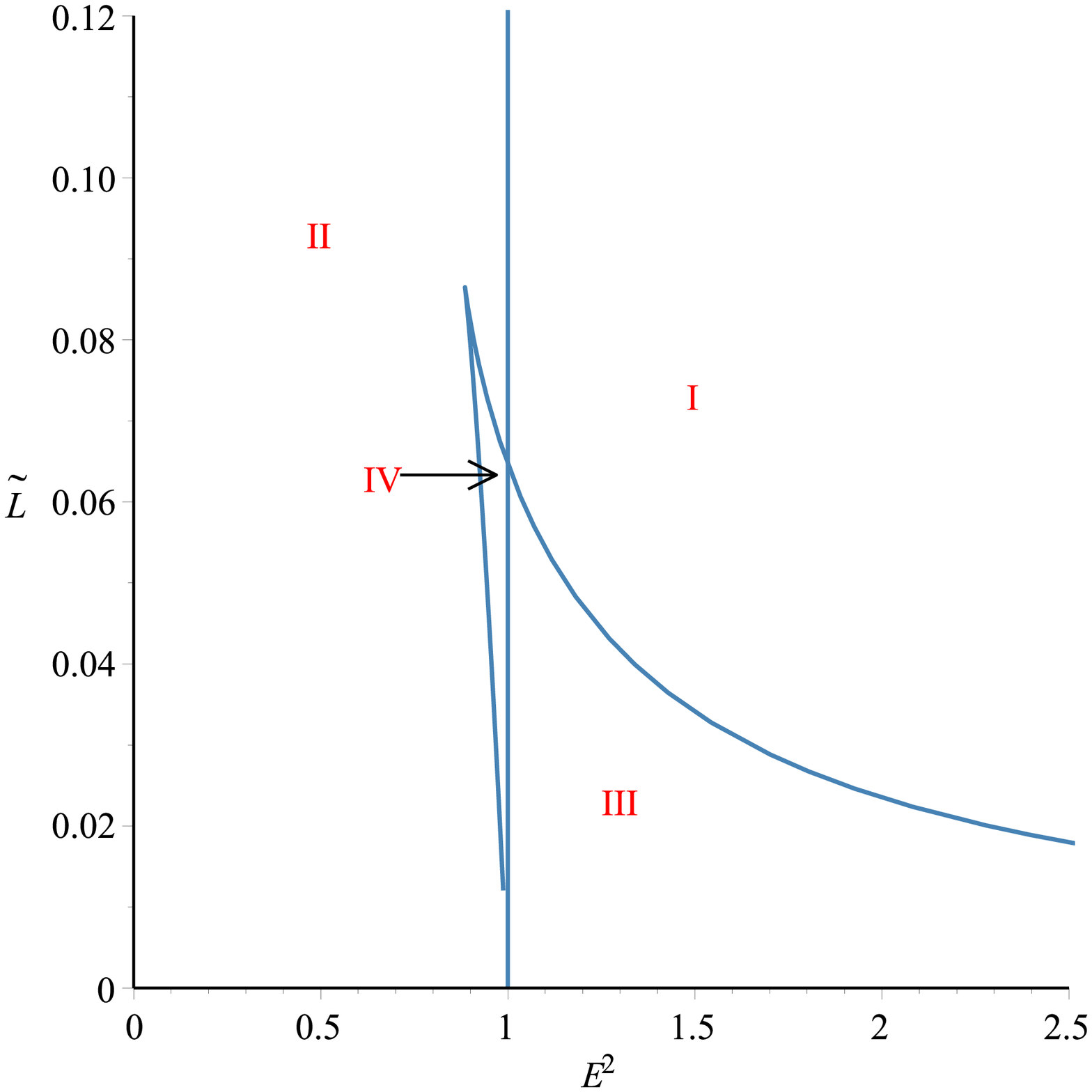}
		}    
		\caption{\footnotesize Plots of $L-E^2$ diagram and regions of different types of geodesic motion for a 
			black hole surrounded by radiation field with the parameters $\tilde{N}=0.12$ and $\tilde{Q}=\sqrt{0.25}$ 
			corresponding to table \ref{tab:RR} 
			for (a): Null ($\epsilon=0$) and (b): Timelike geodesics ($\epsilon=1$). 
			The numbers of positive real zeros in these regions are: I=1, II=2, III=3, IV=4.}
		\label{pic:RRb}
	\end{figure}
	%\clearpage 
	%%%%%%%%%%%%%%%%%%%%%%%%%
	\begin{figure}[!ht]
		\centering
		\subfigure[ $\epsilon=1$, $\tilde{N}=0.12$, $\tilde{Q}=\sqrt{0.25}$, $\tilde{L}=0.065$, $E=\sqrt{1.5}$]{
			\includegraphics[width=0.17\textwidth]{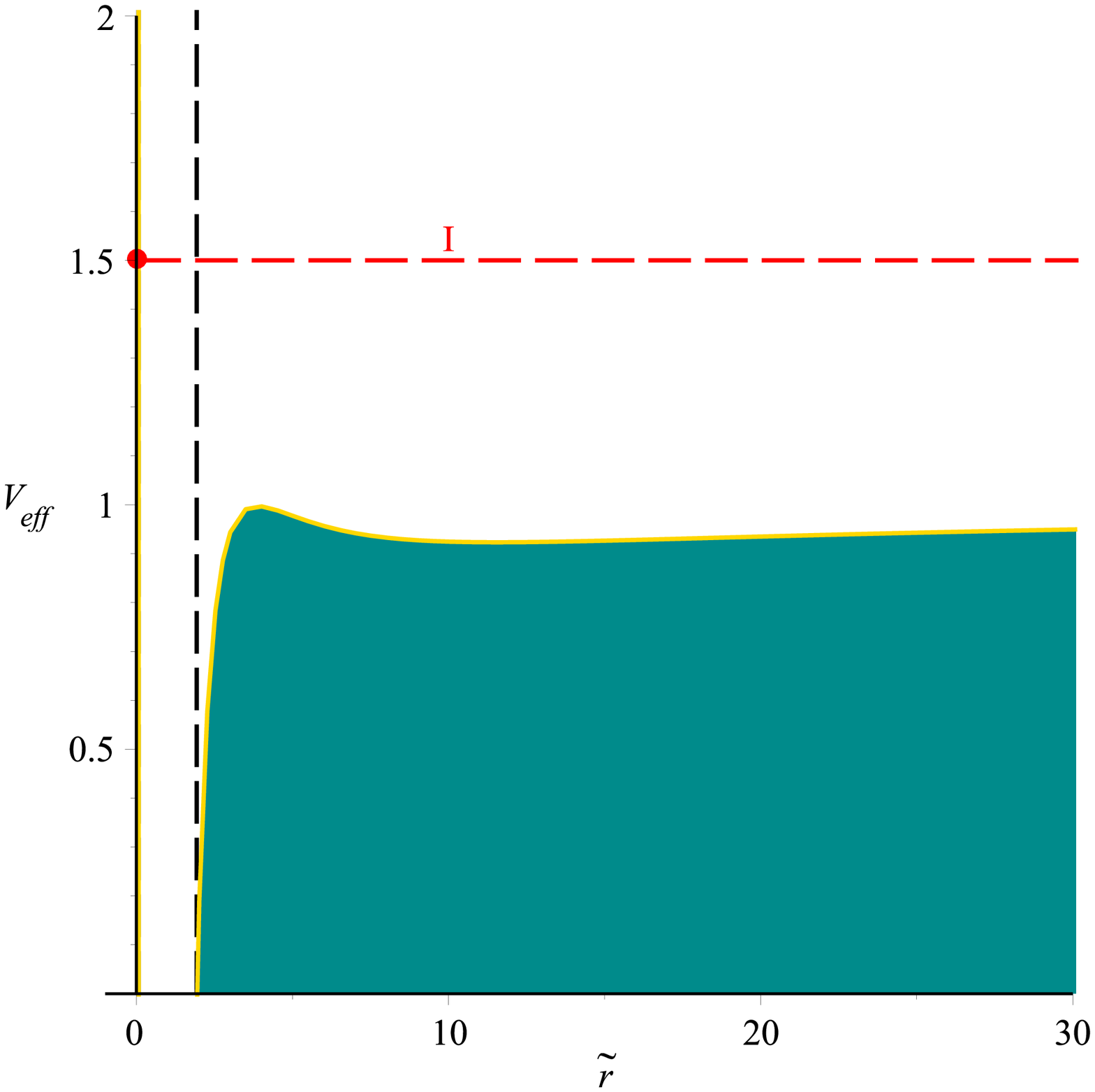}
		}
		\subfigure[ $\epsilon=1$, $\tilde{N}=0.12$, $\tilde{Q}=\sqrt{0.25}$, $\tilde{L}=0.05$, $E=\sqrt{0.5}$]{
			\includegraphics[width=0.17\textwidth]{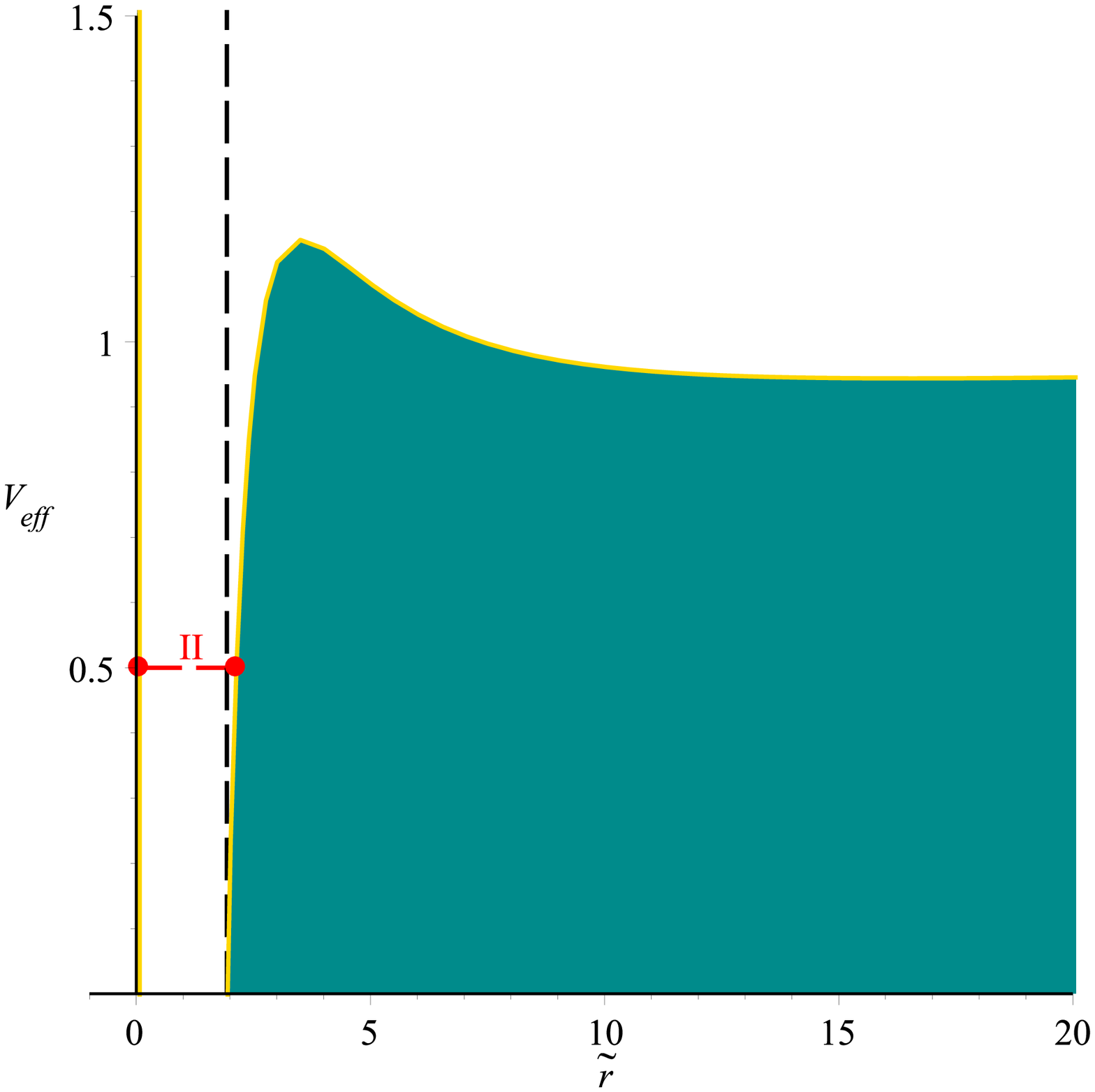}
		}
		\subfigure[ $\epsilon=1$, $\tilde{N}=0.12$, $\tilde{Q}=\sqrt{0.25}$, $\tilde{L}=0.02$, $E=\sqrt{1.65}$]{
			\includegraphics[width=0.17\textwidth]{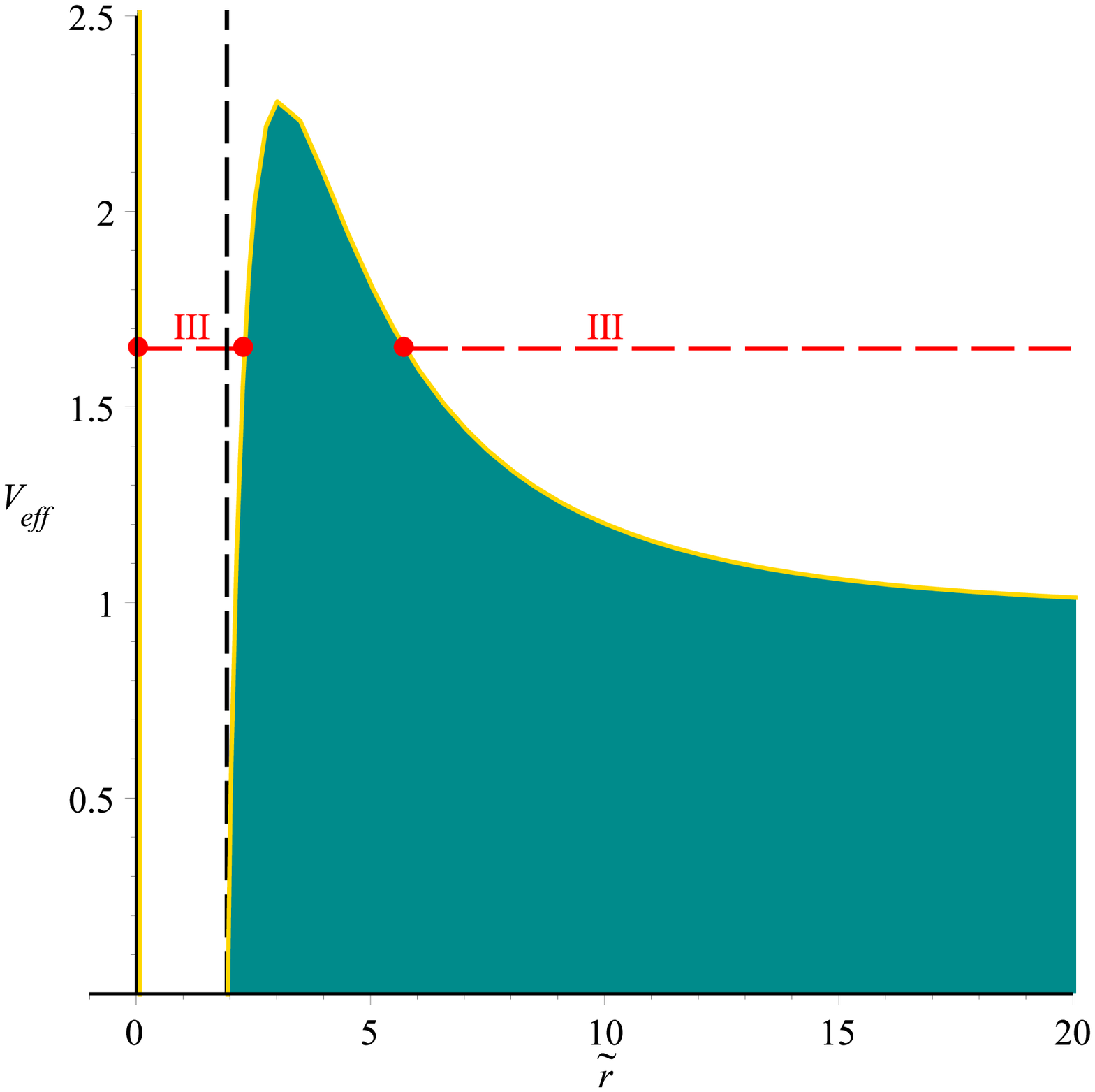}
		}
		\subfigure[ $\epsilon=1$, $\tilde{N}=0.12$, $\tilde{Q}=\sqrt{0.25}$, $\tilde{L}=0.075$, $E=\sqrt{0.92}$]{
			\includegraphics[width=0.17\textwidth]{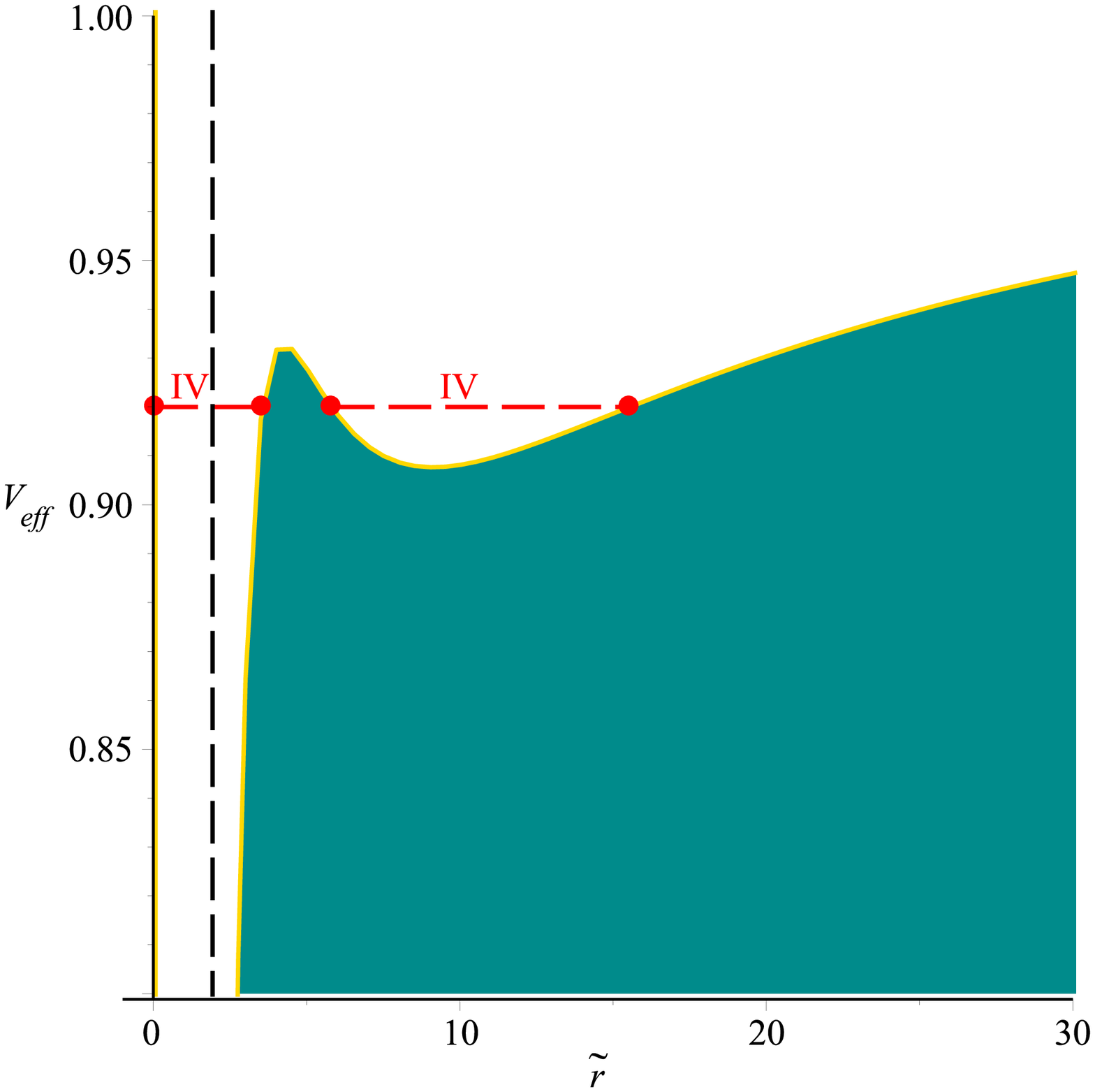}
		}
		\subfigure[ $\epsilon=0$, $\tilde{N}=0.12$, $\tilde{Q}=\sqrt{0.25}$, $\tilde{L}=0.06$]{
			\includegraphics[width=0.17\textwidth]{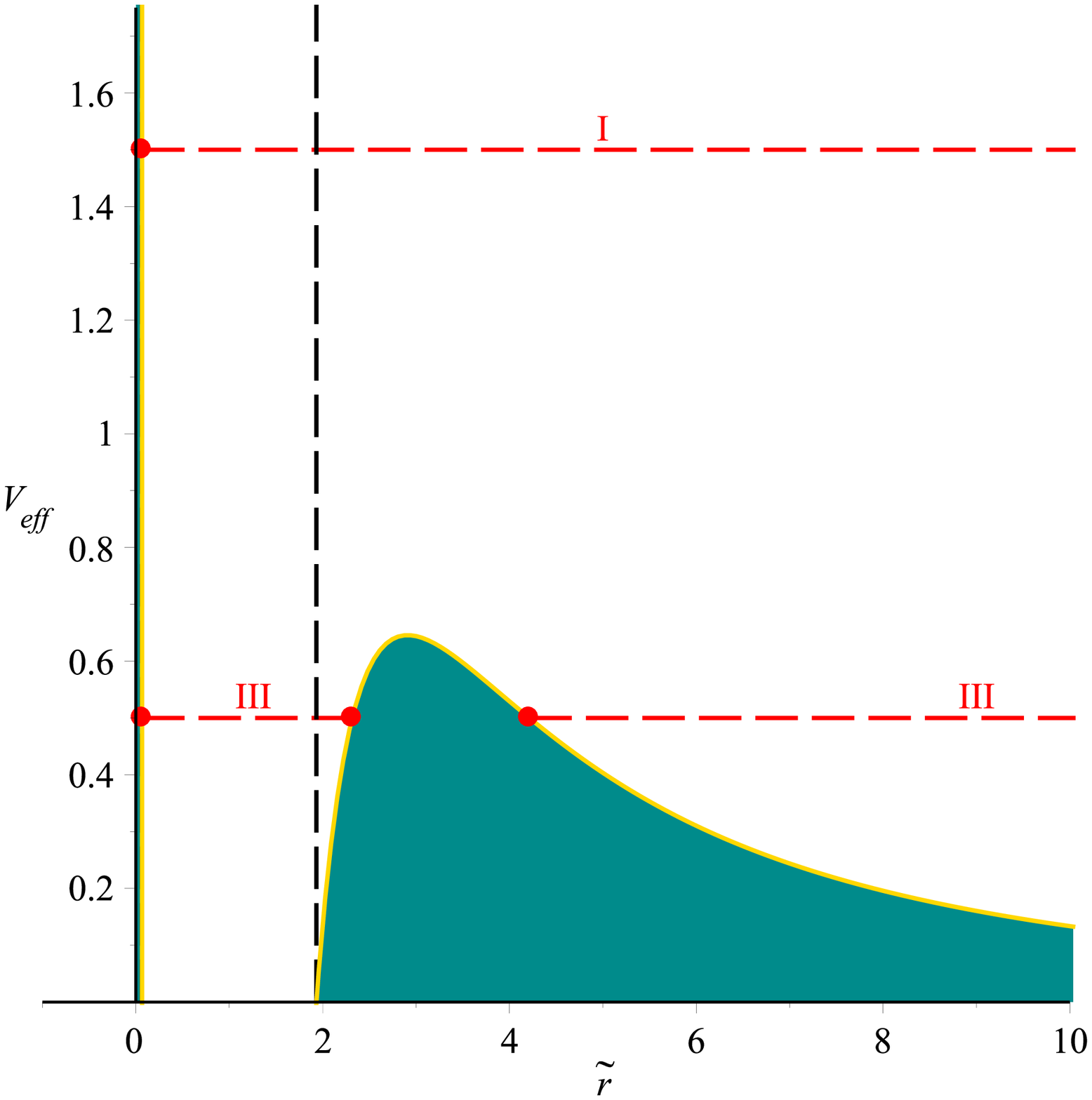}
		}
		\caption{\footnotesize Plots of the effective potential for the different orbit types of Table \ref{tab:RR}, 
			for the case of radiation surrounding field. 
			The horizontal red-dashed lines denotes the squared energy parameter $E^2$. 
			The vertical black dashed lines show the position of the horizons. 
			The red dots marks denote the zeros of the polynomial $R$, which are the 
			turning points of the orbits. In the cyan area, no motion is possible since $\tilde{R} < 0$ .}
		\label{pic:RRa}
	\end{figure}
	%\clearpage

	\clearpage
	%%%%%%%%%%%%%%%%%%%%%%%%%% 
	%%%%%%%%%%%%%%%%%%%   Phantom(LE2-Table-potential)   %%%%%%%%%%%%%%%%%%%%
	%\paragraph{Phantom surrounding field (with $k\lambda=4$) }
	\item Phantom surrounding field (with $k\lambda=4$)
		
		\begin{table}[!ht]
		\begin{center}
			\begin{tabular}{|l|l|c|l|}
				%{|lccll|}
				\hline
				region & pos.zeros & range of $\tilde{r}$ &  orbit \\
				\hline\hline
				I & 1 &
				$|$$--$$\lVert$$\bullet$$\textbf{--------}$$\lVert$$\textbf{-----------------------}$%$\dashrightarrow$
				
				& TEO
				\\  \hline
				II & 2 &
				$|$$--$$\lVert$$\bullet$$\textbf{--------}$$\lVert$$\textbf{---}$$\bullet$$-------$%$\dashrightarrow$
				& MBO
				\\ \hline
				III & 3 &
				$|$$--$$\lVert$$\bullet$$\textbf{--------}$$\lVert$$\textbf{---}$$\bullet$$--$$\bullet$$\textbf{--------------}$%$\dashrightarrow$
				& MBO, EO
				\\ \hline
				IV & 4 &
				$|$$--$$\lVert$$\bullet$$\textbf{--------}$$\lVert$$\textbf{---}$$\bullet$$--$$\bullet$$\textbf{-----}$$\bullet$ $---$
				& MBO, BO
				\\ \hline
			\end{tabular}
			\caption{Types of orbits of the phantom surrounding field. The lines represent the range of the orbits. The dots
				show the turning points of the orbits. The positions of the two horizons are marked by a vertical double line. The
				single vertical line indicates the singularity at $\tilde{r}=0$.}
			\label{tab:RP}
		\end{center}
	\end{table}

	\begin{figure}[!ht]
		\centering
		\subfigure[]{
			\includegraphics[width=0.2\textwidth]{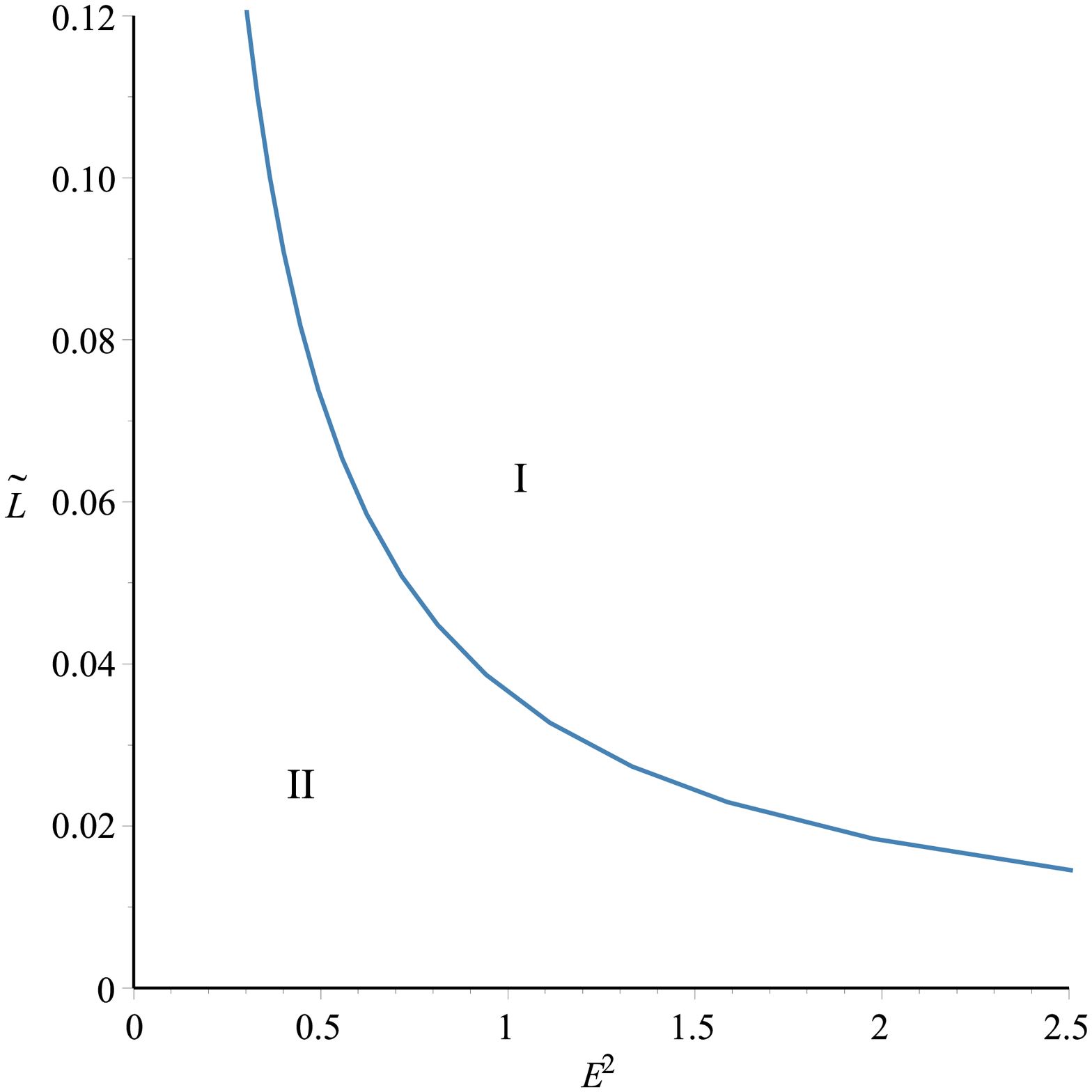}
		}
		\subfigure[]{
			\includegraphics[width=0.2\textwidth]{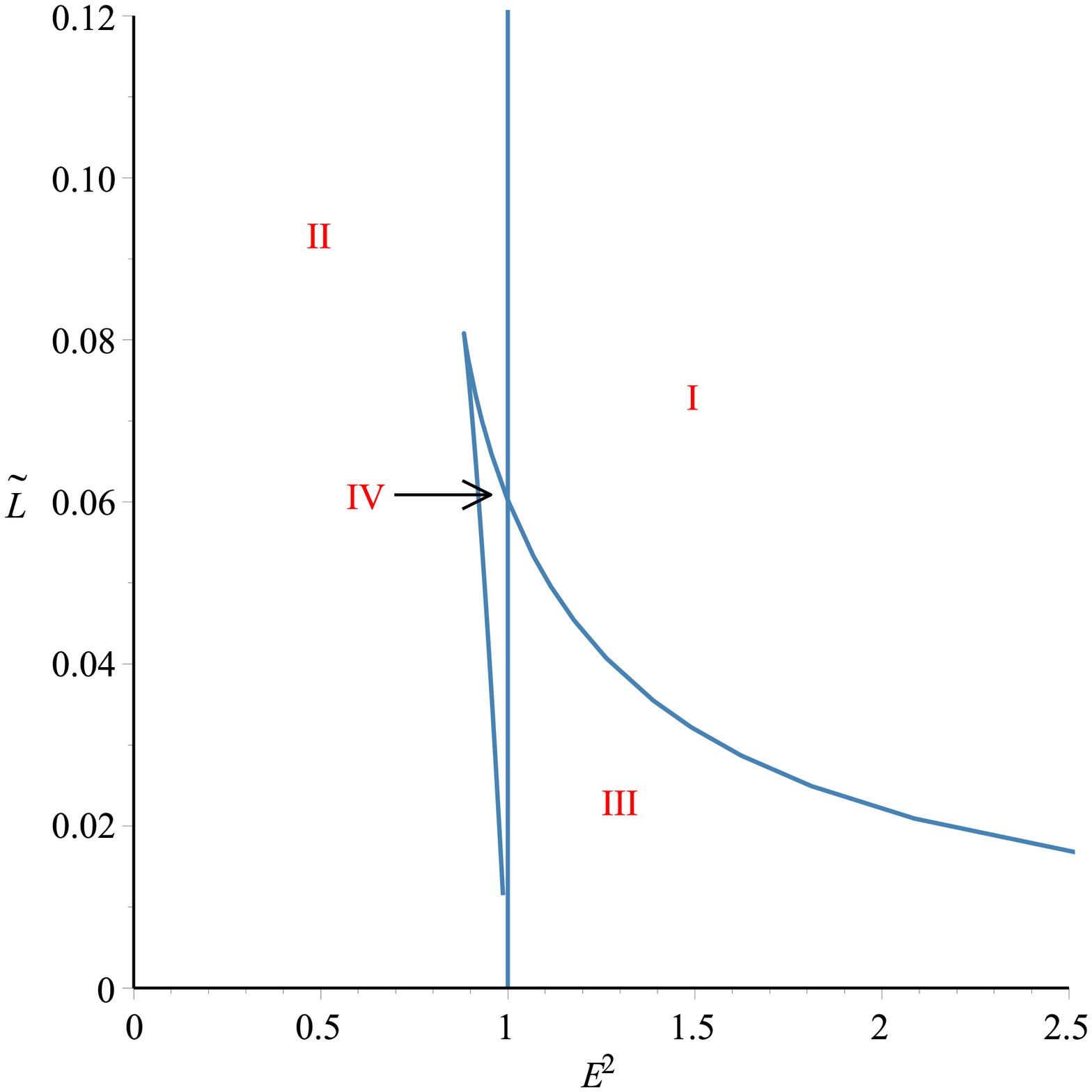}
		}
		\caption{ Plots of $L-E^2$ diagram and region of different types of geodesic motion 
			in the phantom surrounding field with the parameters $k\lambda=4$, $\tilde{N}=0.1$ 
			and $\tilde{Q}=\sqrt{0.25}$ (a): Null and (b): Timelike geodesics. 
			The numbers of positive real zeros in these regions are: I=1, II=2, III=3, IV=4.    
		}
		\label{pic:RPb}
	\end{figure}
	%\clearpage 
	%%%%%%%%%%%%%%%%%%%%
	\begin{figure}[!ht]
		\centering
		\subfigure[$\epsilon=1$, $\tilde{N}=0.1$, $\tilde{Q}=\sqrt{0.25}$, $\tilde{L}=0.07$, $E=\sqrt{1.5}$]{
			\includegraphics[width=0.17\textwidth]{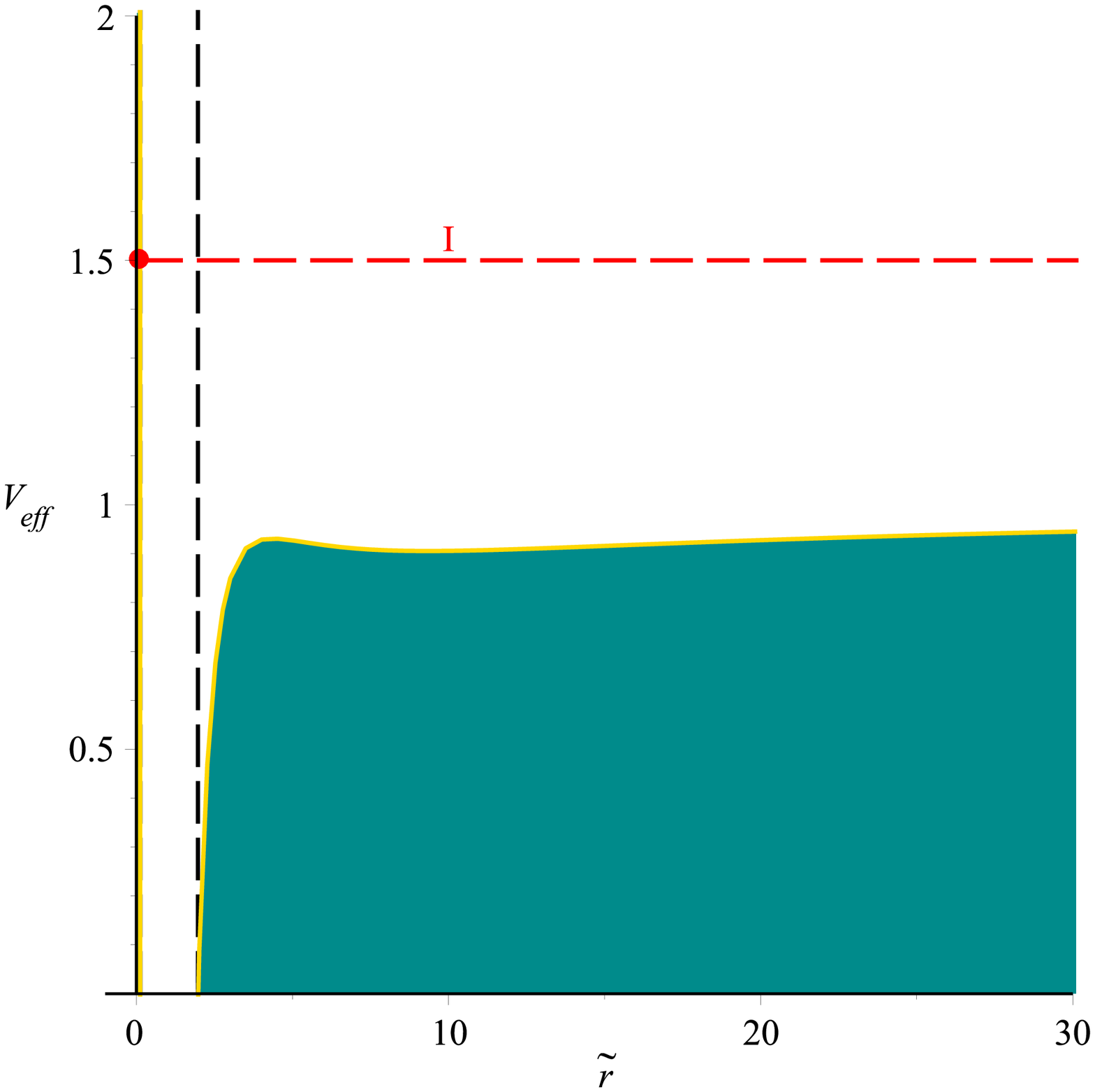}
		}
		\subfigure[$\epsilon=1$, $\tilde{N}=0.1$, $\tilde{Q}=\sqrt{0.25}$, $\tilde{L}=0.075$, $E=\sqrt{0.7}$]{
			\includegraphics[width=0.17\textwidth]{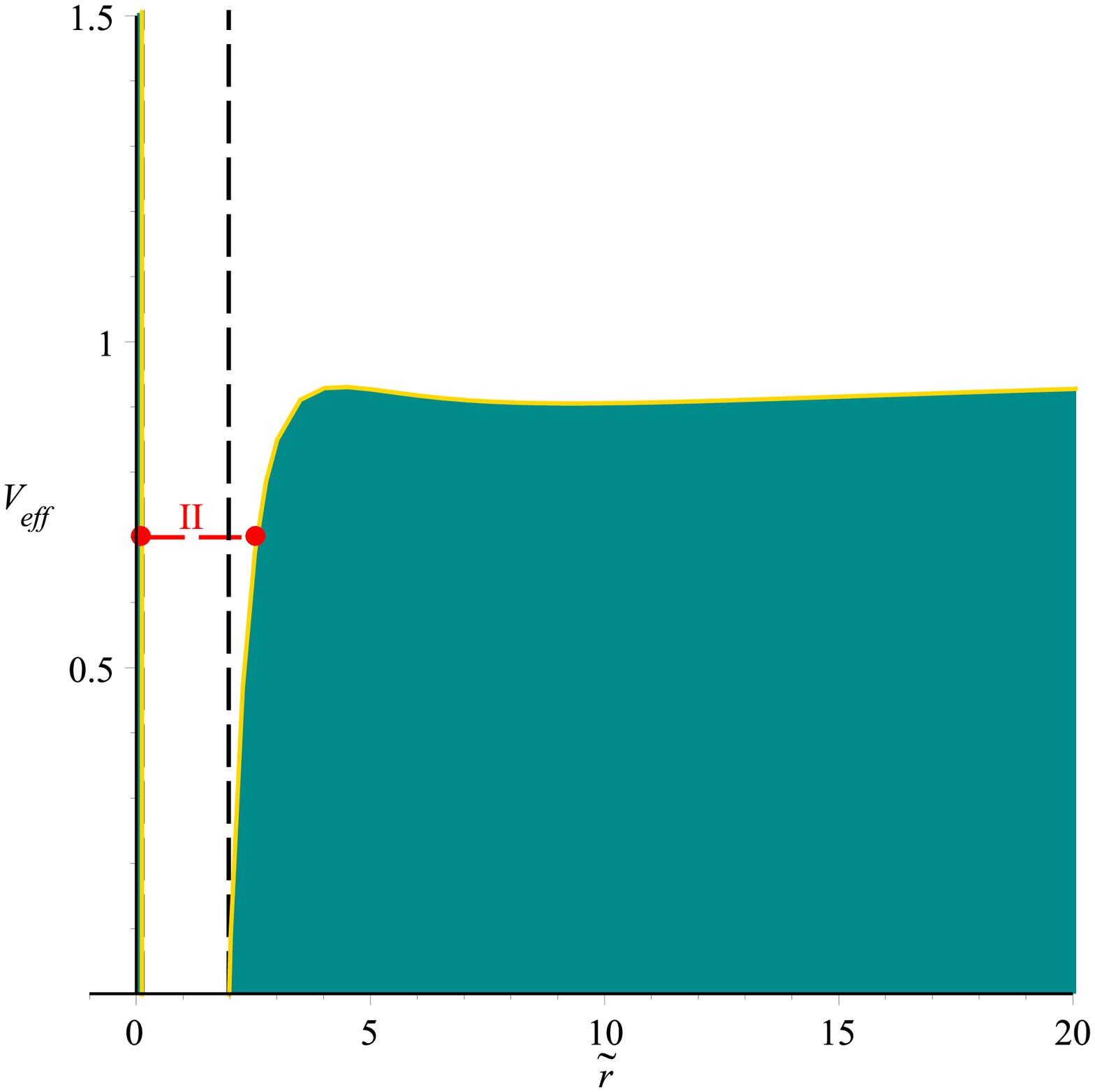}
		}
		\subfigure[$\epsilon=1$, $\tilde{N}=0.1$, $\tilde{Q}=\sqrt{0.25}$, $\tilde{L}=0.01$, $E=\sqrt{1.75}$]{
			\includegraphics[width=0.17\textwidth]{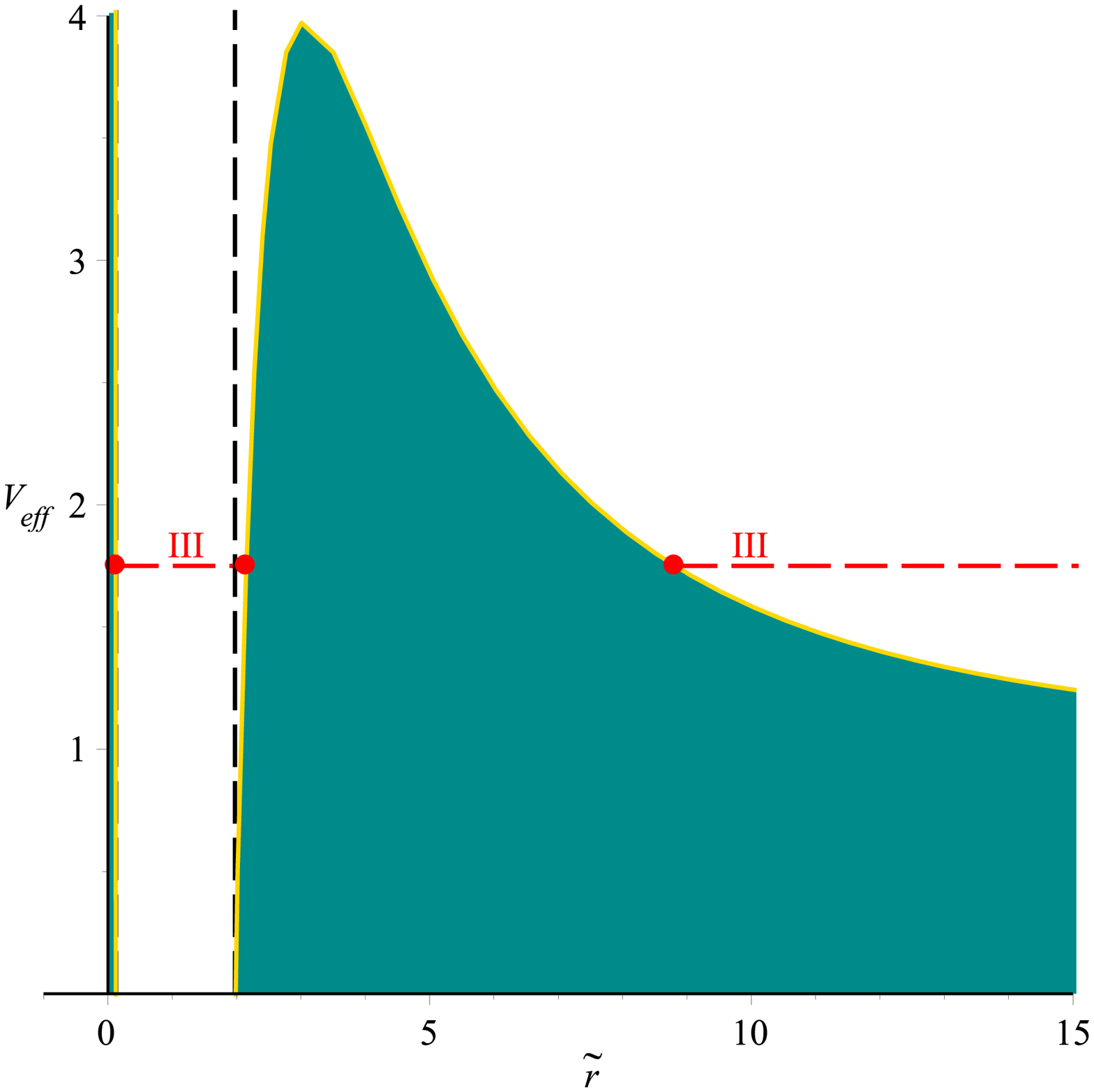}
		}
		\subfigure[$\epsilon=1$, $\tilde{N}=0.1$, $\tilde{Q}=\sqrt{0.25}$, $\tilde{L}=0.06$, $E=\sqrt{0.93}$]{
			\includegraphics[width=0.17\textwidth]{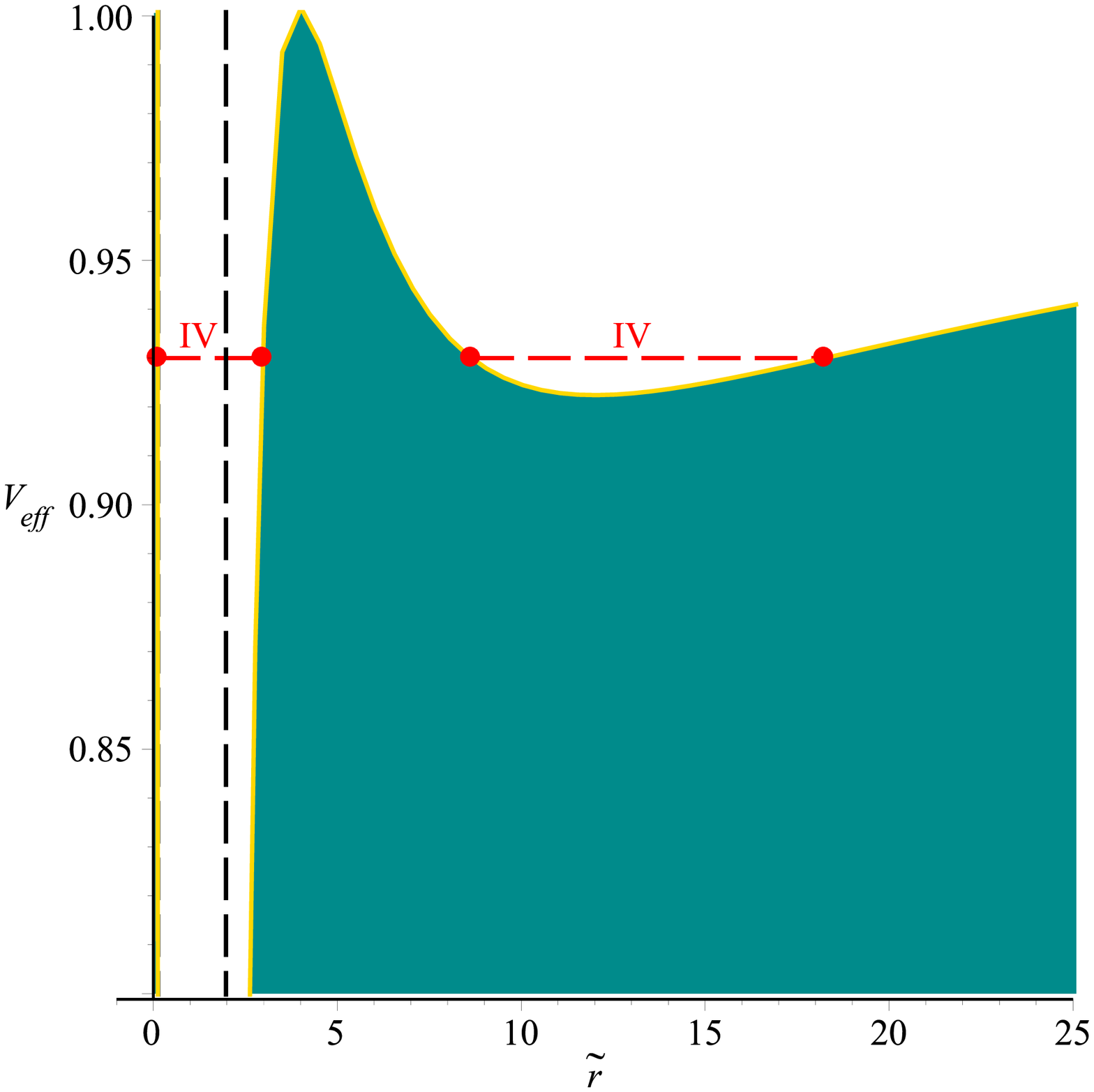}
		}
		\subfigure[$\epsilon=0$, $\tilde{N}=0.1$, $\tilde{Q}=\sqrt{0.25}$, $\tilde{L}=0.06$]{
			\includegraphics[width=0.17\textwidth]{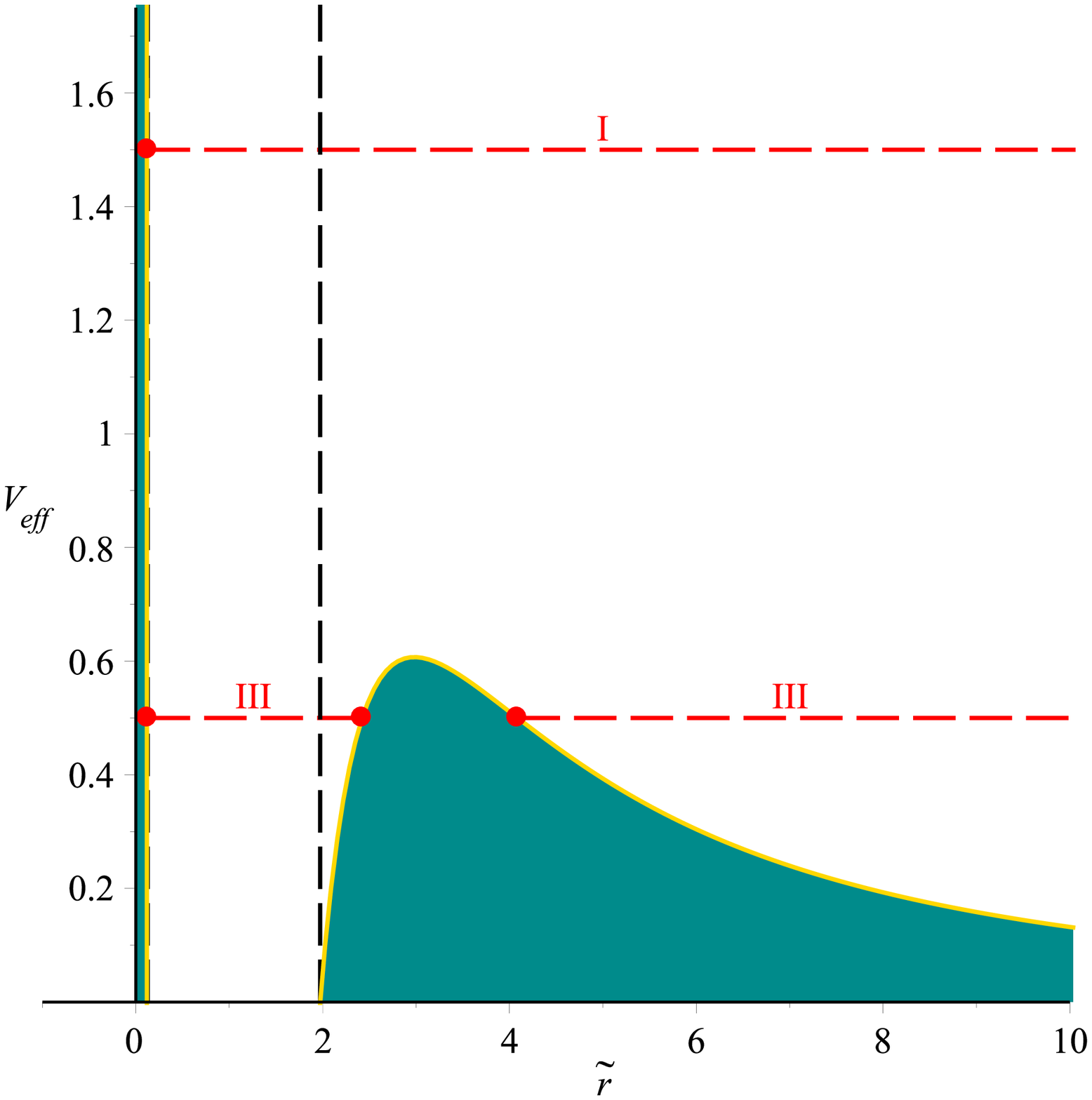}
		}
		\caption{\footnotesize Plots of the effective potential for the different orbit types of Table \ref{tab:RP}, 
			for the case of phantom surrounding field with the parameters $k\lambda=4$. 
			The horizontal red-dashed lines denotes the squared energy parameter $E^2$. 
			The vertical black dashed lines show the position of the horizons. 
			The red dots marks denote the zeros of the polynomial $R$, which are the 
			turning points of the orbits. In the cyan area, no motion is possible since $\tilde{R} < 0$ .}
		\label{pic:RP2}
	\end{figure}
	\clearpage
	%%%%%%%%%%%%%%%%%%%%%%%

	%\clearpage
\end{enumerate}
%%%%%%%%%%%%%%%%%%%%%%%%%%%%%%%%%%%%%%  Orbits
%%%%%%%%%%%%%%%%%%%le2rn
In all surrounding fields (quintessence, dust, radiation, cosmological constant and phantom fields), 
when Rastall geometric parameter becomes zero, the results are reduce to a Reissner-Nordstr\"om black hole
(see Fig. \ref{pic:R0} (a) and table \ref{tab:RN}) and when both electric charge and Rastall geometric
parameter become zero, the metric and results are same as a 
Schwarzschild black hole (see Fig. \ref{pic:R0} (b) and table \ref{tab:Sh}) as our expectation. 
In addition the possible types of orbits for a Reissner-Nordstr\"om black hole 
are BO, EO, TEO and MBO while for a Schwarzschild black hole are TO, EO and BO. 
However by comparing between table \ref{tab:Sh} and tables of all other cases 
(\ref{tab:RQ1}-\ref{tab:RN}), we can be seen that terminating orbit (TO) has appeared, in which  
test particles come from certain point and fall into singularity of a Schwarzschild black hole,
while this orbit cannot appear for all other discussed cases included electrical charged. 
%%%%%%%%%%%%%%%%%%
\begin{figure}[!ht]
	\centering
	\subfigure[]{
		\includegraphics[width=0.18\textwidth]{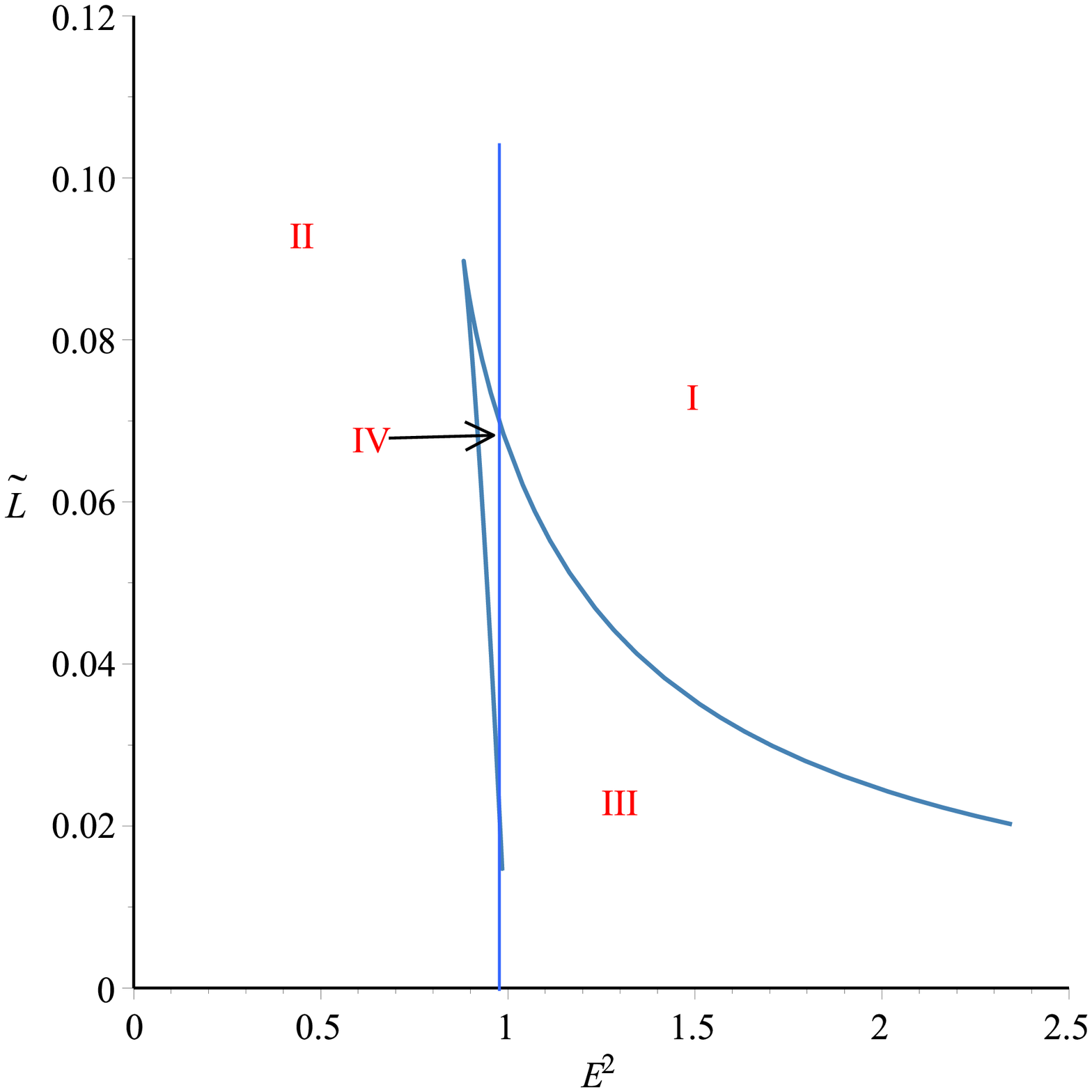}
	}
	\subfigure[]{
		\includegraphics[width=0.18\textwidth]{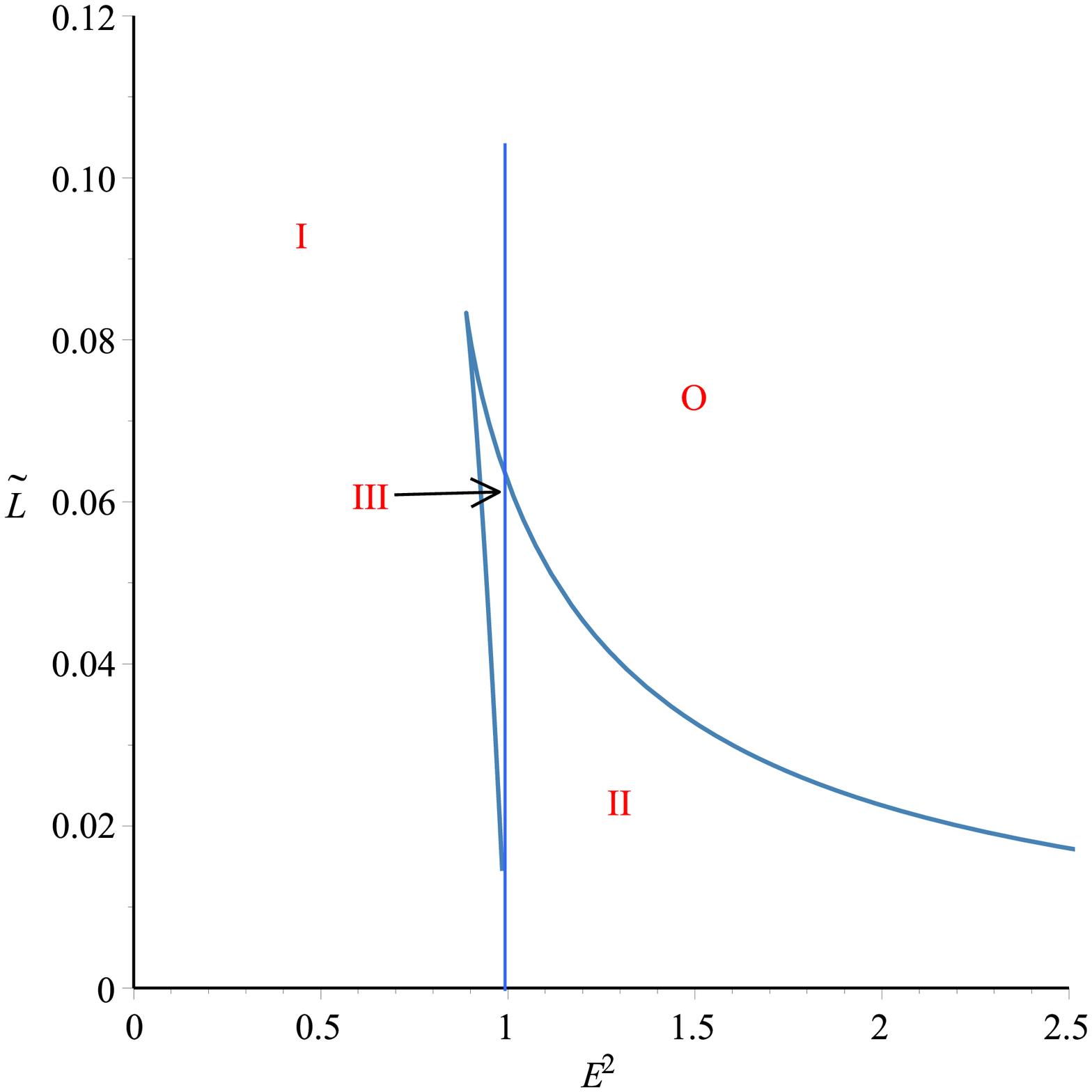}
	}
	\caption{ Plots of $L-E^2$ diagram and region of different types of geodesic motion 
		(a): Reissner-Nordstr\"om ($N=0$) and (b): Schwarzschild ($N=Q=0$) geodesics. 
		The numbers of positive real zeros in these regions are: O=0, I=1, II=2, III=3, IV=4.    
	}
	\label{pic:R0}
\end{figure}
%\clearpage
%%%%%%%%%%%%%%%%new tab 2
\begin{table}[!ht]
	\begin{center}
		\begin{tabular}{|l|l|c|l|}
			%{|lccll|}
			\hline
			region & pos.zeros & range of $\tilde{r}$ &  orbit \\
			\hline\hline
			I & 1 &
			$|$$--$$\lVert$$\bullet$$\textbf{--------}$$\lVert$$\textbf{-----------------------}$%$\dashrightarrow$
			
			& TEO
			\\  \hline
			II & 2 &
			$|$$--$$\lVert$$\bullet$$\textbf{--------}$$\lVert$$\textbf{---}$$\bullet$$-------$%$\dashrightarrow$
			& MBO
			\\ \hline
			III & 3 &
			$|$$--$$\lVert$$\bullet$$\textbf{--------}$$\lVert$$\textbf{---}$$\bullet$$--$$\bullet$$\textbf{--------------}$%$\dashrightarrow$
			& MBO, EO
			\\ \hline
			IV & 4 &
			$|$$--$$\lVert$$\bullet$$\textbf{--------}$$\lVert$$\textbf{---}$$\bullet$$--$$\bullet$$\textbf{-----}$$\bullet$ $---$
			& MBO, BO
			\\ \hline
		\end{tabular}
		\caption{Types of orbits of a Reissner-Nordstr\"om black hole. The lines represent the range of the orbits. The dots
			show the turning points of the orbits. The positions of the two horizons are marked by a vertical double line. The
			single vertical line indicates the singularity at $\tilde{r}=0$.}
		\label{tab:RN}
	\end{center}
\end{table}
%\clearpage
%%%%%%%%%%%%%%%%new tab 2
%%%%%%%%%%%%%%%%new tab 1
\begin{table}[!ht]
	\begin{center}
		\begin{tabular}{|l|l|c|l|}
			%{|lccll|}
			\hline
			region & pos.zeros & range of $\tilde{r}$ &  orbit \\
			\hline\hline
			O & 0 &
			$|$$\textbf{--------------------------------}$%$\dashrightarrow$
			
			& TO
			\\  \hline
			I & 1 &
			$|$$\textbf{------}$$\lVert$$\textbf{---}$$\bullet$$--------$ %$\dashrightarrow$
			& TO
			\\ \hline
			II & 2 &
			$|$$\textbf{------}$$\lVert$$\textbf{---}$$\bullet$$----$$\bullet$$\textbf{---------}$
			& TO, EO
			\\ \hline
			III & 3 &
			$|$$\textbf{------}$$\lVert$$\textbf{---}$$\bullet$$----$$\bullet$$\textbf{-----}$$\bullet$$--$
			& TO, BO
			\\ \hline
		\end{tabular}
		\caption{Types of orbits of a Schwarzschild black hole. The lines represent the range of the orbits. The dots
			show the turning points of the orbits. The positions of the two horizons are marked by a vertical double line. The
			single vertical line indicates the singularity at $\tilde{r}=0$.}
		\label{tab:Sh}
	\end{center}
\end{table}
\clearpage
%%%%%%%%%%%%%%%%%%%%%%
\subsection*{Examples of trajectory of particle motion and light}

\clearpage 
%%%%%%%%%%%%%%%%%%%%
%%%%%%%%%%%%%%%%%%%%%%%%%%%%%%%%%%%%%%%%%%%%%%
\section{CONCLUSIONS}\label{conclusions}
In this paper, the analytical solution of geodesic motion for massless and massive test particles 
in the vicinity of a black hole space-time surrounded by perfect fluid in the context of Rastall gravity has been presented. 
We have studied the timelike and null geodesics equations of motion for the black hole surrounded by cases quintessence, 
dust, radiation, cosmological constant and phantom fields in detail. 
In each case, we obtain the analytical solutions %of geodesic motion 
by regarding and constraint on effective state parameter $\omega_{eff}$ and 
considering some possible values of Rastall coupling constant $k\lambda$, 
so that equation of motion %$f(r)$ in Eq. (\ref{fsr}) have 
included integer powers of $r$ and also % Eq. (\ref{r5}) 
can be solved analytically. \\
%%%%%%%%%%%%%%%%%%%%%%%%%
For a black hole surrounded by quintessence field with $\omega_q=-\frac{2}{3}$, 
two possible cases of $k\lambda$ ($k\lambda=\frac{1}{4}$, $k\lambda=-2$) analysed. 
For the case $k\lambda=-2$, the equation of motion have included the term $\frac{N_q}{r}$ 
as Rastall’s correction term in which here 
the field structure parameter $N_q$ can play the role of the mass of black hole,
%%%%%%%%%%%%%%%
but with the case $k\lambda=\frac{1}{4}$, the term $N_qr^2$ has appeared in which 
the filed structure parameters $N_q$, can describe the acceleration of the Universe.
%parameter $N_s$ To $N_s+2$ 
%if change field structure parameter $N_s\rightarrow N_s+2$ equation of motion in GR 
%%%%%%%%%%%%%%%%%%%%%%
When dust field has considered as a background fluid with $\omega_d=0$, 
also two possible cases of $k\lambda$ ($ k\lambda=\frac{2}{9}, k\lambda=\frac{1}{4}$) 
has been analysed, in which for the case $k\lambda=\frac{2}{9}$, 
the equation of motion has contained Rastall correction term $N_d r$, in which here 
Rastall geometric parameter $N_d$, can play the role of small and 
large-scale physical evidence for the ranges of scalar curvature \cite{Soroushfar:2015wqa},
and for the case $k\lambda=\frac{1}{4}$, the Rastall’s correction term $N_d r^2$ 
has appeared in which field structure constant $N_d$, 
similar to the same case in quintessence filed, play the role of
accelerating expansion of the Universe. %$(N_dr,\;N_dr^2)$.
%%%%%%%%%%%%%%%%%%%%
%For a black hole surrounded by a radiation field 
When radiation is considered as a background field with $\omega_r=\frac{1}{3}$,
the metric is Reissner-Nordstr\"om metric of a black hole,
which the field structure parameter $N_r$ with the electric charge $Q$, 
behave role of the effective charge of this black hole $(\sqrt{Q^2-N_r})$. %($\frac{N_r+Q^2}{r^2}$).\\ %same as in GR. \\
%%%%%%%%%%%%%%%%%%%%%%%%
When we consider a black hole is surrounded by cosmological constant field with $\omega_c=-1$, 
the metric is same as Kieslev metric in GR which was achieved previously. %obtained already in GR by Kieslev
Therefore the term $N_c r^2$ as Rastall correction term will appear, 
in which the Rastall geometric parameter %field structure parameter 
$N_c$ causes the role of accelerating expansion of the Universe, 
%same as in GR equations theories 
same as quintessence and also dust field.  \\
%the Rastall’s correction term $(N_c)$
%%%%%%%%%%%%%%%%%%%%%%
Finally for a black hole surrounded by the phantom field with $\omega_p=-\frac{4}{3}$, 
three cases of $k\lambda$ ($k\lambda=\frac{1}{4}, k\lambda=\frac{2}{3},k\lambda=4$) 
are possible values which have been analysed.  
For two cases ($k\lambda=\frac{1}{4}, k\lambda=\frac{2}{3}$), 
Rastall correction term has appeared in terms of $N_p r$ and $N_p r^2$,
which in the term $N_p r$, Rastall geometric parameter $N_p$ can play the role of small and 
large-scale physical evidence for the ranges of scalar curvature \cite{Soroushfar:2015wqa} 
and in the term $N_p r^2$, field structure parameter $N_p$ play the role of accelerating expansion of the Universe,
but for the case $k\lambda=4$, field structure parameter $N_p$ behaves 
mass role i.e. $\frac{1}{r}$ same as the quintessence field. 
After reviewing the space-time and the corresponding equations of motion,
we classified the complete set of orbit types for massive and 
massless test particles moving on geodesics for each case. 
In addition, it has shown in table \ref{tab:RN} when Rastall geometric parameter  
vanish, the metrics reduce to Reissner-Nordstr\"om black holes, 
while it has shown in table \ref{tab:Sh} when the electric charge of a black hole 
becomes zero the results decrease to Schwarzschild. \\ 
%%%%%%%%%%%%%%%%%%%%%%%%%%
%%%%%%%%%%%%%%%%%%%%%%%%%%
The geodesic equations solved analytically by
Weierstrass elliptic and derivatives of hyperelliptic Kleinian sigma functions. 
We also considered all possible types of orbits.
%%%%%%%%%%%%%%%%%%%%%%
Using effective potential techniques and parametric diagrams,
the possible types of orbits were derived.
For null geodesics EO, TEO and MBO are possible,
while for timelike geodesics EO, TEO, BO and MBO are possible.

%\begin{acknowledgments}
%This work has been supported financially by the Research Institute for Astronomy and
%Astrophysics of Maragha (RIAAM).
%\end{acknowledgments}

%\bibliography{sample-paper}
%\bibliographystyle{prsty}

\end{document}